\documentclass[%
 reprint,
 amsmath,amssymb,
 aps,
]{revtex4-2}

\usepackage{graphicx}
\usepackage{dcolumn}
\usepackage{bm}

\newcommand{\RNum}[1]{\uppercase\expandafter{\romannumeral #1\relax}}
\newcommand{\bef}{\begin{figure}}      
\newcommand{\eef}{\end{figure}}      
\newcommand{\bea}{\begin{eqnarray}}    
\newcommand{\eea}{\end{eqnarray}}      
\newcommand{\be}{\begin{equation}}      
\newcommand{\ee}{\end{equation}}

\graphicspath{{./}{figures/}}

\begin{document}


\title{The Hidden Role of Anisotropies in Shaping Structure Formation in Cosmological N-Body Simulations}

\author{Francesco  Sylos Labini}
\affiliation{Centro  Ricerche Enrico Fermi, Via Panisperna 89a, 00184 Rome, Italy} 
\affiliation{Istituto Nazionale Fisica Nucleare, Unit\`a Roma 1, Dipartimento di Fisica, Universit\'a di Roma ``Sapienza'', 00185 Rome, Italy}
\email{sylos@cref.it}
\date{\today} 

\begin{abstract}
Initial conditions in cosmological $N$-body simulations are typically generated by displacing particles from a regular cubic lattice using a correlated field derived from the linear power spectrum, often via the Zel'dovich approximation. While this procedure reproduces the target two-point statistics (e.g., the power spectrum or correlation function), it introduces subtle anisotropies due to the underlying lattice structure.
These anisotropies, invisible to angle-averaged diagnostics, become evident through directional measures such as the Angular Distribution of Pairwise Distances. Analyzing two Cold Dark Matter simulations with {  varying resolutions and box sizes}, we show that these anisotropies are not erased but are amplified by gravitational evolution. They seed filamentary structures that persist into the linear regime, remaining visible even at redshift $z = 0$.
Our findings demonstrate that such features are numerical artifacts --- emerging from the anisotropic coupling between the displacement field and the lattice --- not genuine predictions of an isotropic cosmological model. These results underscore the importance of critically reassessing how initial conditions are constructed, particularly when probing the large-scale, quasi-linear regime of structure formation.
\\

\end{abstract}


\maketitle


In cosmological $N$-body simulations of Cold Dark Matter (CDM) models, the present-day matter distribution is highly inhomogeneous on small scales --- such as those of galaxies and galaxy clusters ---  but becomes statistically homogeneous and isotropic on sufficiently large scales \citep{Davis_etal_1985}.
A central and longstanding question in cosmology is whether the large-scale structures observed in galaxy redshift surveys --- extending across tens to hundreds of megaparsecs --- are consistent with the predictions of $\Lambda$CDM $N$-body simulations \citep{Huchra_Geller_1982,Gott_etal_2005,DeMarzo_etal_2021,Lopez_etal_2022,Lopez_etal_2024}.

Numerous studies have claimed that CDM simulations, under the standard $\Lambda$CDM framework, successfully reproduce the large-scale structures observed in the Universe. The evolved mass distribution exhibits a rich and interconnected network of voids, walls, clusters, and filaments --- collectively known as the cosmic web \citep{White_etal_1987,Springel_etal_2005,Park_etal_2012,Sawala_etal_2025}. On scales exceeding several hundred Mpc, the distribution becomes increasingly featureless, approaching statistical homogeneity.

In both simulations and observations, large-scale structures are commonly identified using Friends-of-Friends algorithms \citep{Huchra_Geller_1982,Davis_etal_1985,Lopez_etal_2024,Sawala_etal_2025}, which group particles or galaxies into structures based on a fixed linking length. This algorithm, while simple and widely used, introduces a scale dependence in structure detection that warrants careful interpretation. Within this framework, it has been argued that structures spanning several hundred Mpc are consistent with CDM predictions --- a conclusion that appears puzzling in light of the underlying physics.

On small scales ($r \lesssim 10\,\mathrm{Mpc}/h$), non-linear gravitational collapse leads to the formation of virialized, quasi-spherical halos \citep{Navarro_etal_1997}. In this regime, the non-linear dynamics erase the memory of the initial conditions (ICs), so that the resulting structures predominantly reflect the bottom-up gravitational assembly characteristic of CDM-like structure formation, rather than preexisting features in the primordial density field.

On larger scales ($r \gtrsim 10\, \mathrm{Mpc}/h$), the dynamics remain in the linear or mildly non-linear regime, where structures such as filaments can form through the coherent amplification of initial density correlations. Notably, cosmological simulations exhibit filaments that extend far beyond 10 Mpc, in some cases reaching scales comparable to the entire simulation volume --- up to several hundred megaparsecs \citep{White_etal_1987,Springel_etal_2005,Park_etal_2012}. Despite their prominence, the physical origin of these extended structures is often not scrutinized in detail, raising important questions about their relation to ICs, finite-size effects, and numerical artifacts.

In standard CDM cosmology, particles typically travel no more than $\sim 10\, \mathrm{Mpc}/h$ from their initial positions during the simulation. This implies that structures larger than this scale must emerge from correlations already present in the ICs. At early times, the two-point correlation function $\xi(r)$ of the CDM density field exhibits a clear transition: it follows a power-law decay at small scales, while at intermediate scales ($10 \lesssim r \lesssim 100\, \mathrm{Mpc}/h$) it approaches an exponential decay \cite{book}. The characteristic scale of this exponential behavior --- usually defined as the correlation length in statistical physics \cite{book} --- is approximately $r_c \approx 18\, \mathrm{Mpc}/h$ \citep{SylosLabini_2011}. Beyond this scale, $\xi(r)$ {   has } negligible amplitude, and no long-range correlations or coherent structures of even small amplitude are expected to be amplified under linear gravitational evolution.

The two-point correlation function $\xi(r)$ crosses zero at approximately $120\; \mathrm{Mpc}/h$ and remains negative beyond this scale. This alternation between positive and negative correlations reflects the super-homogeneous nature of the distribution \cite{Gabrielli_etal_2002,Gabrielli_etal_2003,Torquato+Stillinger_2003,book,Torquato_2018,Philcox+Torquato_2023}. {  Super-homogeneous systems are characterized by suppressed mass fluctuations compared to a random (Poisson) distribution: the variance of mass within a spherical region grows proportionally to the surface area rather than the volume. Examples of super-homogeneity include all perfect crystals, perfect quasicrystals, and exotic amorphous states of matter  \cite{Torquato_2018}.}

While standard analyses typically focus on the non-linear amplification of $\xi(r)$ at small scales, comparatively little attention has been paid to the regime in which $\xi(r)$ remains positive but exhibits an exponential decay --- corresponding to the linear growth phase. Yet, this is precisely the range where filamentary structures are frequently observed in simulations. Although some of these features may result from chance alignments, potentially producing structures that extend far beyond the expected correlation length, their systematic appearance across different simulations suggests the presence of a more fundamental underlying mechanism.

We argue that such large-scale filamentary structures on scales larger than 10-20 Mpc$/h$ are artifacts of how the ICs are generated. In most $N$-body codes, ICs are produced by displacing particles from a regular lattice (the so-called pre-ICs) using a correlated Gaussian displacement field derived from the initial power spectrum (PS), often via the Zel'dovich approximation (see, e.g., \cite{Carlberg+Couchman_1989,Baugh_etal_1995,Couchman_etal_1995,Crocce_etal_2006} and references therein). {  While this procedure ensures that spherically averaged statistics, such as the PS $P(k)$ or the two-point correlation function $\xi(r)$, match the theoretical model reasonably well over a finite range of spatial scales, it simultaneously introduces subtle but non-trivial anisotropies into the particle distribution. These anisotropies arise from the intrinsic spatial correlations and geometric symmetries of the regular grid used as the pre-initial configuration. Crucially, they are not fully erased by the small-amplitude, correlated displacement field applied during the generation of the  ICs.
}
These residual anisotropies are typically not captured by angle-averaged statistics. However, they become evident when analyzed using directional statistics, such as the Angular Distribution of Pairwise Distances (ADPD), which is sensitive to anisotropic clustering. The ADPD provides a more complete diagnostic of the initial density field, revealing directional features that may seed spurious structures during gravitational evolution.

The assumption of statistical isotropy is central to the $\Lambda$CDM model. Yet, we show that commonly used methods for generating ICs subtly violate this assumption in a manner that impacts structure formation. In this work, we demonstrate that anisotropies arising from the coupling between an isotropic displacement field and the underlying lattice are not erased during evolution; instead, they persist and are amplified into filamentary structures that remain prominent even in the linear regime at the present epoch. These features are not transient fluctuations but long-lived imprints of the initial configuration. As they do not arise from the theoretical model, they must be regarded as numerical artifacts.

We emphasize that our focus is on a distinct feature --- namely, anisotropic patterns in the ICs --- as opposed to the issues more commonly addressed in the literature, which pertain to different kinds of discreteness effects. These effects concern how $N$-body simulations approximate the dynamics governed by the coupled Vlasov-Poisson equations in the limit of a large number of particles. Theoretically, dark matter is treated as a collisionless fluid, whose evolution is described by the Vlasov-Poisson system. In practice, however, $N$-body simulations represent this fluid using a finite set of discrete \textquotedblleft macroparticles\textquotedblright whose trajectories are evolved under gravitational interactions. The fundamental assumption is that this discrete particle system provides an accurate sampling of the continuous fluid behavior. Several discreteness-related issues have been explored in the literature, including correlations induced by the initial particle sampling \cite{Joyce+Marcos_2007a,Joyce+Marcos_2007b}, 
{  impact of particle noise on cosmological N-body simulations \cite{Romeo_etal_2008}, } two-body relaxation effects \cite{Binney+Knebe_2002,El-Zant_2006}, and dynamical discreteness effects captured by particle linear theory \cite{Marcos_etal_2006}. (In this context, \cite{Garrison_etal_2016} proposed corrections to mitigate small-scale discreteness effects by modifying the ICs themselves.)

{ 
The impact of artificial anisotropies introduced by pre-ICs has been systematically investigated in several recent studies. \cite{Masaki_etal_2021} quantified these effects by comparing cosmological simulations initialized from a regular grid with those based on a glass configuration. Their results demonstrated that glass-based pre-ICs yield more stable and accurate outcomes across a wide range of scales, consistently outperforming the standard grid-based approach.

Building on this, \cite{Racz_etal_2021} examined the anisotropy inherent in gravitational force calculations arising from the 3-torus topology imposed by periodic boundary conditions. To isolate this effect, they employed glass-like ICs, thereby eliminating grid-related artifacts. Their analysis suggests that the initial grid not only seeds small-scale preferential directions but may also amplify the intrinsic anisotropic features of the periodic gravitational field.

More recently, \cite{Yu_etal_2025} studied the residual anisotropy sourced by cubic lattice pre-ICs and showed that such anisotropies are not fully erased by nonlinear gravitational evolution. While the amplitude of the effect remains modest, the anisotropy signal was found to be statistically significant in the orientations of haloes and filaments. Taken together, these studies demonstrate that lattice-induced anisotropies can leave measurable imprints on the large-scale structure, even at late times. Moreover, they collectively support the conclusion that glass-based pre-ICs are more effective in mitigating such artificial anisotropies, reinforcing the case for their adoption in high-precision cosmological simulations.

In this work, we employ the ADPD as our primary tool to detect and quantify anisotropic features and we demonstrate that the anisotropy induced by cubic lattice pre-initial conditions persists over time and plays a role in the formation of large-scale structures.}  While the ADPD  has not yet been widely adopted in cosmological analyses, it directly quantifies directional anisotropies in pairwise correlations --- a statistical aspect that has been extensively studied using other approaches in the literature. For example, \cite{Sousbie_2011} introduced an algorithm based on persistent homology for detecting filamentary structures, where directional anisotropy is a central feature, whereas \cite{Tempel_etal_2012} analyzed pair orientations to study the alignment of galaxies within the large-scale structure. Similarly, \cite{Hamilton_1998} examined anisotropies in redshift-space distortions, providing foundational tools for analyzing directional correlations and \cite{Lee+Penn_2000} investigated angular correlations between galaxy spin axes, conceptually related to pairwise angular statistics. For a broader overview of techniques used to identify anisotropic features in simulations, see \cite{AragonCalvo_etal_2007,Colberg_etal_2008,Libeskind_etal_2017} who introduce various numerical methods for filament detection and the study of structure orientation.

{  This paper is structured as follows. In Sect.~\ref{lin_non_lin}, we begin by reviewing the key statistical properties that characterize CDM-like models at early times, with a particular focus on the distinction between linear and non-linear regimes. In Sect.~\ref{gen_ic}, we outline the commonly used procedures for generating ICs in cosmological \(N\)-body simulations, emphasizing the role of lattice-based pre-initial configurations and their implications. Sect.~\ref{methods} is devoted to describing our methodology for identifying anisotropies in point distributions; we also validate our approach using a series of controlled, artificial test cases. In Sect.~\ref{Anis_IC}, we turn to the quantification of the local anisotropy theoretically expected in cosmological ICs drawn from Gaussian random fields. We then apply this formalism in Sect.~\ref{results} to analyze two CDM simulations and present the main results of our study. Finally, in Sect.~\ref{conclusions}, we summarize our conclusions and discuss the broader implications of our findings for the generation of ICs and the interpretation of structure formation in cosmological simulations.
}

\section{Theoretical models}
\label{lin_non_lin}

\begin{figure*} 
\includegraphics[width=0.32\textwidth]{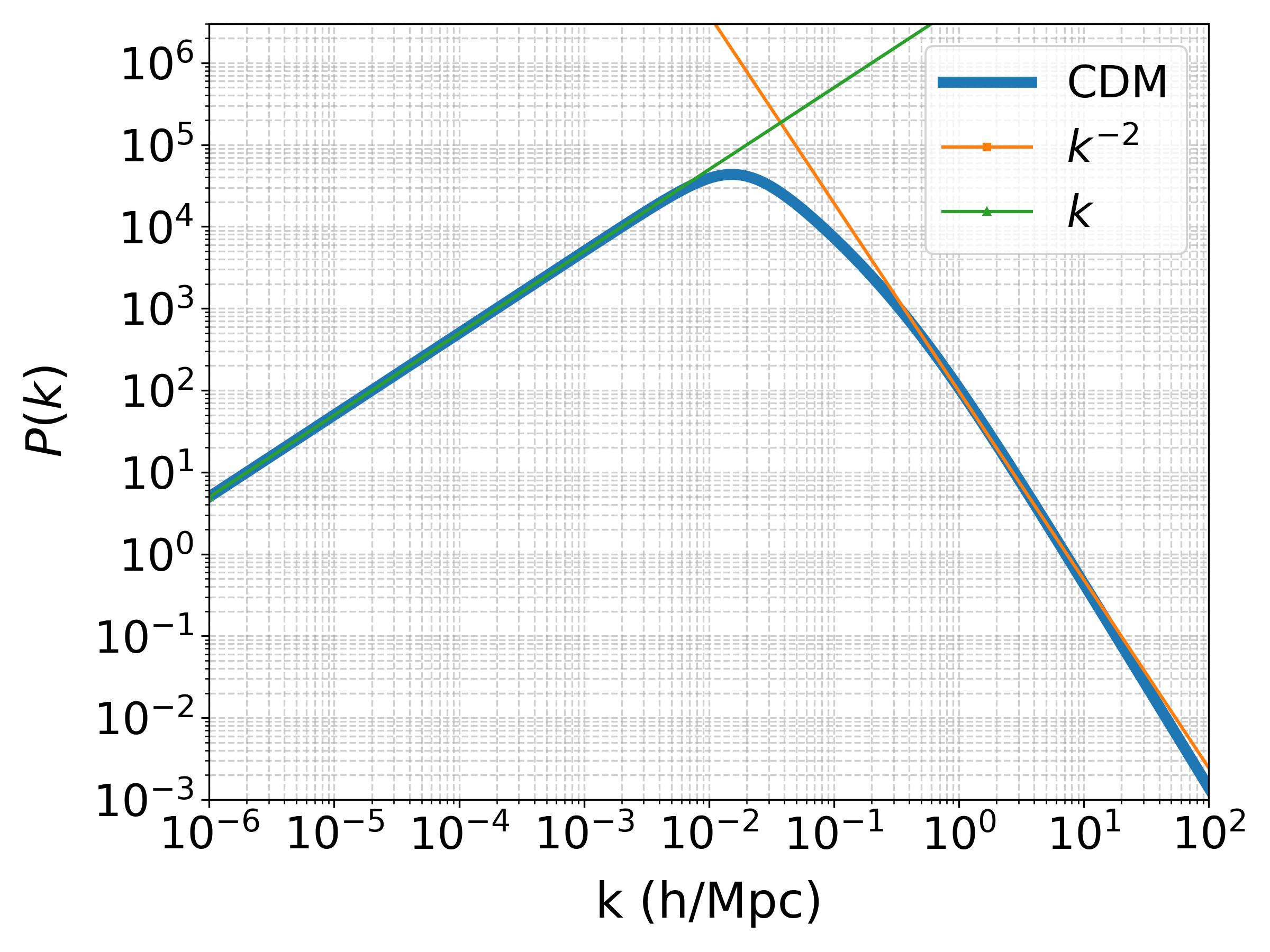}    
\includegraphics[width=0.32\textwidth]{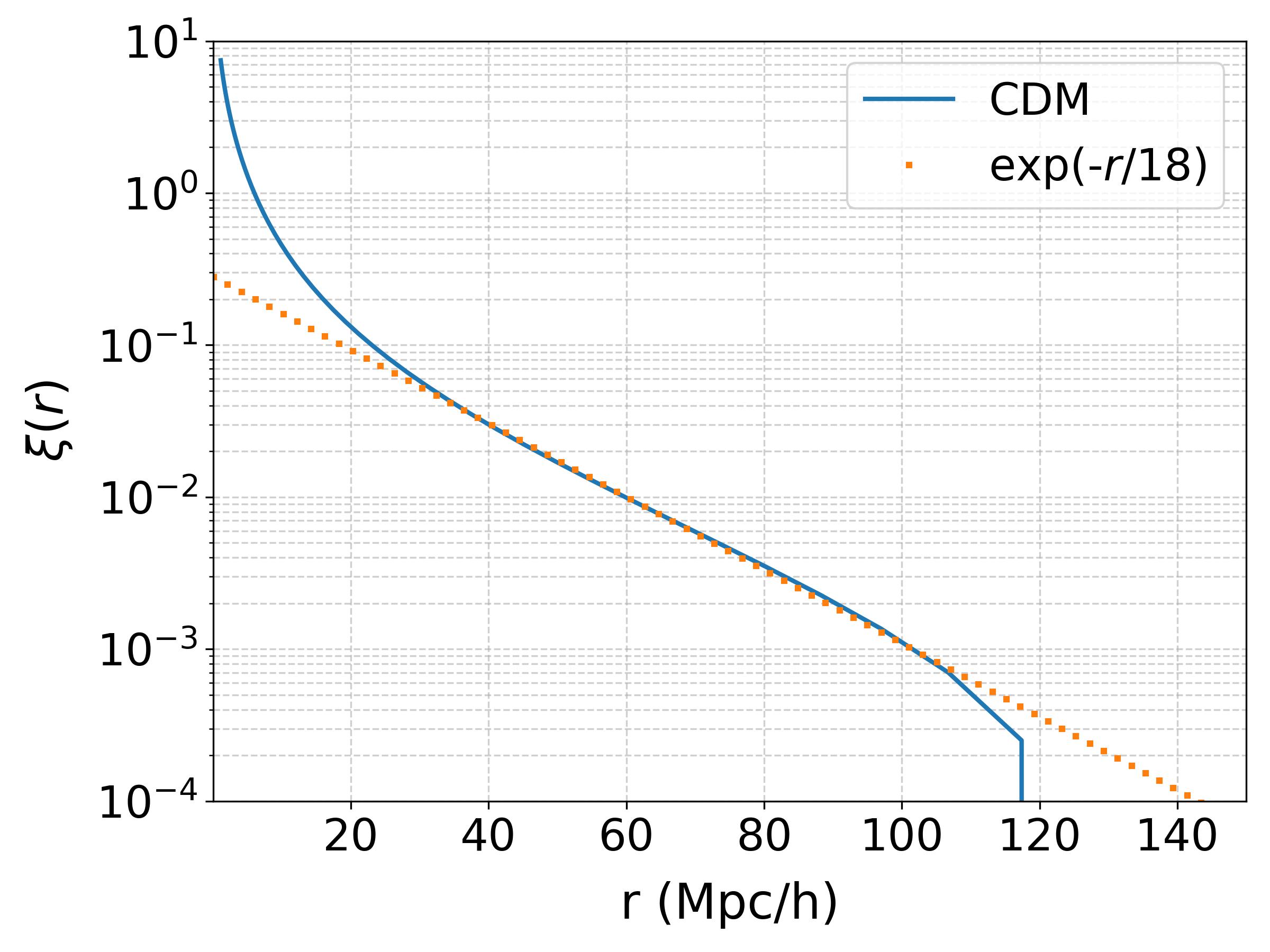}  
\includegraphics[width=0.32\textwidth]{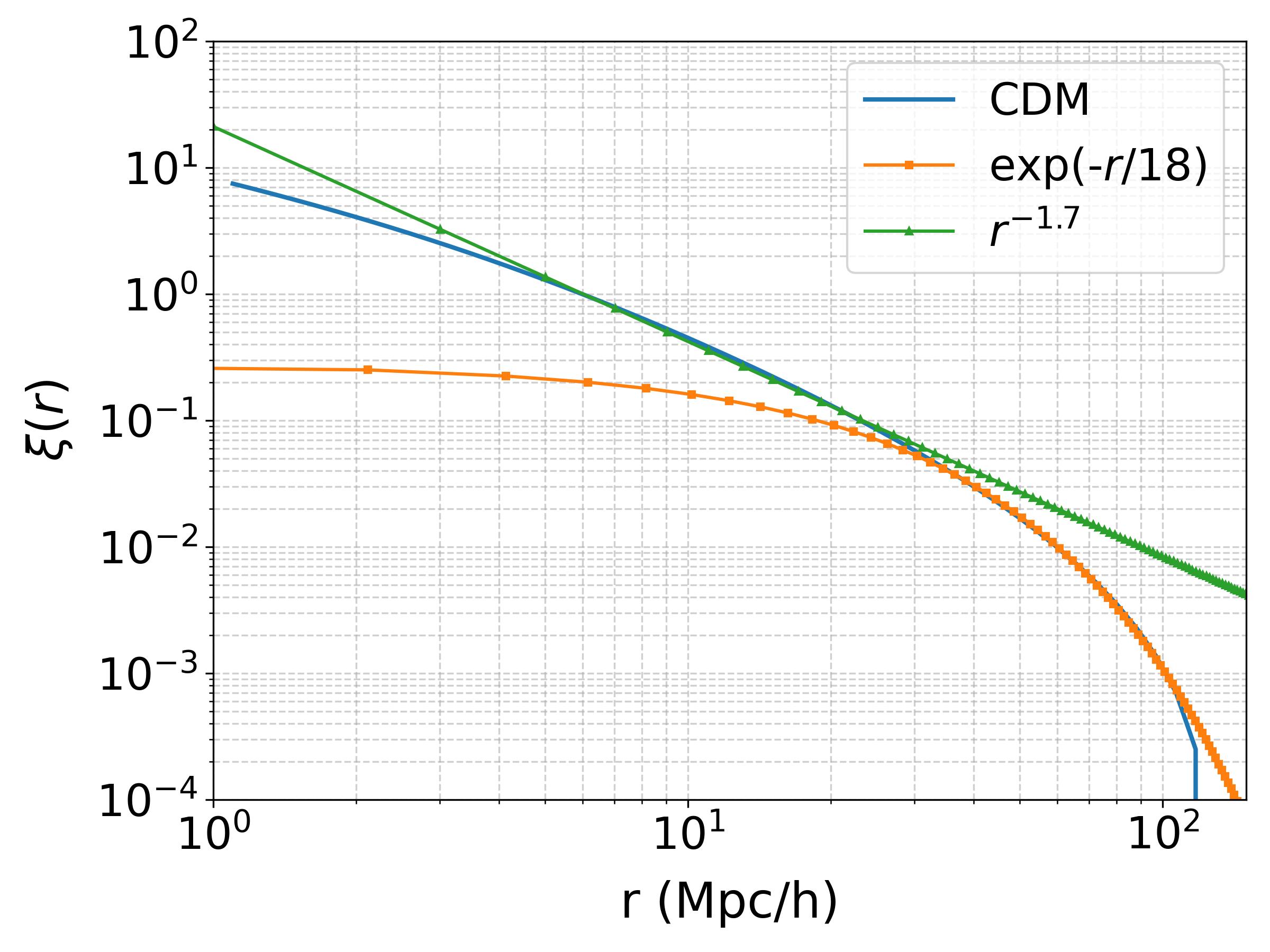}    
\caption{    
Power spectrum (left) and corresponding two-point correlation function in linear-log (middle) and log-log (right) plots for a typical early-time Cold Dark Matter model (see Eq.~\ref{pk}). The vertical red line marks the zero-crossing scale $r_t \approx 120 \mathrm{Mpc}$/$h$. Amplitudes are shown in arbitrary units.
} 
\label{vanilla} 
\end{figure*}

The correlation function of a typical early-time CDM model is shown in Fig.~\ref{vanilla}, along with the corresponding PS.  
The PS exhibits an approximate behavior given by \citep{SylosLabini_2011} 
\be
\label{pk} 
P(k) = A \frac{k}{1 + \left( \frac{k}{k_t} \right)^m},
\ee
{  where $A$ is a normalization constant, }the turnover scale is $k_t \approx 0.05$ $h$/Mpc, and the exponent at small scales (i.e., large $k$) is $m \approx 2.5$.  
{  
For the sake of clarity, we have adopted a PS without baryon acoustic oscillations (BAO), as these features are not essential to the main arguments of our work. Including BAO would merely introduce small-amplitude wiggles in the correlation function at scales around $100$~Mpc$/h$, without affecting our  conclusions.
}

At the initial time, the correlation function --- i.e., the Fourier conjugate of the PS $P(k)$ --- satisfies $\xi(r) < 1$ for all $r$, indicating that density fluctuations are small and remain in the linear regime. The functional form of the correlation function corresponding to a PS of the type given in Eq.~\ref{pk} can be computed numerically (see Fig.\ref{vanilla}). In summary {  it behaves } as follows: within the range $\sim 10$--$100\,\mathrm{Mpc}$/$h$, it exhibits an exponential decay of the form
\begin{equation}
\xi(r) \propto \exp\left(-\frac{r}{r_c}\right),
\end{equation}
with a characteristic scale of $r_c \approx 18\,\mathrm{Mpc}$/$h$. This scale corresponds to the typical size of correlated structures and it is usually defined to be the \textquotedblleft correlation length\textquotedblright in statistical physics (see \cite{book} and references therein).
On smaller scales ($r < 10\,\mathrm{Mpc}$/$h$), it follows a power-law behavior of the type
\[
\xi(r) \propto r^{-\gamma}
\]with 
$1< \gamma <2$.

At $r_t \approx 120\,\mathrm{Mpc}$/$h$, a scale corresponding approximately to $\sim k_t^{-1}$, 
 the correlation function crosses zero and remains negative at larger scales, so that it satisfies the overall constraint 
\begin{equation} 
P(0) = \int_0^\infty \xi(r)\, r^2\, \mathrm{d}r = 0 \;.
\end{equation}
Note that the condition $P(0) = 0$ is sometimes incorrectly attributed to mass conservation. In reality, it reflects a global constraint on the nature of the correlations, indicating that the overall distribution of CDM-like fluctuations is \emph{super-homogeneous} \citep{Gabrielli_etal_2002,Gabrielli_etal_2003,book} --- or \emph{hyper-uniform} 
\citep{Torquato+Stillinger_2003,Torquato_2018}.

{  Physically,  the meaning of the concept of super-homogeneous can be clarified by considering a general classification of stochastic processes in terms of their spectral properties. }
In a general classification of stochastic processes, consider the case where the PS behaves as $P(k) \propto k^n$ in the limit $k \rightarrow 0$. The corresponding mass variance, defined as
\[
\sigma^2(r) = \frac{\langle \Delta N \rangle^2}{\langle N \rangle^2} \propto r^{-\gamma},
\]
obeys the relation $\gamma = 3 + n$ for $-3 < n < 1$, and $\gamma = 4$ for $n > 1$.
The random (Poisson) case corresponds to a PS with $n = 0$, i.e., $P(k) \propto \text{const.}$, which leads to a mass variance that scales as $\sigma^2(r) \propto r^{-3}$. In contrast, the Harrison-Zel'dovich spectrum --- characteristic of the small-$k$ behavior of CDM-like models --- corresponds to $n = 1$, for which the mass variance behaves as $\sigma^2(r) \propto \log(r) \, r^{-4}$. 

{  
This behavior corresponds to the fact that mass fluctuations in a super-homogeneous system scale with the surface area of the sampling region, rather than its volume. 
}
This is in stark contrast to the Poisson case, where fluctuations scale with volume. The suppressed large-scale fluctuations in CDM-like models reflect the long-range order and enhanced uniformity characteristic of super-homogeneous fields. Thus, the correlation properties of the CDM theoretical model at early times can be summarized as follows:
\begin{enumerate} 
\item For $r < 10,\mathrm{Mpc}/h$, there are positive, but small in amplitude, power-law correlations: $P(k) \propto k^{-2}$ and $\xi(r) \propto r^{-1}$. In this regime, during time evolution up to the present epoch  non-linear gravitational clustering distorts the initial shape of $\xi(r)$ when its amplitude becomes of order unity, leading to a  change in the power-law index.
\item In the range $10 \lesssim r \lesssim 100\,\mathrm{Mpc}/h$, correlations remain positive but decay more rapidly than a power law and $\xi(r)<1$ up to the present epoch. An exponential function provides a good approximation of the behavior, with the characteristic scale of linear structures given by the correlation length $r_c \approx 18\,\mathrm{Mpc}/h$. As fluctuations remain small on these scales, gravitational dynamics only linearly amplify the initial density field and its correlations.
\item For $r \gtrsim 120\,\mathrm{Mpc}/h$, where the PS behaves as $P(k) \propto k$, the correlation function follows $\xi(r) \propto -r^{-4}$, indicating an anti-correlated regime and, consequently, the absence of large-scale structure. This range of scales also remain in the linear regime during the whole evolution.
\end{enumerate}

To provide a rough estimate of the size of non-linear structures formed in simulations, let us assume that a particle has a mean velocity of $250\alpha$ km s$^{-1}$ and that the timescale is approximately $10\beta$ Gyr.  
The typical distance a particle travels in such a time is
\[
\lambda \sim \frac{250 \alpha \times 10 \beta \times 3.15 \times 10^{16}}{3.1 \times 10^{19}} \sim 2.5\alpha\beta \;\; \mathrm{Mpc}.
\]
If we take $\beta = 1$ and $\alpha = 2$, which represents an upper limit, we find $\lambda \approx 5$ Mpc.

This length scale sets the size of a typical void, i.e., a region devoid of particles. It corresponds to the scale of non-linearity: for $r > \lambda$, the average density is well defined and fluctuations are small.  
Therefore, for $r > \lambda$, the correlation properties reflect those of the ICs, modulated only by a linear amplification due to gravitational clustering in an expanding universe.


\section{Generation of initial conditions}
\label{gen_ic}

\subsection{The Zel'dovich approximation}

The generation of ICs for cosmological $N$-body simulations typically involves two distinct steps~\citep{Efstathiou_etal_1985,Jenkins_1998}. 
First, a uniform distribution of particles is established to represent the unperturbed universe.  Second, density fluctuations with the desired statistical properties are imposed on top of this initial configuration.

By neglecting the correlations and fluctuations inherent in the pre-initial particle distribution, any desired linear fluctuation spectrum can, in principle, be generated using the \textit{Zel'dovich approximation} (see, e.g., \cite{Efstathiou_etal_1985} and references therein). 
In this framework, a displacement field $\vec{u}(\vec{r})$ is applied to a uniform initial background density field with mean density $\rho(\vec{r}) = \rho_0$. Assuming mass conservation under this transformation, the continuity equation yields:
\begin{equation}
\rho(\vec{x}) \, d^3x = \rho_0 \, d^3q,
\end{equation}
where $\vec{x} = \vec{q} + \vec{u}(\vec{q})$ defines the mapping from the Lagrangian (initial) coordinate $\vec{q}$ to the Eulerian (displaced) coordinate $\vec{x}$. 
This relation allows one to compute the evolved density field $\rho(\vec{x})$ in terms of the Jacobian determinant of the transformation, ultimately connecting the initial displacement field to the resulting density fluctuations:
\begin{equation}
\label{eq:continuity}
\rho(\vec{r}) - \rho_0 + \vec{\nabla} \cdot \left( \rho_0 \vec{u}(\vec{r}) \right) = 0 \;.
\end{equation}

Assuming statistical homogeneity and isotropy of the displacement field $\vec{u}(\vec{r})$, one can transform Eq.~\eqref{eq:continuity} into Fourier space.
Taking the expectation value of the square modulus of the density fluctuations, i.e., the PS, one obtains:
\[
P(k) = \langle |\delta_k|^2 \rangle = k^2 \langle |\vec{u}(\vec{k})|^2 \rangle = k^2 P_u(k) \;.
\]
Since the PS $P_u(k)$ must be integrable at small $k$, i.e., it cannot diverge faster than $k^{-3}$, we must have:
\[
\lim_{k \rightarrow 0} P(k) \propto k^n \quad \text{with} \quad n > -1 \;.
\]
This implies that by applying a suitably correlated displacement field to a uniform density background, one can construct a fluctuation field with a PS
 of the form $P(k) \propto k^n$ for $n \le -1$.

\subsection{Pre-initial conditions}

\noindent
In the generation of ICs for cosmological $N$-body simulations, fluctuations and correlations associated with the \textit{pre-initial} configuration are often overlooked, or implicitly assumed to be negligible relative to those introduced by the imposed perturbations. 

As discussed in the introduction, a number of studies have investigated the so-called ``discreteness effects''  
{ 
\cite{Baertschiger_SylosLabini_2002,Binney+Knebe_2002,Joyce_etal_2005,Marcos_etal_2006,El-Zant_2006,Joyce+Marcos_2007a,Romeo_etal_2008,Garrison_etal_2016,Masaki_etal_2021,Racz_etal_2021,Yu_etal_2025}}.
In this work, we focus on a related but distinct issue: whether, from the perspective of statistical characterization, the ICs provide a fair realization of the underlying theoretical model. We show that the answer to this problem depends critically on the statistical diagnostics employed. While standard measures such as the PS and two-point correlation function offer angle-averaged information, they may fail to capture directional features that violate statistical isotropy. Our aim is to assess these subtle anisotropies using more sensitive directional statistics.

Specifically, we focus on the widely adopted choice for the pre-initial particle distribution, in which an unperturbed universe is represented by a regular cubic lattice of particles. Although this configuration appears spatially uniform, a perfect lattice is not devoid of structure: it exhibits non-trivial, anisotropic correlations and fluctuations across all scales. These intrinsic features of the lattice can interact nonlinearly with the imposed displacement field, potentially introducing systematic artifacts into the ICs.

In particular, it can be shown that the unconditional mass variance within spheres of radius $r$ in a perfect lattice in three dimensions scales as \cite{book}:
\begin{equation}
\sigma^2(r) \propto \frac{1}{r^4},
\end{equation}
a behavior markedly different from that of a Poisson distribution, for which $\sigma^2(r) \propto 1/r^3$.  This slow decay of fluctuations in the lattice reflects its long-range order --- i.e., it is a super-homogeneous distribution --- and its inherent directional coherence. As we will show, these features can introduce spurious anisotropies into the ICs and significantly impact the subsequent evolution of structure in $N$-body simulations.

Moreover, the two-point correlation function for a lattice satisfies 
\[
\xi(\vec{r}_1, \vec{r}_2) = \xi(\vec{r}_1 - \vec{r}_2) \neq \xi(|\vec{r}_1 - \vec{r}_2|) \;, 
\] 
indicating that it lacks rotational invariance~\citep{book}. In other words, a lattice explicitly breaks spatial isotropy.

{  As we will discuss below, the PS of ICs generated from a regular cubic grid typically agrees well with the theoretical prediction for wave-numbers smaller than the Nyquist frequency, \( k_N = \pi n_0^{1/3} \), where \( n_0 \) is the average number density.}
However, it is important to emphasize that the PS of a perfect 3D cubic lattice is non-zero only at specific discrete wave-vectors (including some below the Nyquist frequency) and zero elsewhere. As a result, for certain wave-vectors, the agreement with the theoretical spectrum is not well satisfied. This discrepancy is often overlooked because the PS  is  computed through spherical averaging, which implicitly assumes statistical isotropy. In contrast, a cubic lattice possesses only discrete translational invariance and exhibits strong anisotropies in Fourier space \citep{book}. Therefore, spherical averaging can obscure underlying directional artifacts that are present in the ICs due to the symmetry of the lattice. 

On the other hand, the agreement of the actual two-point correlation function with that of the theoretical model depends on the amplitude of the displacement field applied to the initial particle distribution, and hence on the redshift at which the ICs are generated. When the average initial displacement $\langle \delta \rangle$ exceeds the lattice spacing $a$, the measured two-point correlation function $\xi(r)$ closely matches the theoretical expectation. In contrast, when $\langle \delta \rangle \ll a$, the correlation function exhibits oscillatory features induced by the underlying lattice structure, indicating a residual imprint of the discretized initial configuration.

Thus, in general, the problem with the procedure described above is that it neglects the fluctuations and correlations present in the \textit{pre-initial} configuration. This neglect is justified only if such fluctuations play no role in the subsequent evolution of the system, under the assumption that one seeks to isolate the growth of structure arising solely from the imposed perturbations on a uniform density field.

Indeed, the rationale behind using a lattice (or \textquotedblleft glassy\textquotedblright configuration --- see below) is that these are special distributions in which the net 
gravitational force is (effectively) zero. For example, a Poisson distribution will evolve and form structures even in the absence of additional perturbations, whereas a perfect lattice will not. 

However, it is far from clear that using a cubic lattice or glass configurations as a pre-ICs entirely resolve this issue. Both represent unstable point configurations under gravity, and even a small applied perturbation may result in an evolution that depends sensitively on the residual correlations or fluctuations present in the unperturbed configuration.
In principle, such discreteness effects can be minimized by refining the discretization, --- i.e., increasing the number of particles to better sample the underlying density field. In practice, however, $N$-body simulations often probe the non-linear regime  of structure formation at scales comparable to the discretization length. Under these conditions, the influence of pre-initial correlations becomes, at the very least, problematic \citep{splinter_1998,Baertschiger:2001eu,Gabrielli_etal_2003}.

As previously noted, a lattice is not an isotropic system: intrinsic anisotropies of the lattice persist in the particle distribution even after the displacement field has been applied
{  \citep{Masaki_etal_2021,Racz_etal_2021,Yu_etal_2025}.}  The question we aim to explore below is whether these residual anisotropies have any dynamical effects during the subsequent gravitational evolution.

Note that  an alternative method for generating the pre-initial particle distribution for an N-body simulation consists in the construction of a \textit{glass-like} configuration \citep{White_1994}: this is  generated  by starting from a Poisson distribution and running an N-body code with the sign of the acceleration reversed 
(see \cite{Crocce_etal_2006} and reference therein). The main motivation for using a glass configuration instead of a grid is to eliminate the symmetries inherent to a lattice. 
The resulting distribution is highly isotropic but still exhibits long-range correlations akin to those of a lattice~\citep{Gabrielli_etal_2003}.
In this work, we do not consider the case of glass-like pre-ICs; a more detailed study of this scenario will be presented in a forthcoming paper.


\section{Methods} 
\label{methods}

\subsection{Correlation function and power spectrum}

We will determine the $n$-dimensional correlation function $\xi_{n\text{D}}(r)$ and PS $P_{n\text{D}}(k)$, where $n = 2, 3$, for the distribution of particles in a simulation box of side length $L$, assuming periodic boundary conditions. For the three-dimensional case ($n = 3$), we consider the full volume of the box.   We also analyze quasi-two-dimensional slices, i.e., slabs with vertical thickness $\Delta z$ much smaller than the box side length, extending across both the $x$ and $y$ directions. Within each slice, we compute the two-dimensional correlation function $\xi_{2\text{D}}(r)$ and PS $P_{2\text{D}}(k)$. Anisotropies introduced by the lattice used in the pre-ICs are still present even in these two-dimensional slices. Since the numerical analysis in two dimensions is simpler and faster than in three dimensions, we will primarily present results in 2D. We note that, for both simulations, the results presented correspond to a single two-dimensional slice. However, we have repeated the analysis across multiple independent slices, finding  similar results in all cases. This confirms that the observed features are robust and not specific to a particular projection.  As the statistical properties averaged over spherical (or circular) regions have been extensively studied in the literature, we do not repeat such an analysis in this work.


\subsection{Angular Distribution of Pairwise Distances }

Estimations of the two-point correlation function or the PS are typically based on the assumption that the underlying distribution is isotropic. However, it is precisely this assumption that we aim to test. To determine whether a point distribution is isotropic or exhibits preferred orientations, one can analyze spatial anisotropies using a range of statistical and geometric methods. We have focused on one particular statistical measure after testing several possibilities. This choice does not imply that it is necessarily the most suitable or optimal diagnostic; rather, it demonstrates that this statistic is effective in identifying anisotropic features in the point distribution.
 
The ADPD methodology is conceptually linked to directional correlation functions, pair orientation statistics, and filament detection in point distributions. It is computed by measuring the angles formed between all pairs of points. If the distribution of angles is uniform, the system can be considered isotropic. Conversely, clustering of angles around specific directions signals the presence of anisotropy and possibly underlying filamentary or ordered structures.

As mentioned above, given the geometry of the system, we focus on {  two-dimensional slices with finite thickness}. Although structure formation is inherently a fully three-dimensional dynamical process, it is important to consider that particles typically move only a few Mpc over the entire time evolution. Consequently, from a statistical perspective, it is sufficient to isolate a two-dimensional slice of thickness greater than $\sim 5-10$ Mpc in order to track approximately the same subsample of particles throughout the evolution.

Moreover, the symmetry of the lattice used as a pre-ICs justifies reducing the analysis of spatial anisotropies to a two-dimensional slice. This approach is not only numerically more efficient but also retains the essential features of directional correlations that are relevant to the problem.

To implement this, we compute the angle for each pair of points $(x_i, y_i)$ and $(x_j, y_j)$ using:
\[
\label{eq:theta}
\theta_{ij} = \tan^{-1}\left( \frac{y_j - y_i}{x_j - x_i} \right) \;.
\]

We then construct a histogram of all $\theta_{ij}$ values. This histogram has a straightforward interpretation: if it is approximately flat, the system is isotropic. On the other hand, if it exhibits peaks at specific angles, the system has a preferred orientation.

{  Note that in simulations with periodic boundary conditions, the computation of pairwise distances is typically limited to scales up to \(L/2\), where \(L\) is the side length of the cubic simulation box. This limitation arises from the application of the \emph{minimum image convention}, which ensures that the distance between two particles is computed as the shortest possible vector connecting them across the periodic replicas of the box. For any pair of points, their separation in each coordinate is given by the minimum of \(|x_i - x_j|\) and \(L - |x_i - x_j|\), such that the maximum allowed distance in any coordinate direction is \(L/2\). As a result, the maximum Euclidean separation considered is effectively constrained to \(L/2\). This convention avoids ambiguities in distance calculations and ensures that statistical estimators such as the correlation function or the ADPD remain consistent and free of periodic artifacts. Furthermore, considering scales larger than \(L/2\) would introduce artificial correlations due to the finite box size and would not yield physically meaningful results in the context of a spatially periodic system.}

{  The generalization of the ADPD to quantify angular anisotropies in a three-dimensional point distribution, 
is straightforward. The direction of a separation vector is specified by two angles in spherical coordinates: the azimuthal angle $\phi \in [0, 2\pi)$,
and the polar angle $\theta \in [0, \pi]$. The unit sphere is divided into $N_{\Omega}$ angular bins uniformly in $(\theta, \phi)$.

To probe anisotropy at a specific scale, one selects a spherical shell of width $\Delta r$ around a chosen radius $r$. Only point pairs separated by distances in the interval $[r - \Delta r/2, r + \Delta r/2]$ are considered.
For each pair of points $(\mathbf{x}_i, \mathbf{x}_j)$, the separation vector is computed $\mathbf{r}_{ij} = \mathbf{x}_j - \mathbf{x}_i$. 
If  $|\mathbf{r}_{ij}|$ falls within the chosen shell, the direction of the normalized vector $\hat{\mathbf{r}}_{ij} = \mathbf{r}_{ij} / |\mathbf{r}_{ij}|$ is converted to spherical coordinates $(\theta, \phi)$ and the corresponding angular bin is incremented.
After all valid pairs are counted, the angular histogram is normalized by the total number of contributing pairs. The resulting distribution $p_i(r)$ represents the probability of finding a pair in the $i$-th angular bin at scale $r$.

The degree of anisotropy is quantified by the angular variance:
\begin{equation}
\sigma^2_\Omega(r) = \sum_{i=1}^{N_{\Omega}} \left(p_i(r) - \frac{1}{N_{\Omega}}\right)^2,
\end{equation}
where $p_i(r)$ is the normalized count in bin $i$. For a statistically isotropic distribution, $p_i(r) \approx 1/N_{\Omega}$ and $\sigma^2_\Omega(r)$ is close to zero. Larger values indicate significant directional anisotropies.

}

\subsubsection{Identifying Filaments Using the ADPD}

To investigate the scale dependence of anisotropy, we compute the ADPD  in bins of pairwise separation $r$. This allows us to examine how directional correlations vary with scale. A particularly effective diagnostic is the two-dimensional heatmap representation of $\text{ADPD}(\theta, r)$, where the $x$-axis corresponds to the angle $\theta$, the $y$-axis to the separation $r$, and the color scale indicates the amplitude of $\text{ADPD}(\theta, r)$. In this format, coherent anisotropic structures manifest as vertical ridges --- i.e., features at fixed $\theta$ --- that persist across multiple radial bins. Such ridges correspond to excesses of pairs aligned along specific directions, indicative of filamentary structures.

For example, in a point distribution concentrated along a narrow, elongated bar (a quasi-one-dimensional structure in two dimensions), the ADPD exhibits a pronounced peak at the angle $\theta_0$ corresponding to the bar's orientation. This peak is strongest for separations $r$ smaller than the bar's length and gradually diminishes for $r > \ell$, where $\ell$ is the bar's physical extent. The angular width of the peak reflects the bar's straightness: a perfectly straight structure yields a narrow peak, while a curved or irregular shape results in a broader angular feature.

\begin{figure}[htbp]
  \begin{minipage}[t]{0.49\textwidth}
    \centering
    \includegraphics[width=\linewidth]{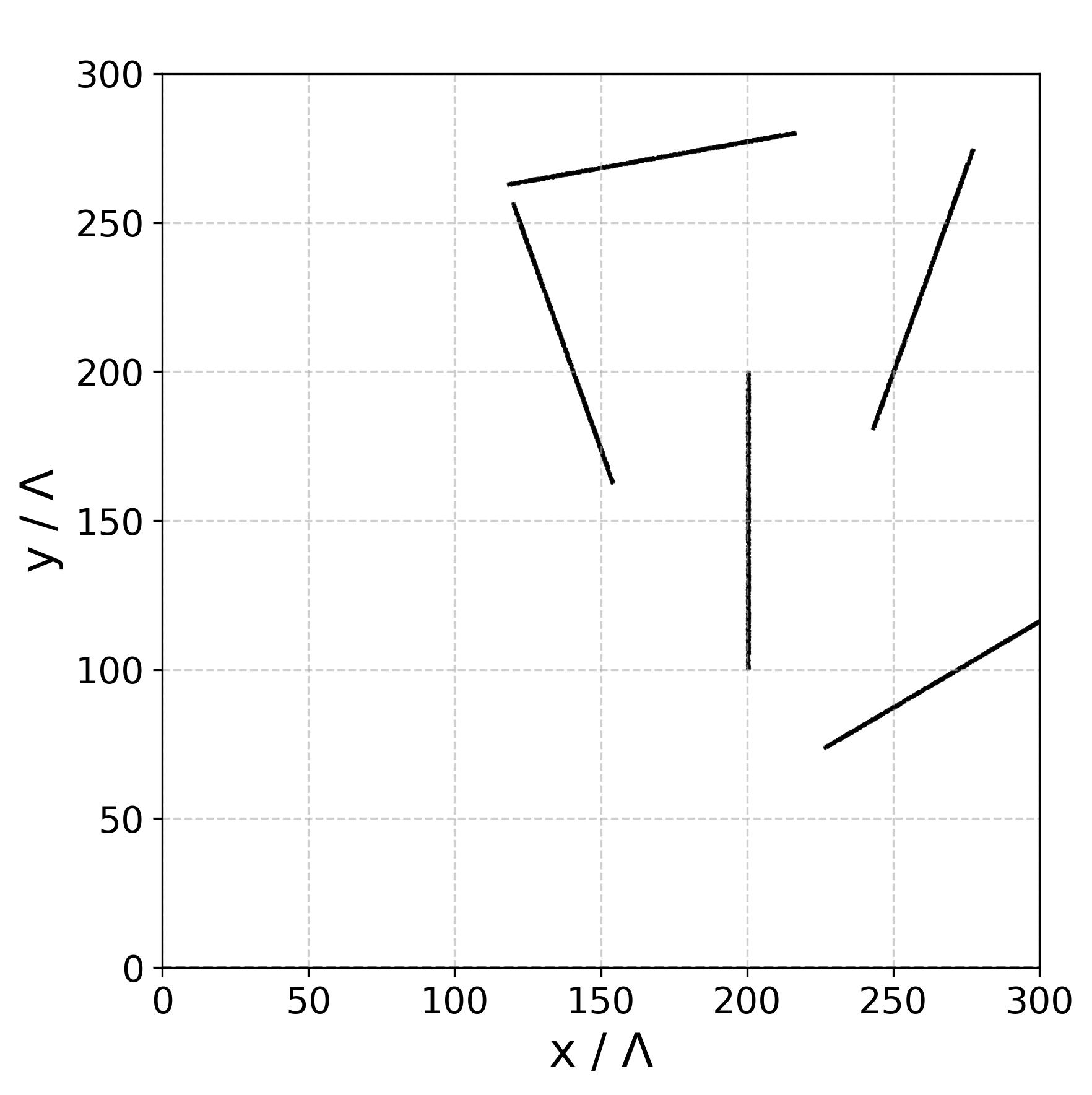}
      \end{minipage}
  \hfill
  \begin{minipage}[t]{0.49\textwidth}
    \centering
    \includegraphics[width=\linewidth]{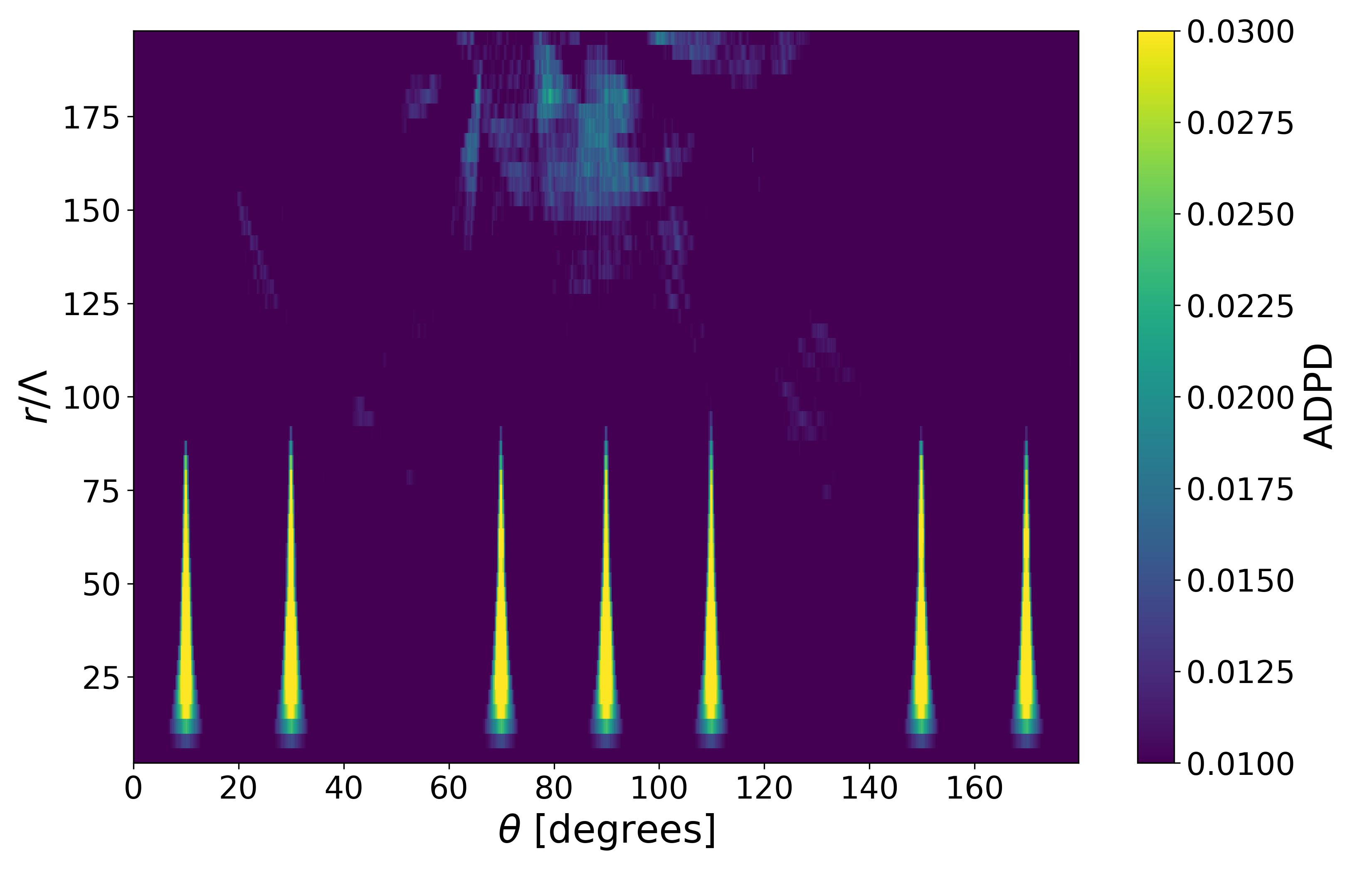}
     \end{minipage}
\caption{  
Point distribution (upper panel) and heatmap of the Angular Distribution of Pairwise Distances (ADPD, lower panel) for a system composed of multiple bars. Each bar is characterized by an orientation angle $\theta_i$ relative to the $x$-axis and a finite length $\ell_i$. Length scales are expressed in units of the average distance between nearest neighbors $\Lambda$.
}
\label{ManyBars} 
\end{figure}

If the distribution contains multiple bars or filaments, each with orientation $\theta_i$ and length $\ell_i$, the ADPD heatmap displays a set of ridges at angles $\theta_i$, extending up to separations $r \approx \ell_i$ (see Fig.\ref{ManyBars}). When orientations are similar, the ridges may merge or broaden; if they are randomly distributed, the heatmap approaches isotropy, and directional features fade.

This representation thus provides a direct and scale-resolved view of anisotropic structures in the point distribution, making it a powerful tool to identify and characterize directional features such as filaments and bars.

\subsubsection{Analysis lattice's like structures} 

Let us move to the case of a regular lattice in 2D. As a reference we consider the case of a random Poisson distribution; in addition we consider shuffled lattices  (SL) with different levels of shuffling $\delta$. The SL configurations are generated by randomly displacing each particle from its position on a perfect lattice by a quantity $\delta \le 1$, expressed in units of the lattice spacing $a$. We refer the interested reader to \cite{book} for a comprehensive analytical and numerical discussion of the two-point correlation function and the  PS in such systems.
\begin{figure*} 
\includegraphics[width=0.49\textwidth]{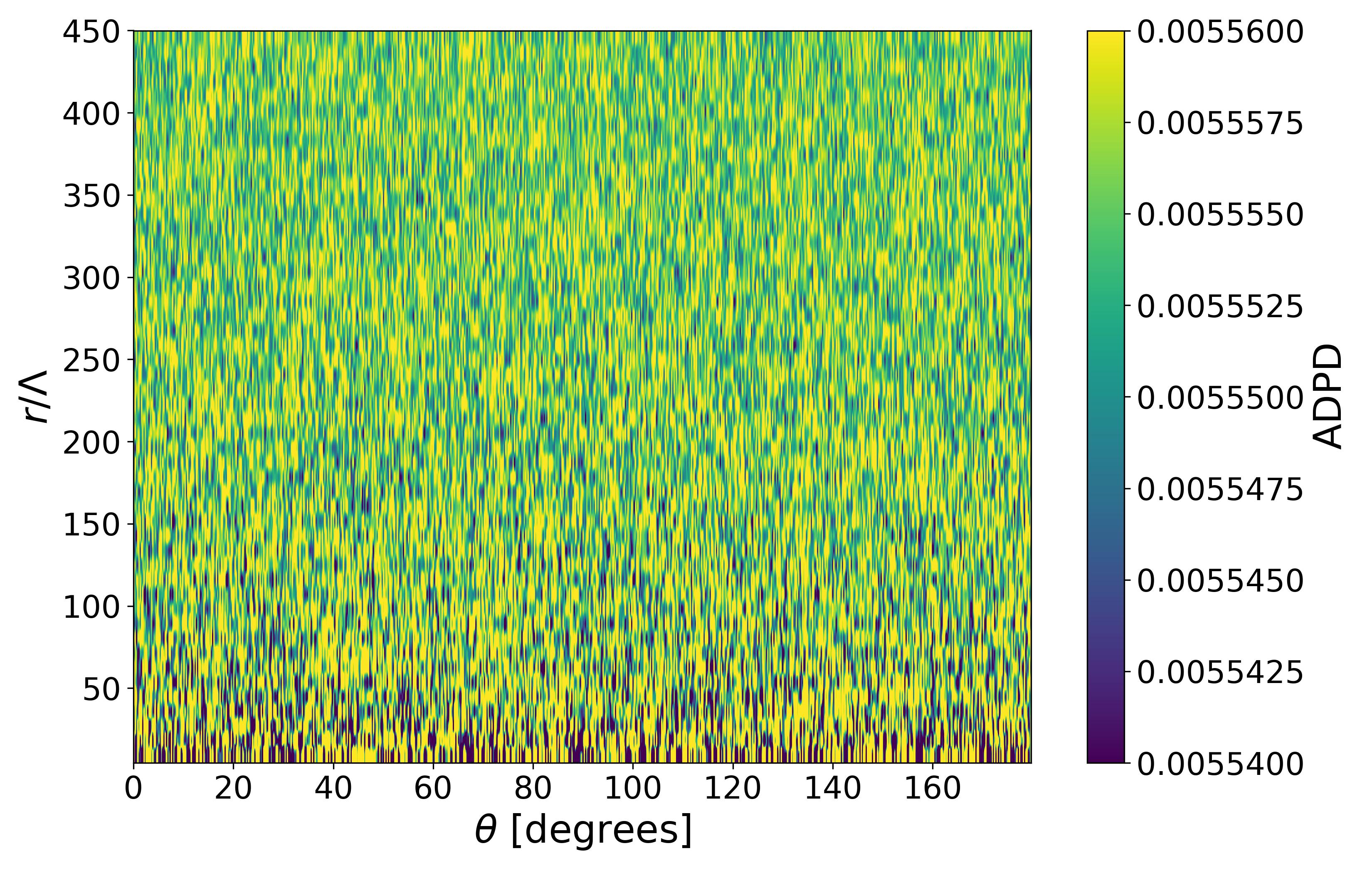}  
\includegraphics[width=0.49\textwidth]{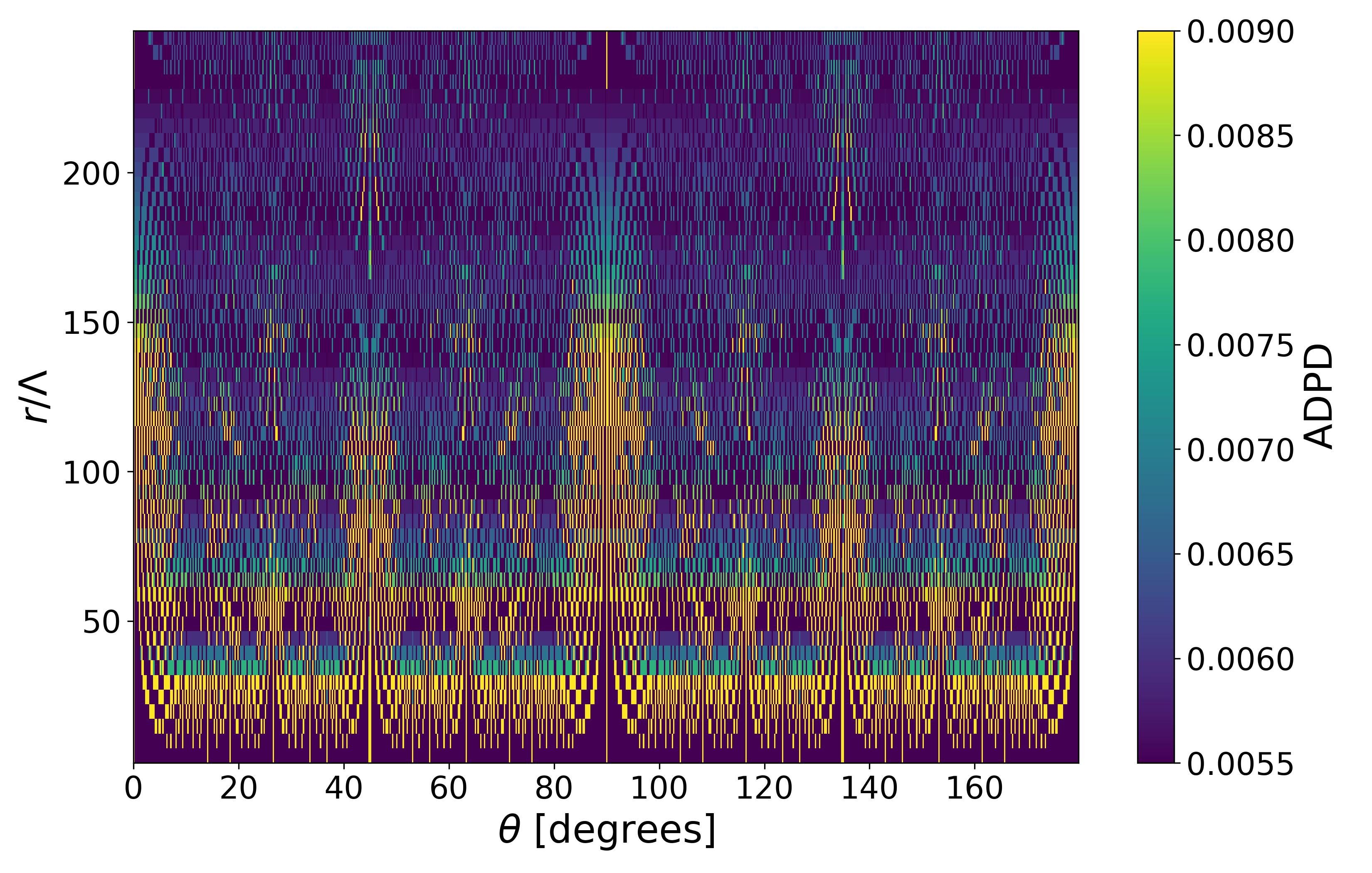}  
\includegraphics[width=0.49\textwidth]{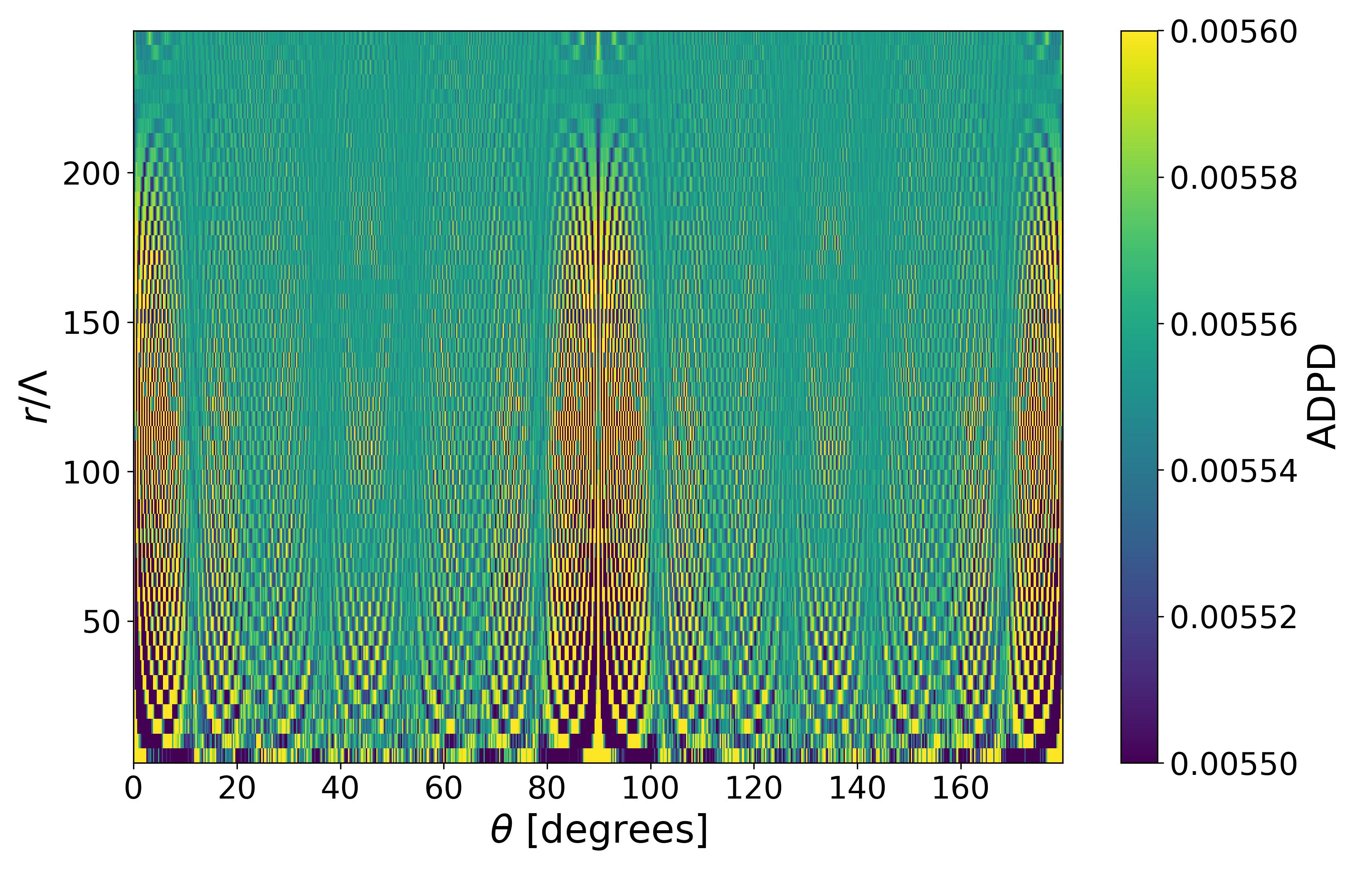}  
\includegraphics[width=0.49\textwidth]{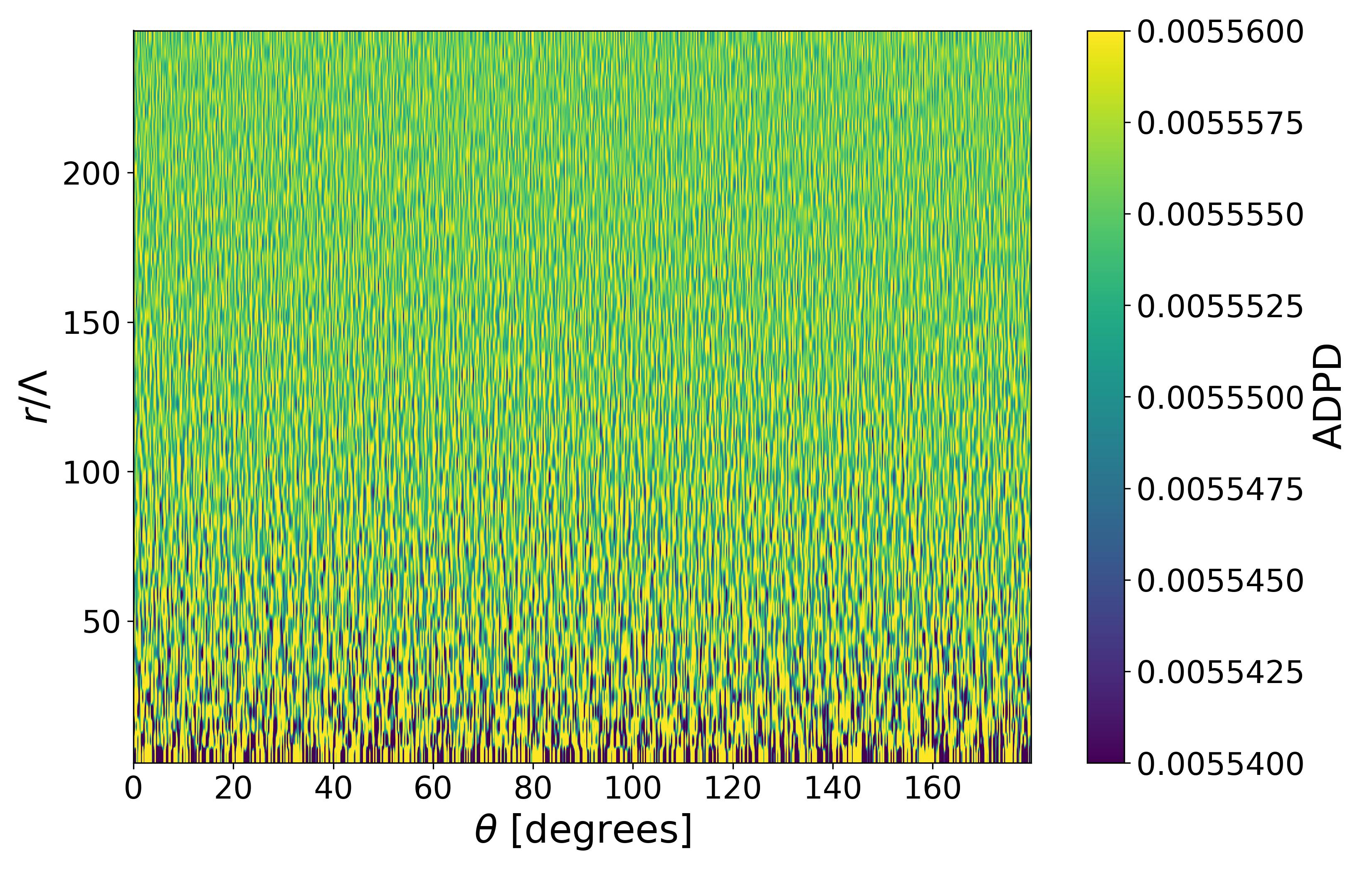} 
\caption{  
Angular Distribution of Pairwise Distances (ADPD) for four configurations: a Poisson distribution (top left), a perfect cubic lattice (top right), a shuffled lattice with displacement amplitude $\delta = 0.6$ (bottom left), and with $\delta = 1.0$ (bottom right).  For improved visual interpretation, the color scale in each panel has been individually adjusted to enhance contrast and highlight relevant structures in the ADPD. Length scales are expressed in units of the average distance between nearest neighbors $\Lambda$.}
\label{adpd_heatmap_SL} 
\end{figure*}

The ADPD for a random two-dimensional distribution, such as an isotropic Poisson process, is shown in the top-left panel of Fig.~\ref{adpd_heatmap_SL}. One observes the that there are no preferred directions, i.e. the ADPD is uniform in angle for any fixed radius $r$.
If the ADPD is normalized such that the integral over the interval $\theta \in [0^\circ, 180^\circ]$ equals 1, then:
\begin{equation}
\label{adpd_random}
\text{ADPD}(\theta, r) = \frac{1}{180^\circ} = 0.00\overline{5} \;.
\end{equation}
In real data, finite sampling and discreteness effects introduce fluctuations around the mean. However, in a truly random field, these deviations are expected to be uncorrelated and consistent with statistical noise.

If the point distribution forms a regular two-dimensional lattice (e.g., a square grid), then the ADPD has an amplitude larger than Eq.\ref{adpd_random} and it exhibits distinct features, as illustrated in the top-right panel of Fig.~\ref{adpd_heatmap_SL}: the pairwise distances between points are concentrated at discrete values, such as $a$, $\sqrt{2}a$, $2a$, etc., where $a$ is the lattice spacing. More importantly, the separation vectors between point pairs are strongly aligned along the symmetry axes of the lattice. These typically include:
    $
    \theta = 0^\circ \text{ (horizontal)}$, $\quad \theta = 90^\circ \text{ (vertical)}$, 
    $\quad \theta = 45^\circ,\, 135^\circ \text{ (diagonals)},
    $
    along with their symmetric counterparts.
These peaks persist across all values of $r$ for which the lattice-induced distances are sampled. Consequently, the ADPD$(\theta, r)$ heatmap displays vertical stripes at the corresponding angles, indicating strong angular anisotropy. These stripes extend over a broad range of $r$, depending on the extent of the lattice's long-range order. These peaks are always statistically significant. This can be verified by considering the uncertainty in the ADPD, which arises from finite sampling and is estimated as $\sqrt{N_{\mathrm{bin}}}$, where $N_{\mathrm{bin}}$ is the number of counts in each angular bin. At certain separations, the probability density in specific directions is enhanced relative to a random isotropic distribution, while in others it is suppressed --- an imprint of the lattice's anisotropic and ordered structure.

The bottom left panel of Fig.~\ref{adpd_heatmap_SL} shows the ADPD of a shuffled lattice with $\delta = 0.6$. Although the coherent lattice structure is partially disrupted, directional peaks remain clearly visible across all pairwise separations. The persistence of narrow features, particularly at $\theta = 0^\circ$ and $\theta = 90^\circ$, even with significantly reduced amplitude compared to the perfect lattice, reflects the residual one-dimensional alignment inherited from the original configuration. Relative to the unperturbed case, the peaks are broader and weaker, indicating a partial loss of coherence. Nevertheless, the presence of these anisotropic signatures at both small and large separations suggests that directional ordering persists on large scales despite the imposed perturbations.

A simple $\chi^2$ test confirms that the observed peaks both in this case and in the simulations discussed in the following section are  statistically significant, demonstrating that the ADPD deviates from a uniform isotropic distribution. As the perturbation amplitude increases ($\delta \gtrsim 0.5$), the peak structure becomes progressively suppressed at all scales. This trend is illustrated in the bottom right panel of Fig.~\ref{adpd_heatmap_SL}, where $\delta = 1.0$: in this case, the heatmap closely resembles that of a Poisson distribution, with no statistically significant angular features.

{  
The key message of Fig.~\ref{adpd_heatmap_SL} can be summarized as follows: a lattice or a distribution close to a lattice produces regular, anisotropic patterns in the ADPD heatmap, whereas a random (Poisson) distribution does not. In particular, the top-left panel of Fig.~\ref{adpd_heatmap_SL} is clearly one of the noisiest, as it corresponds to a Poisson distribution and is thus dominated by random fluctuations. The bottom-right panel, showing a shuffled lattice with $\delta = 1.0$, also closely resembles a Poisson distribution, as expected. }

\subsection{Angular Variance of the ADPD and Comparison with Radial Number Variance}

The ADPD quantifies the distribution of angles $\theta$ formed by point pairs separated by a fixed radial distance $r$. The \emph{angular variance} of the ADPD at a given scale $r$ is defined as
\begin{equation}
\sigma^2_\theta(r) = \sum_{i=1}^{N_\theta} \left( p_i(r) - \frac{1}{N_\theta} \right)^2,
\end{equation}
where $p_i(r)$ is the normalized pair count in the $i$-th angular bin (with total $N_\theta$ bins). This variance captures directional anisotropies: in an ideal isotropic distribution, the ADPD is flat and $\sigma^2_\theta(r) = 0$ in the infinite-sample limit. 

For a uniform Poisson distribution of \( N \) particles in a two-dimensional square box of side \( L \), the expected amplitude of the angular variance of the ADPD 
at a given scale \( r < L \) is governed by the number of particle pairs at that scale, the angular resolution, and the effects of finite-sample fluctuations.

The number of distinct particle pairs separated by a distance \( r \) within a radial bin of width \( \Delta r \) is approximately given by:
\begin{equation}
\bar{n}(r) \approx N \cdot 2\pi r \Delta r \cdot \frac{N}{L^2} = \frac{2\pi N^2 r \Delta r}{L^2} \,.
\end{equation}
This quantity determines the statistical sampling available at each scale \( r \).

\begin{figure} 
\includegraphics[width=0.49\textwidth]{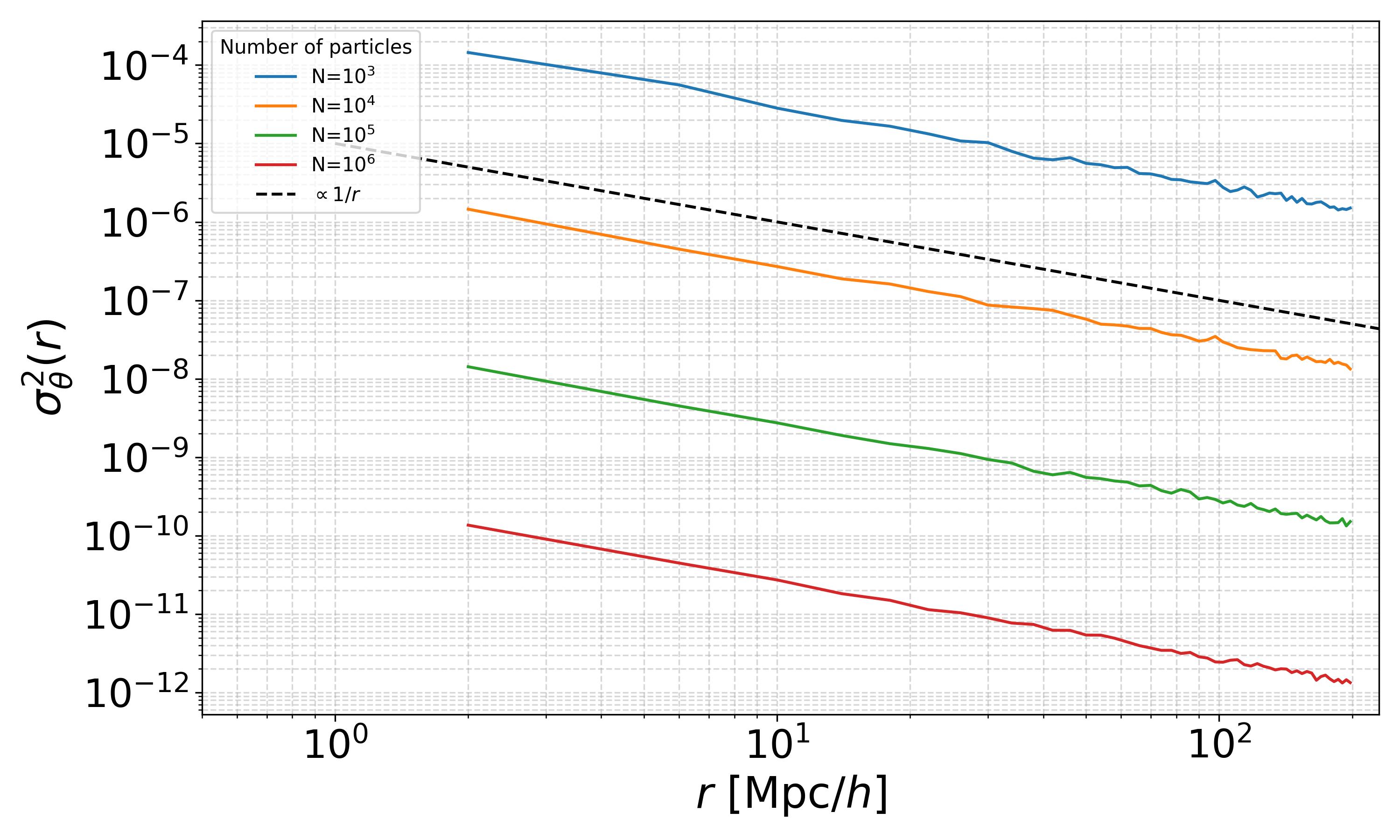} 
\caption{     
Angular variance of the Angular Distribution of Pairwise Distances  for Poisson distributions
in a box of  side $L=400$ Mpc/$h$ with the different number of particles (see labels). 
A  line $\propto 1/r$ (see Eq.\ref{ang_var_adpd_Poisson}) is also shown as a reference.}
\label{fig_ang_var_adpd_Poisson} 
\end{figure}

When the angle between particle pairs is binned into \( N_\theta \) equal intervals, the expected probability per bin is \( p = 1/N_\theta \). Due to statistical fluctuations in a finite sample, the normalized angular histogram will exhibit a non-zero variance. Specifically, the expected angular variance of the ADPD is given by:
\begin{equation}
\mathbb{E}[\sigma^2_\theta(r)] \approx \frac{1 - \frac{1}{N_\theta}}{\bar{n}(r)} \,.
\end{equation}
Therefore, the angular variance scales inversely with both the number of particle pairs at distance \( r \) and the number of angular bins:
\begin{equation}
\label{ang_var_adpd_Poisson} 
\sigma^2_\theta(r) \propto \frac{1}{N^2 r \Delta r N_\theta} \cdot L^2 \,.
\end{equation}
This implies that \( \sigma^2_\theta(r) \propto 1/N^2 \), and hence becomes negligible in the limit of large \( N \). As a result, any excess variance above this Poissonian expectation indicates the presence of anisotropic correlations in the particle distribution. Fig.\ref{fig_ang_var_adpd_Poisson}  shows the behavior of Eq.\ref{ang_var_adpd_Poisson} for some realizations of a Poisson distribution with different number of points. 

Importantly, this angular variance is a purely directional statistic and differs from the \emph{number variance in spheres}, which quantifies fluctuations in the number of points $N(R)$ within a circular window of radius $R$:
\begin{equation}
\sigma^2(R) = \frac{\langle N^2(R) \rangle - \langle N(R) \rangle^2}{ \langle N(R) \rangle^2} \;. 
\end{equation}
While the number variance probes isotropic density fluctuations over a region of scale $R$, the ADPD angular variance probes the uniformity of angular pairwise correlations at a fixed radius $r$. The two quantities thus offer complementary information: the former is sensitive to isotropic clustering (e.g., via the two-point correlation function), while the latter is sensitive to anisotropic patterns such as filaments or grid-aligned structures that escape angle-averaged statistics.

\begin{figure} 
\includegraphics[width=0.49\textwidth]{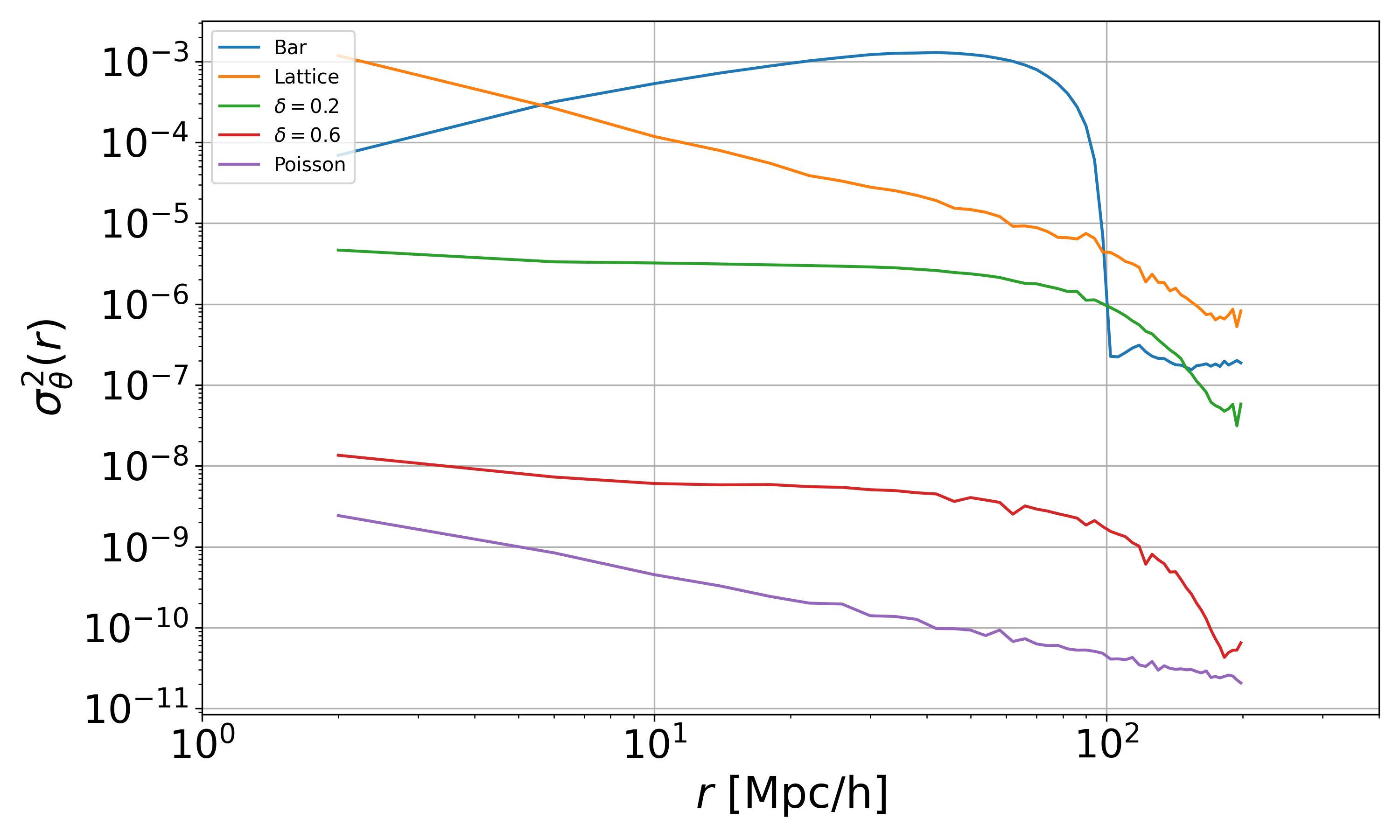} 
\caption{  
Angular variance of the Angular Distribution of Pairwise Distances 
 for four different configurations: a single bar, a perfect cubic lattice, a shuffled lattice with displacement amplitude $\delta = 0.2$, and with $\delta = 0.6$.  
All distributions are generated  in a box of  side $L=400$ Mpc/$h$ and have the same number of points $N= 10^5$.
As reference the angular variance of the ADPD for a Poisson with same number of particles is also reported.
} 
\label{adpd_heatmap_angular_variance_tests} 
\end{figure}

To quantify anisotropies in the point distribution, we measure the angular variance $\sigma_\theta^2(r)$ of the ADPD. In the plots shown in Fig.~\ref{adpd_heatmap_angular_variance_tests}, the $y$-axis reports the angular variance, which quantifies the spread of ADPD$(\theta)$ around its mean at fixed radius. Higher values indicate stronger directional modulations --- i.e., greater anisotropy --- whereas lower values reflect a trend toward isotropy, corresponding to a flatter ADPD. The angular variance is plotted as a function of the pairwise separation $r$, over which it is computed.

In the case where the distribution consists of an almost one-dimensional segment of linear length 100 Mpc/$h$, the angular variance $\sigma_\theta^2(r)$ exhibits a sharp rise followed by a pronounced peak at $r_{\text{peak}} \approx 50$ Mpc/$h$. This behavior indicates that anisotropic structures dominate at small scales, while the peak reflects the scale at which directional coherence is strongest. Beyond this, the variance drops and approaches zero for $r \geq 100$ Mpc/$h$, confirming that the field becomes effectively isotropic on large scales. From this behavior we conclude that the dominant scale of anisotropic structures is approximately $r_{\text{peak}} \approx 50$ Mpc/$h$. The perfect cubic lattice, due to its high degree of symmetry, exhibits the largest angular variance $\sigma^2_\theta(r)$. In contrast, the shuffled lattices with displacement amplitudes $\delta = 0.2$ and $\delta = 0.6$ display progressively lower values of $\sigma^2_\theta(r)$, as the random displacements break the underlying symmetry. The configuration with $\delta = 0.6$ is less symmetric than that with $\delta = 0.2$, and accordingly, it shows a smaller angular variance. However, the angular variance remains higher than that of a random (Poisson) point distribution with the same number of particles, indicating that the system retains a degree of anisotropy exceeding that expected from pure shot noise.
 

\section{Quantifying Local Anisotropy in Cosmological Initial Conditions}
\label{Anis_IC} 
{
  
In cosmological $N$-body simulations, the initial density field is typically modeled as a Gaussian random field that is statistically homogeneous and isotropic. This implies that, in an ensemble sense, the field has no preferred location or direction. However, a {single realization} in a simulation  inevitably contains local fluctuations that deviate from perfect isotropy. These fluctuations give rise to {local anisotropies} in the gravitational field, even though the ensemble-averaged statistics remain isotropic. From a quantitative standpoint, the question of symmetry breaking in structure formation can be posed in terms of these local deviations. Specifically, one wants to quantify  how do random fluctuations in an otherwise statistically isotropic Gaussian field lead to local anisotropic gravitational collapse. 

This can be addressed by analyzing the \emph{deformation tensor}, defined as the Hessian of the gravitational potential. The eigenvalues and eigenvectors of this tensor describe the local tidal field, determining the direction and nature of gravitational collapse. Even though the global field is isotropic on average, the random nature of fluctuations ensures that individual regions will typically exhibit anisotropic deformation, driven by the local configuration of the deformation tensor.
A commonly used statistical descriptor of local anisotropy is the variance of the eigenvalues of the deformation tensor \cite{Forero-Romero_etal_2009,Libeskind_etal_2018}:
\begin{equation}
T_{ij} = \frac{\partial^2 \phi}{\partial x_i \partial x_j},
\end{equation}
where $\phi(\mathbf{x})$ is the peculiar gravitational potential. 

Anisotropic deformation in the linear regime arises when the eigenvalues of the deformation tensor $T_{ij}$ differ significantly. This leads to differential gravitational accelerations along the principal axes: the largest eigenvalue corresponds to the direction with the strongest compression. As a result, the matter distribution undergoes slight anisotropic deformations --- flattening first along one axis (suggestive of forming planar \textquotedblleft pancake\textquotedblright-like features), then along a second axis (resembling filaments), and finally toward more compact configurations. However, it is important to note that these are not actual collapses, but rather small distortions predicted by linear theory. While these deformations may indicate the preferred directions of future nonlinear evolution, this is not necessarily the case. In practice, a number of additional dynamical processes come into play --- most notably, the bottom-up aggregation of particles, which can develop sufficient peculiar velocities pressure to inhibit or even prevent the expected top-down collapse \cite{SylosLabini_CapuzzoDolcetta_2020}.

The amplitude of these anisotropic modes is directly related to the PS of initial density fluctuations. In linear theory, the variance of the deformation tensor is given by \cite{Forero-Romero_etal_2009}:
\begin{equation}
\langle T_{ij}^2 \rangle \propto \int_0^\infty k^4 P(k, z)\, dk,
\end{equation}
where $P(k, z)$ is the PS at redshift $z$. Since the PS evolves as $P(k, z) = D^2(z)\, P(k, 0)$ in the linear regime, the amplitude of anisotropic deformations grows proportionally to the square of the linear growth factor :
$\sigma_{\theta,3D}^2(z,r) \propto D^2(z).$

More precisely, the theoretical expectation for the anisotropy variance smoothed on scale $r$ at redshift $z$ is given by  \cite{Forero-Romero_etal_2009}:
\begin{equation}
\label{sigma_theo}
\sigma_{\theta,3D}^2(z,r) = D^2(z) \times \frac{4}{15\pi^2} \int_0^\infty dk \, k^4 P(k)\, W_{3D}^2(k r),
\end{equation}
where \( P(k) \) is the linear matter PS  at \( z = 0 \), and \( D(z) \) is the linear growth factor. The function \( W_{3D}(k r) \) is a smoothing window function that defines the filtering scale: common choices include a Gaussian window function or a Top-hat window function (in real space).  Eq.\ref{sigma_theo}  expression quantifies how local anisotropies in the deformation tensor depend on the underlying PS and evolve through cosmic time.

Similarly  the 2D anisotropy variance  can be derived from projecting a 3D PS onto a 2D plane and applying a 2D top-hat window function: 
\begin{equation}
\label{sigma_theo_2D}
\sigma_{\theta,2D}^2(z, r) = \frac{D^2(z)}{2\pi} \int_0^\infty dk\, k^3 P(k) W_{2D}^2(kr)
\end{equation}
where $W_{2D}(kr)$ is the Fourier transform of the 2D top-hat window. 

We adopt a parametrized form of the linear matter PS for the $\Lambda$CDM model of Eq.\ref{pk}  with $m = 2.5$  and $k_t = 0.05$, $h$/{Mpc} 
and we employ a top-hat {  (TH)} window function for spatial filtering.
The normalization constant \( A \) is fixed by requiring that the mass variance on a sphere of radius \( R = 8\, h^{-1}\mathrm{Mpc} \) at redshift \( z = 0 \) satisfies 
\cite{Planck2015}
\begin{equation}
 \sigma^2(R=8\, h^{-1}\mathrm{Mpc}, z=0) = \sigma_8^2 = 0.8 \;. 
\end{equation}
Considering that 
\be
\sigma_8^2 = \frac{1}{2\pi^2}\int_0^\infty k^2 P(k)\,W_{\rm TH}^2(kR_8)\,dk,
\ee
{  where $R_8=R=8\, h^{-1}\mathrm{Mpc}$}, 
we find 
\bea
&&
A(z=0) = \frac{ \sigma_8^2 } {  \frac{1}{2\pi^2}\int_0^\infty k^2 \hat{P}(k)\,W_{\rm TH}^2(kR_8)\,dk} 
\\ \nonumber && = 3.14 \times 10^5 \;\;  (\mathrm{Mpc}/h)^3 \;,
\eea
{  where $\hat{P}(k)$ is the shape function. }
This normalization ensures consistency with the amplitude of density fluctuations inferred from Cosmic Microwave Background observations.

In the range of distance and redshift in which the   matter PS linearly evolves with redshift we have 
\begin{equation}
\label{ps_theo}
P(k, z) = D^2(z) \cdot P(k, z=0),
\end{equation}
so that the normalization factor at redshift \( z \) becomes
\begin{equation}
A(z) = D^2(z) \cdot A(z=0),
\end{equation}
where \( D(z) \) is the linear growth factor and \( A(z=0) \) is fixed to match the fluctuation amplitude at the present epoch.

The linear growth factor in a flat \(\Lambda\)CDM cosmology in Eq.\ref{sigma_theo}-\ref{ps_theo} is given by \cite{Peebles_1980}:
\begin{equation}
D(z) = \frac{g(z)}{g(0)} \cdot \frac{1}{1+z}
\end{equation}
where
\begin{equation}
g(z) = \frac{5}{2} \frac{\Omega_m(z)}{\Omega_m(z)^{4/7} - \Omega_\Lambda(z) + \left(1 + \frac{\Omega_m(z)}{2} \right)\left(1 + \frac{\Omega_\Lambda(z)}{70} \right)},
\end{equation}
\bea
&&
\Omega_m(z) = \frac{\Omega_{m,0} (1+z)^3}{E(z)^2}
\;,\;\; \Omega_\Lambda(z) = \frac{\Omega_{\Lambda,0}}{E(z)^2},
\eea
\begin{equation}
E(z) = \sqrt{\Omega_{m,0} (1+z)^3 + \Omega_{\Lambda,0}}.
\end{equation}

\begin{figure} 
\begin{center}
\includegraphics[width=0.49\textwidth]{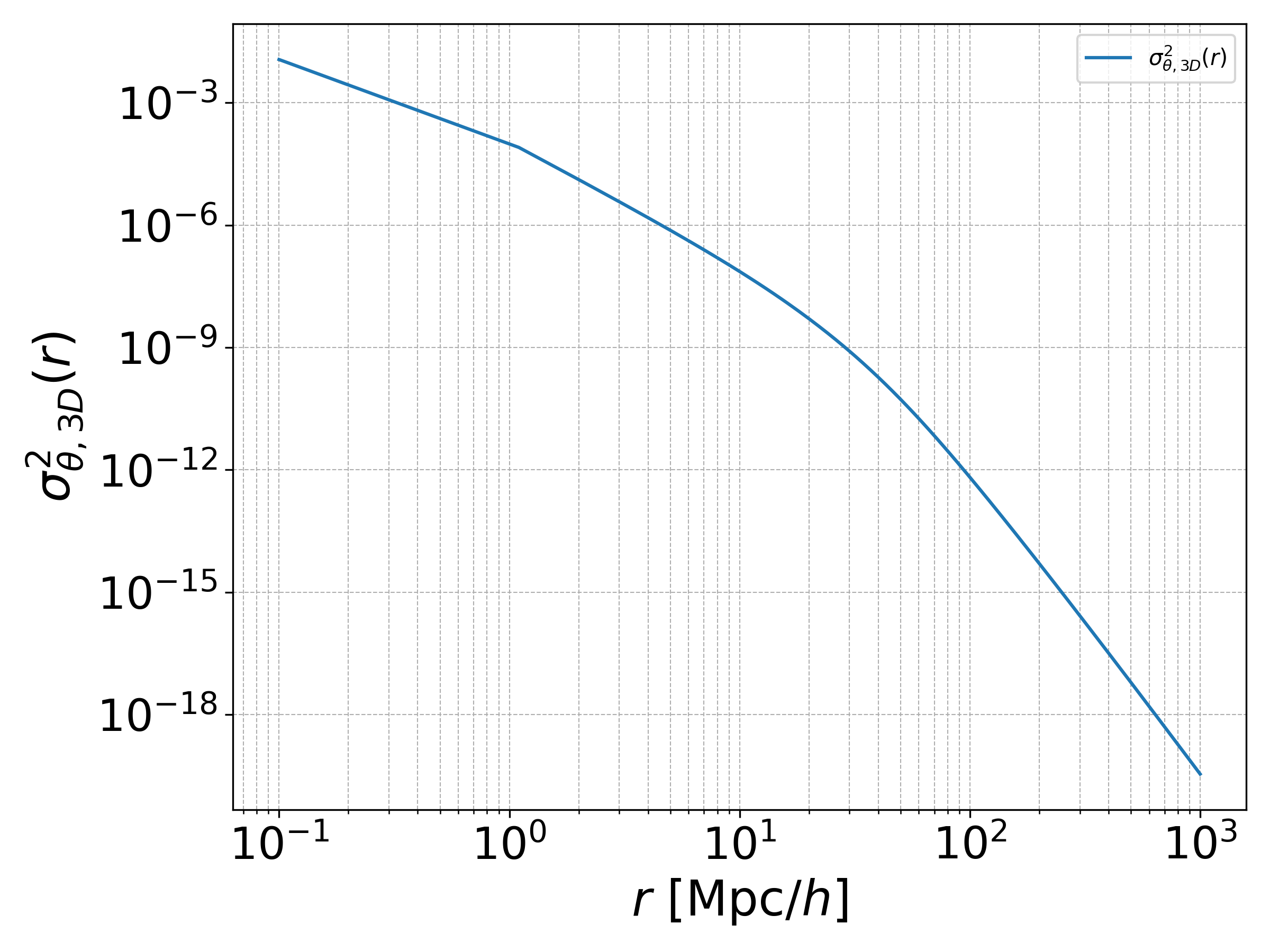} 
\includegraphics[width=0.49\textwidth]{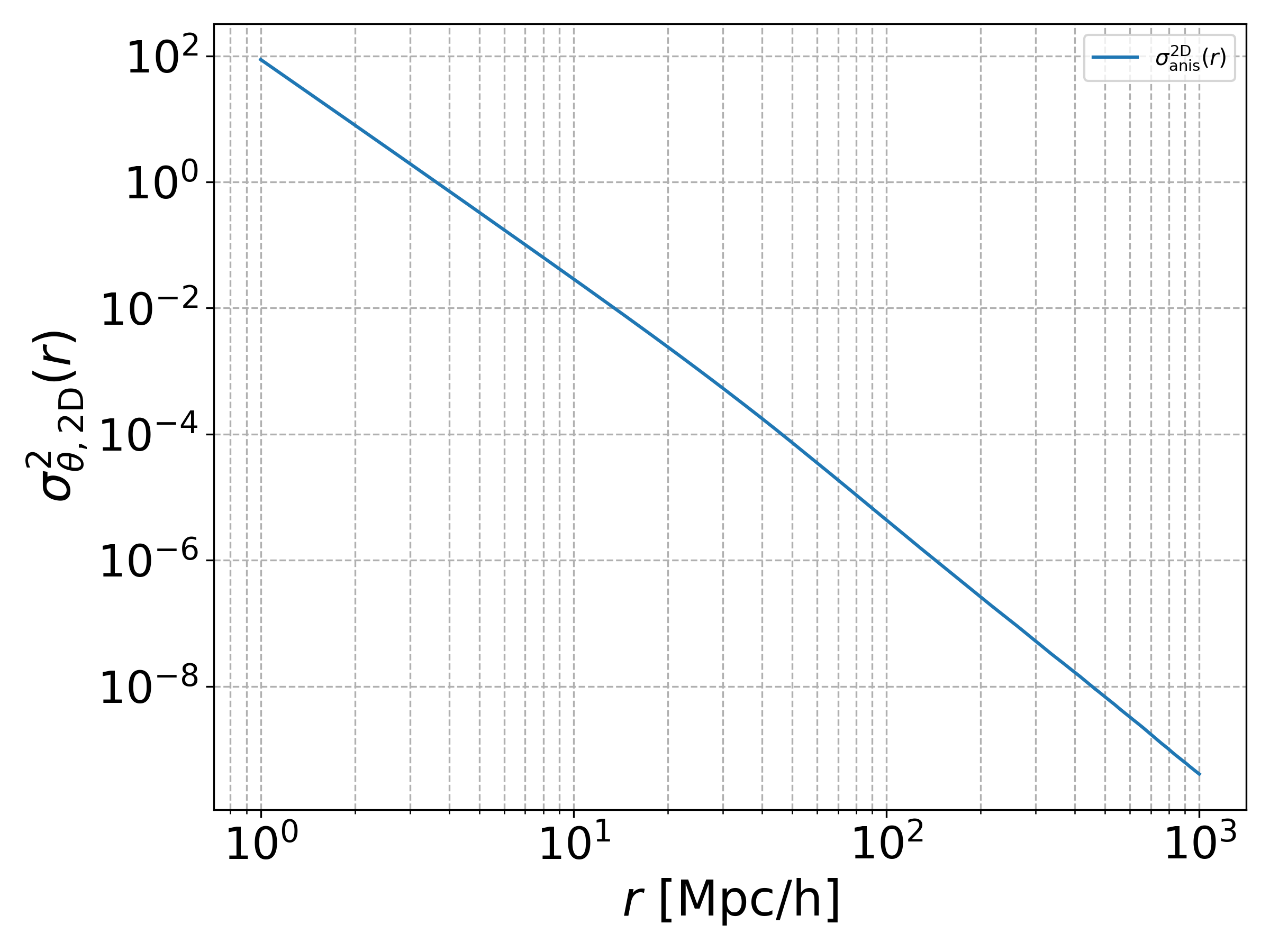} 
\end{center} 
\caption{
Upper panel: theoretical anisotropy variance  with top-hat window (Eq.\ref{sigma_theo}) in function of scale at $z=0$.
Bottom panel: theoretical anisotropy 2D variance at $z=0$ (Eq.\ref{sigma_theo_2D}).}
\label{LGF_LCDM} 
\end{figure}
The upper and bottom panels of Fig.\ref{LGF_LCDM}  show respectively  the theoretical anisotropy variance with top-hat window in function of scale in 3D (Eq.\ref{sigma_theo}) and 2D (Eq.\ref{sigma_theo_2D}). Local anisotropy induced by fluctuations increases with time as the universe evolves, even though the underlying statistical field remains isotropic. One may notice the relatively rapid decline of the theoretical anisotropy variance with top-hat window in function of scale. Specifically, we find that the angular variance decays approximately as 
\be
\label{sigma_theta_2D}
\sigma_{\theta,2D}^2(z, r) \propto r^{-4} \;. 
\ee
This behaviour provides, in principle, a natural reference for comparison with measurements 
of \( \sigma^2_{\theta}(z,r) \) in simulations.  
However, Eq.~\ref{sigma_theta_2D} applies to a continuous field, whereas the angular variance 
measured in a discrete point distribution depends on the specific discretisation adopted, 
and its amplitude cannot be predicted \emph{a priori}.  
By contrast, as long as the point distribution remains uniform and statistically isotropic, 
the asymptotic scaling of \( \sigma^2_\theta(r) \) --- analogous to that of a Poisson 
distribution --- and the corresponding featureless appearance of the ADPD heatmap constitute 
robust, model-independent expectations.
 }


\section{Results in simulations}
\label{results}
{  We consider two CDM simulations described in \cite{Diemer+Kravtsov_2014,Diemer+Kravtsov_2015}. 
The ICs for the simulations were generated using a second-order Lagrangian perturbation theory code   \cite{Crocce_etal_2006} that represents a refinement of the e Zel'dovich  Approximation discussed in Sect.\ref{gen_ic}.}
The first simulation, referred to as the \emph{small box simulation}, consists of $N = 256^3$ particles within a cubic box of side length $L = 62.5\,\mathrm{Mpc}$$/h$ 
{  with an initial redshift $z_{\mathrm{IC}} =49$.   }
The second, the \emph{large box simulation}, contains $N = 1024^3$ particles in a box of side length $L = 1000\,\mathrm{Mpc}$$/h$ (but, for simplicity, we limit the analysis to $L'=400$ Mpc$/h$) and an initial redshift $z_{\mathrm{IC}} =49$  \footnote{Available at https://www.astro.umd.edu/$\sim$diemer/erebos/}. The initial average distance between nearest neighbors is respectively $\Lambda \approx 0.25$ Mpc$/h$ and $\Lambda\approx 1$ Mpc$/h$.

{  The substantial differences in particle number (i.e., resolution) and  box size  between} the two simulations allow us to investigate the impact of finite-size effects on statistical measures such as the two-point correlation function and the ADPD. In both box cases, we select a two-dimensional thin slice of side length $L$ and thickness $\Delta z = 5$ Mpc$/h$ for the small box and $\Delta z = 10$ Mpc$/h$ for the large box.

\subsection{Small box simulation}

\subsubsection{Anisotropies in the Initial Conditions: ADPD Analysis}
\label{sec:adpd_ic}

\begin{figure}[htbp]
  \begin{minipage}[t]{0.43\textwidth}
    \centering
    \includegraphics[width=\linewidth]{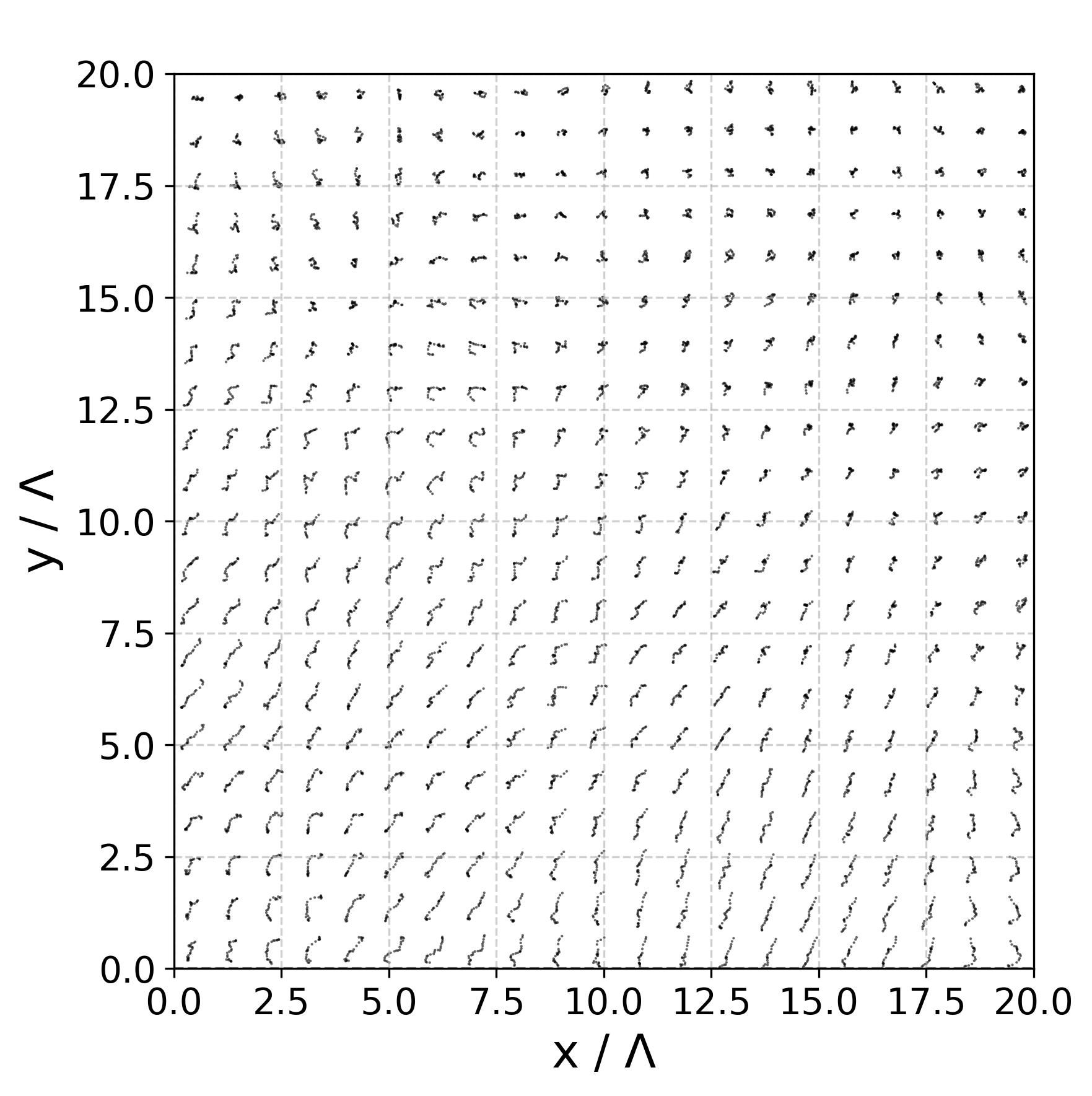}
  \end{minipage}
  \hfill
  \begin{minipage}[t]{0.49\textwidth}
    \centering
    \includegraphics[width=\linewidth]{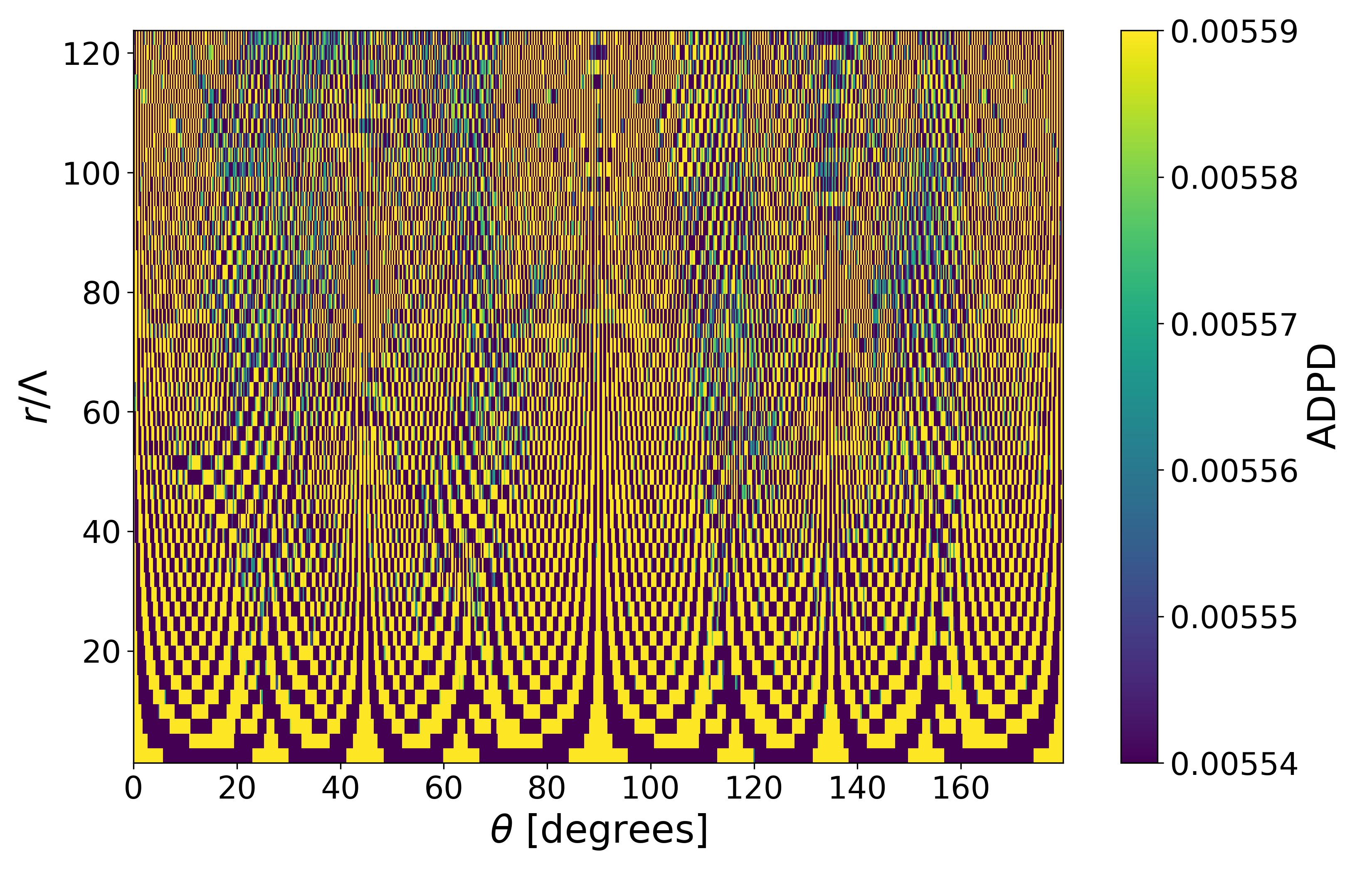}
  \end{minipage}
  \caption{    
{Upper panel:} Point distribution in a zoomed region of the initial conditions  for the small-box simulation at redshift $z_{\mathrm{IC}} = 49$.  {  (units are in Mpc/$h$)}. 
{Bottom panel:} Heatmap of the Angular Distribution of Pairwise Distances (ADPD) corresponding to the same configuration.
Length scales are expressed in units of the average distance between nearest neighbors $\Lambda$.}
\label{adpd_heatmap_small_box_IC} 
\end{figure}

{  
Fig.~\ref{adpd_heatmap_small_box_IC} displays the ADPD heatmap {  for the initial particle configuration at redshift $z_{\mathrm{IC}} =49$}. The distribution is clearly \textit{anisotropic}: as discussed above, the ICs are generated by applying a correlated and statistically isotropic displacement field --- consistent with a $\Lambda$CDM PS --- to particles initially arranged in a perfect cubic lattice. At high redshift (e.g., $z_{\mathrm{IC}} = 49$), the amplitude of density fluctuations is still small, and therefore the typical particle displacement is smaller than the lattice spacing $a$. As a consequence, the particle distribution retains clear imprints of the original grid, and the ADPD is dominated by the directional symmetries inherent to the lattice structure. Specifically, the ADPD heatmap at $z=49$ exhibits prominent vertical striping in angle $\theta$, with strong, repeating peaks at $\theta = 0^\circ$, $90^\circ$, $180^\circ$, and $270^\circ$, along with secondary features at $45^\circ$ intervals (see the upper left panel of Fig.\ref{adpd_heatmap_SL}). These patterns directly reflect the inherent anisotropy of the cubic lattice, which features preferred directions aligned with the Cartesian axes and diagonals. Because the displacements are small, their contribution to the ADPD is negligible at this stage, the anisotropic lattice signature dominates.

These results indicate that the configuration cannot be regarded as a statistically fair sample of an isotropic, correlated CDM initial field. The directional features detected in the ADPD reflect systematic deviations from the expected statistical properties and are likely numerical artifacts.
}

{  Under ideal conditions, CDM initial configurations should satisfy the following:
the fluctuation field should be statistically isotropic;
the two-point correlation function $\xi(r)$ should decay exponentially, with characteristic scale $r_c \sim 15$~Mpc$/h$, and vanish for $r \gg r_c$;
and the normalized ADPD should be flat, i.e., independent of $\theta$, with constant value $\text{ADPD}(\theta) \sim 1/180^\circ$.
In contrast, the ADPD heatmap for this realization shows clear deviations from these expectations. 
}

When analyzing the case of the ICs, we must consider that even for an isotropic random distribution, the angular pair counts in each angular bin fluctuate due to finite sampling. This results in a non-zero angular variance, even though the expected mean over many realizations is flat. In the ICs, we expect to detect finite-size statistical noise, as well as possible residual imprints from the pre-initial lattice configuration.  To quantify the contribution of random noise versus structural anisotropy to the observed angular variance, we compare the ICs to a Poisson reference sample. Specifically, we generate a Poisson random distribution with the same number of points and box size as the ICs. We then compute the angular profile of the ADPD in the same radial range and measure the angular variance as a function of radial scale $r$. This provides a baseline expectation for the variance arising from a purely isotropic distribution under identical conditions.

\begin{figure} 
\includegraphics[width=0.49\textwidth]{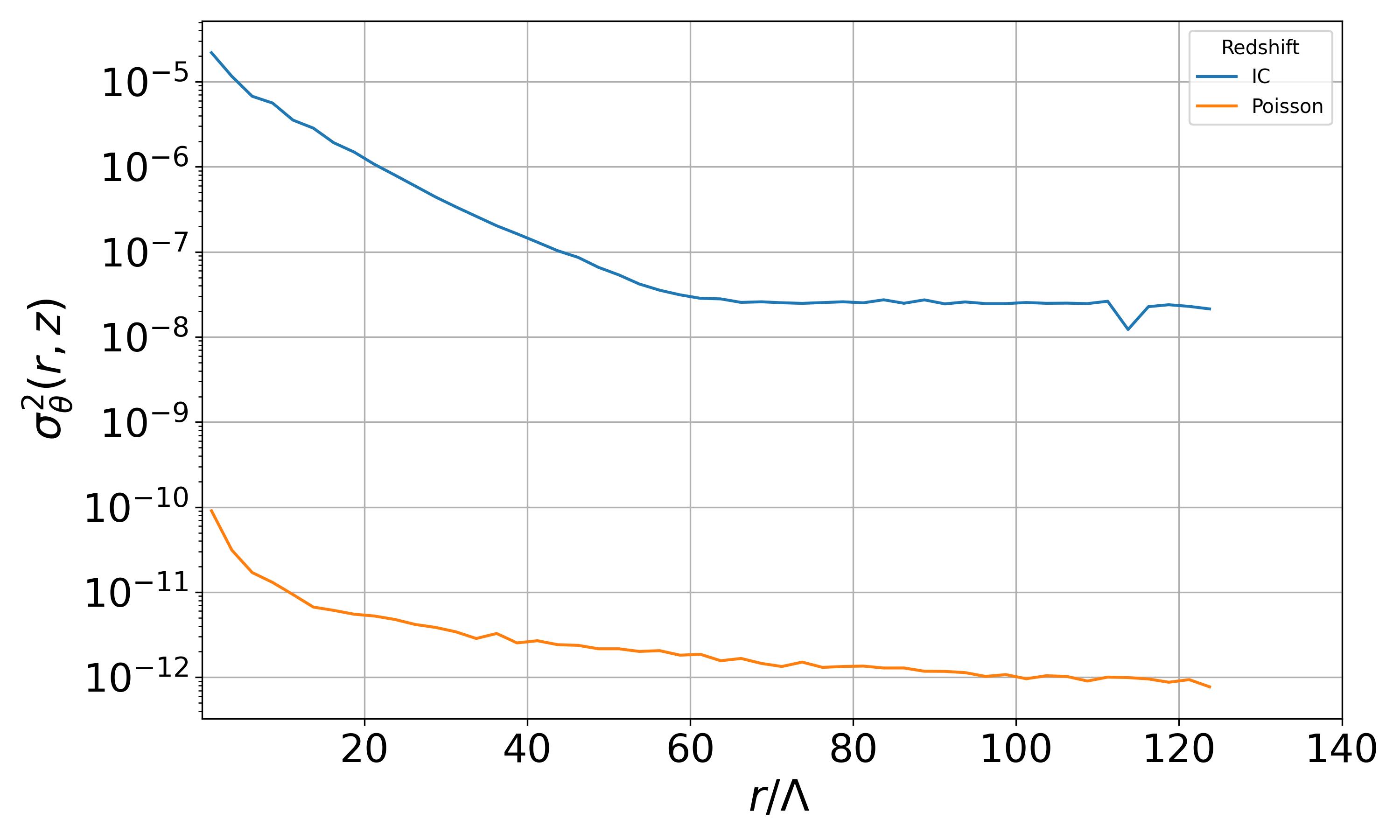} 
\caption{   
Angular variance of the ADPD computed from the initial conditions  of the small-box simulation. This diagnostic reveals the presence of directional anisotropies imprinted by the coupling between the isotropic displacement field and the underlying regular lattice.
Length scales are expressed in units of the average distance between nearest neighbors $\Lambda$.}
\label{variance_poisson_smallbox} 
\end{figure}

By defining
\begin{equation}
\Delta \sigma_\theta^2(r) = \sigma^2_{\theta,\text{IC}}(r) - \sigma^2_{\theta,\text{Poisson}}(r),
\end{equation}
we can assess the presence of excess anisotropy. If $\Delta \sigma_\theta^2(r) > 0$, it indicates that the  ICs exhibit anisotropies beyond what is expected from Poisson noise alone. In the case of  the small box simulation $\sigma^2_{\theta,\text{Poisson}}(r)$ is more than {  5 orders} of magnitude smaller than $ \sigma^2_{\theta,\text{IC}}(r)$ (see Fig.\ref{variance_poisson_smallbox}).
{  In addition, one can observe that, {  at large enough $r$},  $\sigma^2_{\theta,\text{IC}}(r) \approx \text{const.}$, whereas the theoretical expectation scales as $\propto r^{-4}$ (see Eq.~\ref{sigma_theta_2D}).} 

\subsubsection{Time Evolution of Anisotropies in the Small-Box Simulation}
\label{sec:adpd_evolution}

\begin{figure*} 
\includegraphics[width=0.24\textwidth]{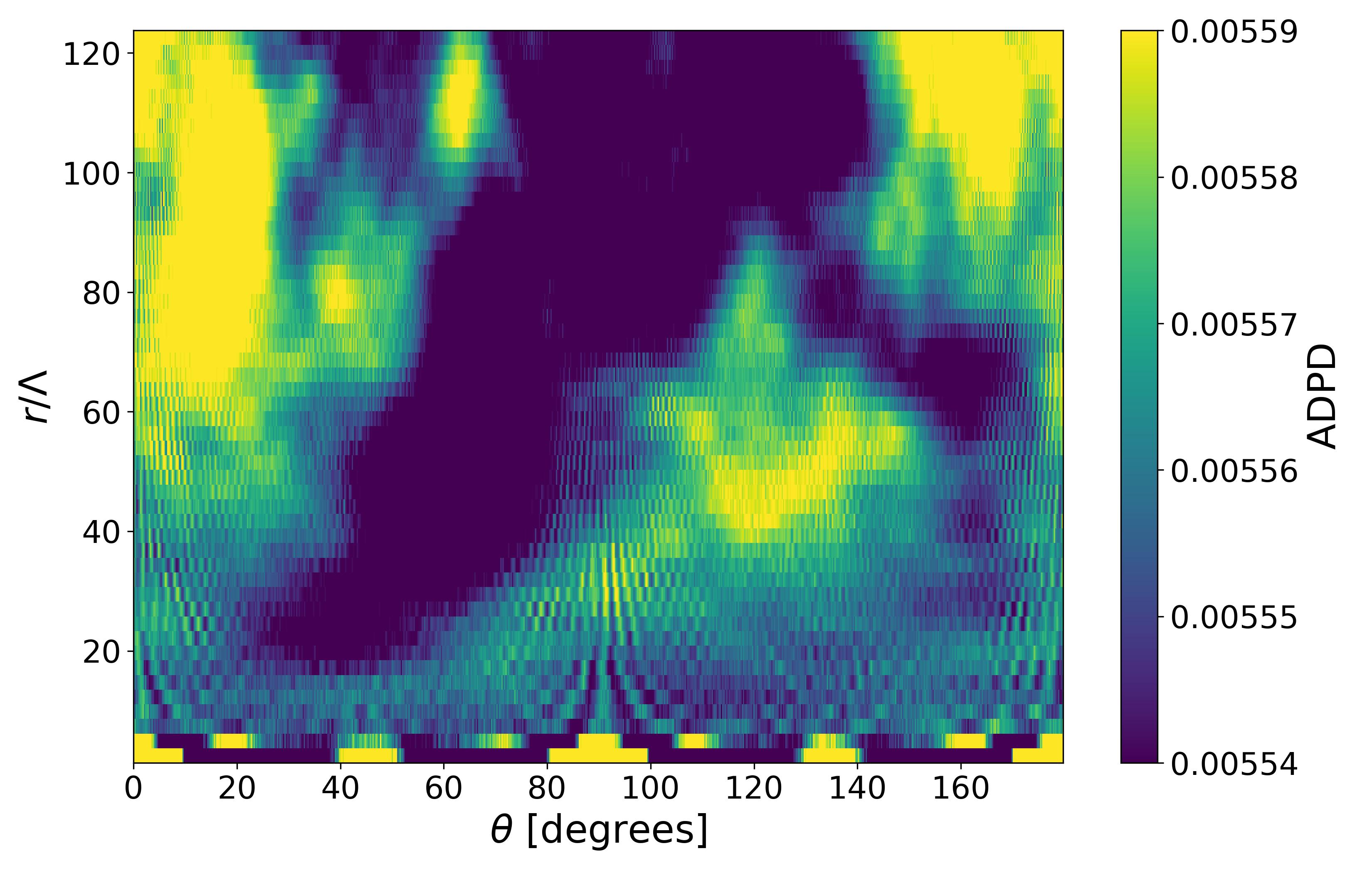} 
\includegraphics[width=0.24\textwidth]{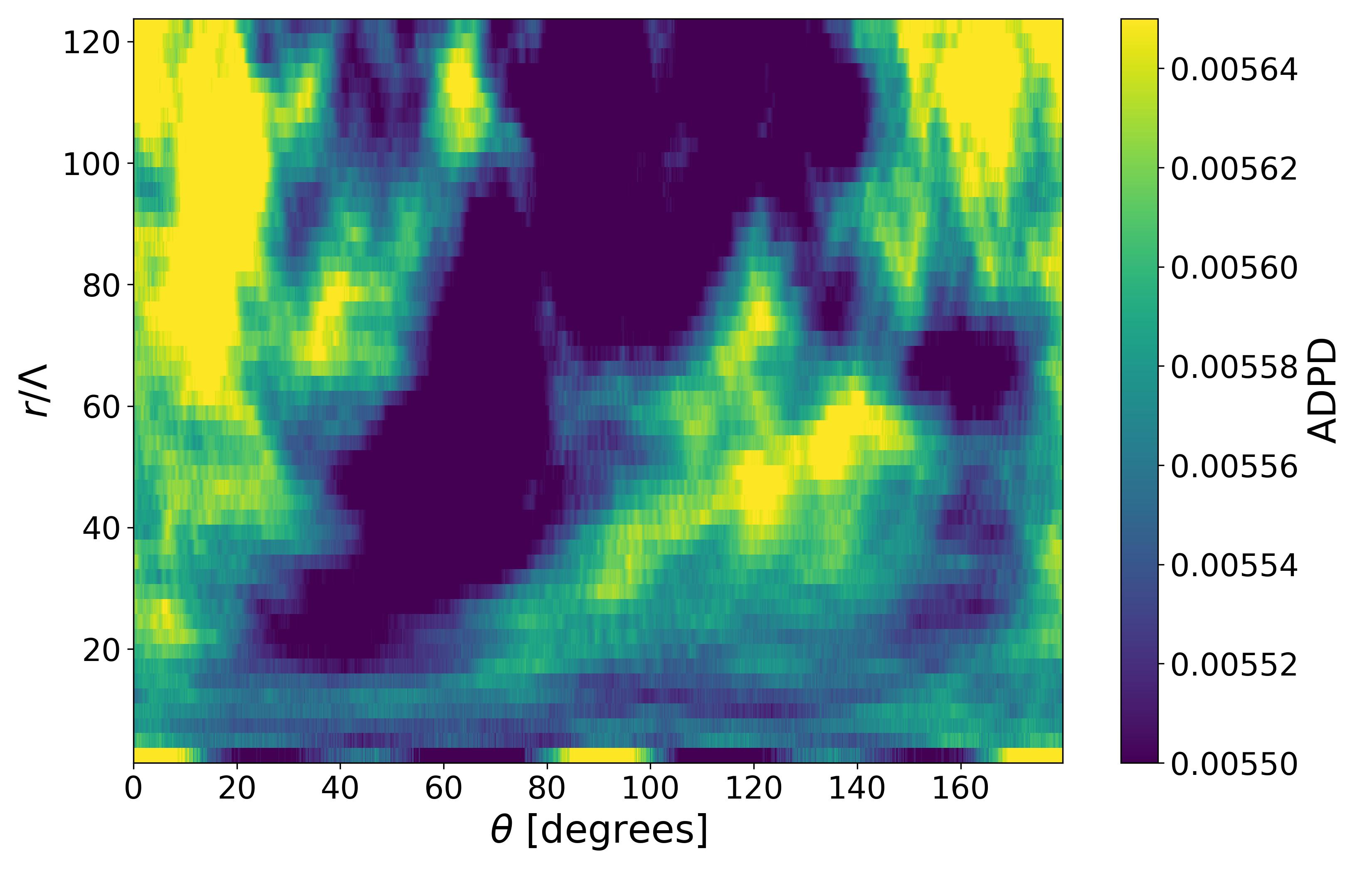} 
\includegraphics[width=0.24\textwidth]{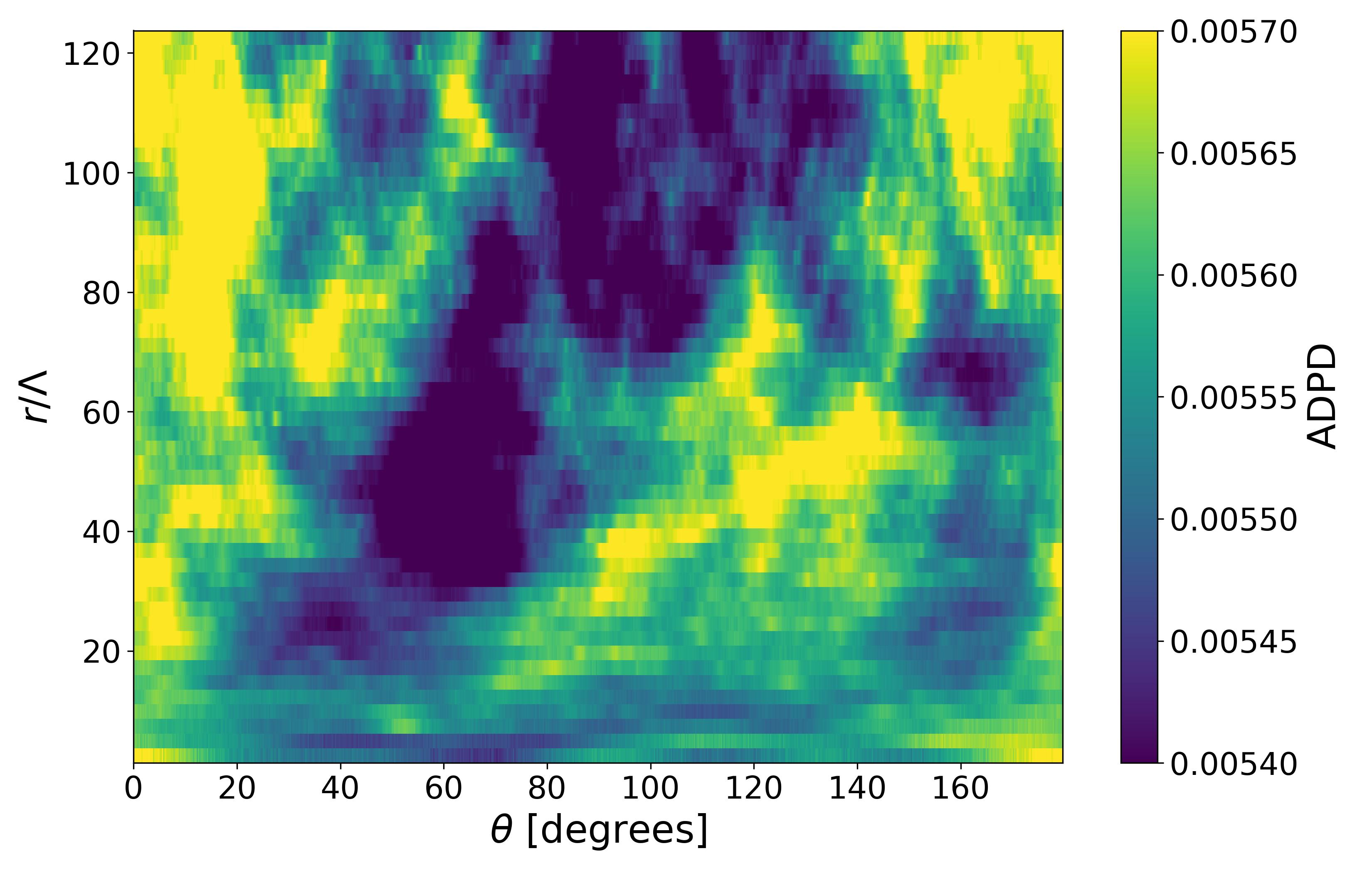} 
\includegraphics[width=0.24\textwidth]{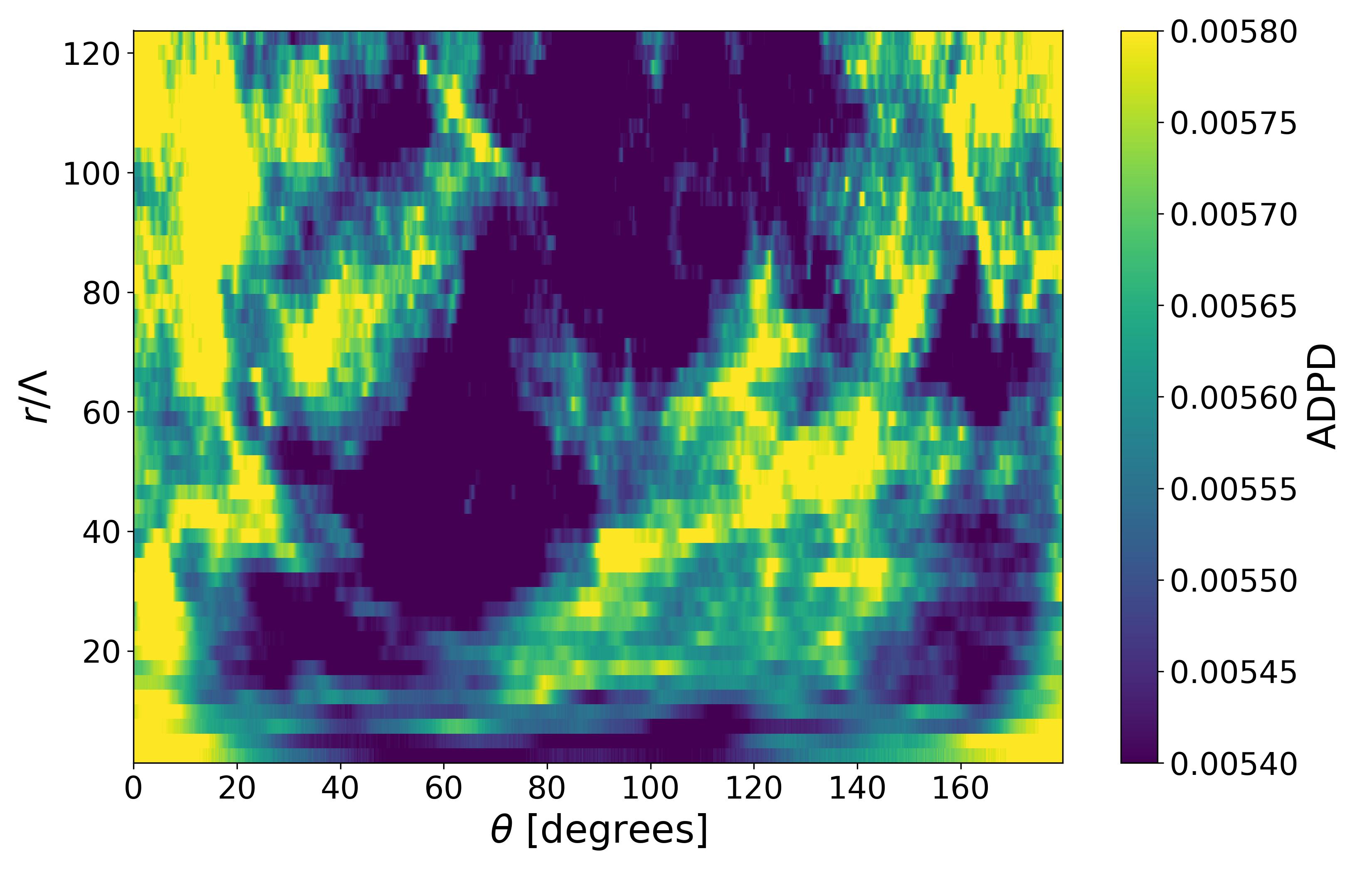} \\
\includegraphics[width=0.24\textwidth]{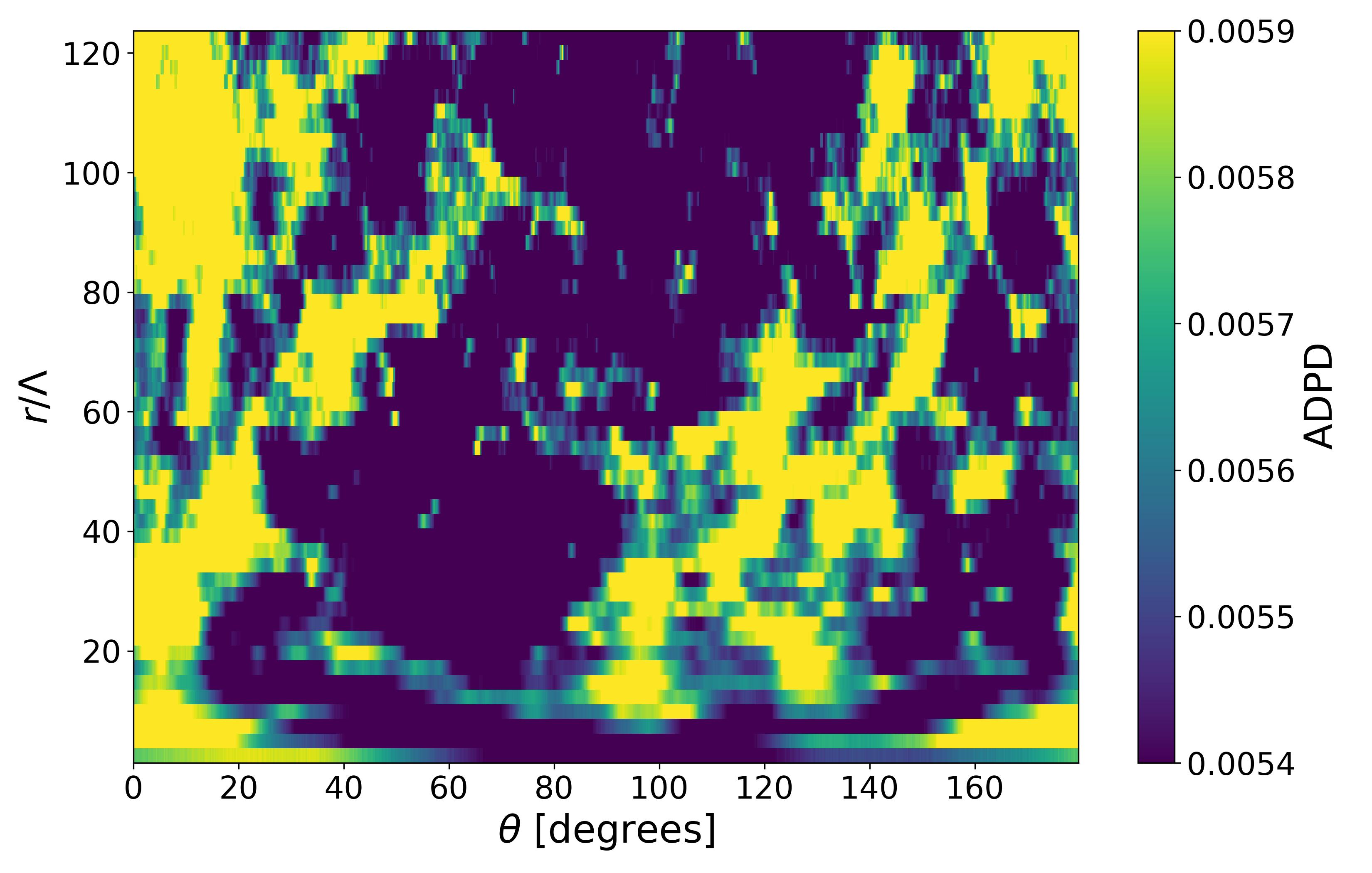} 
\includegraphics[width=0.24\textwidth]{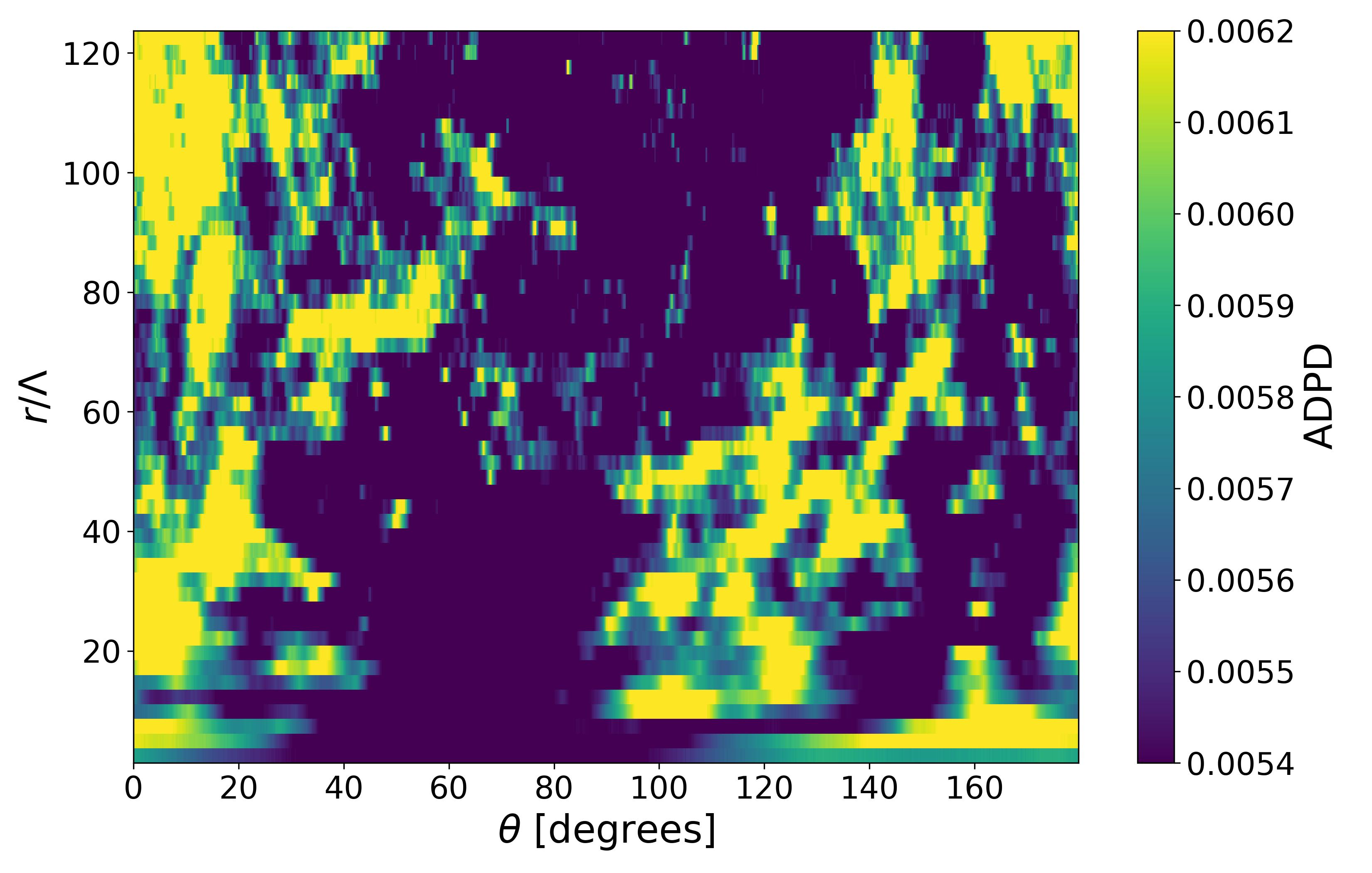} 
\includegraphics[width=0.24\textwidth]{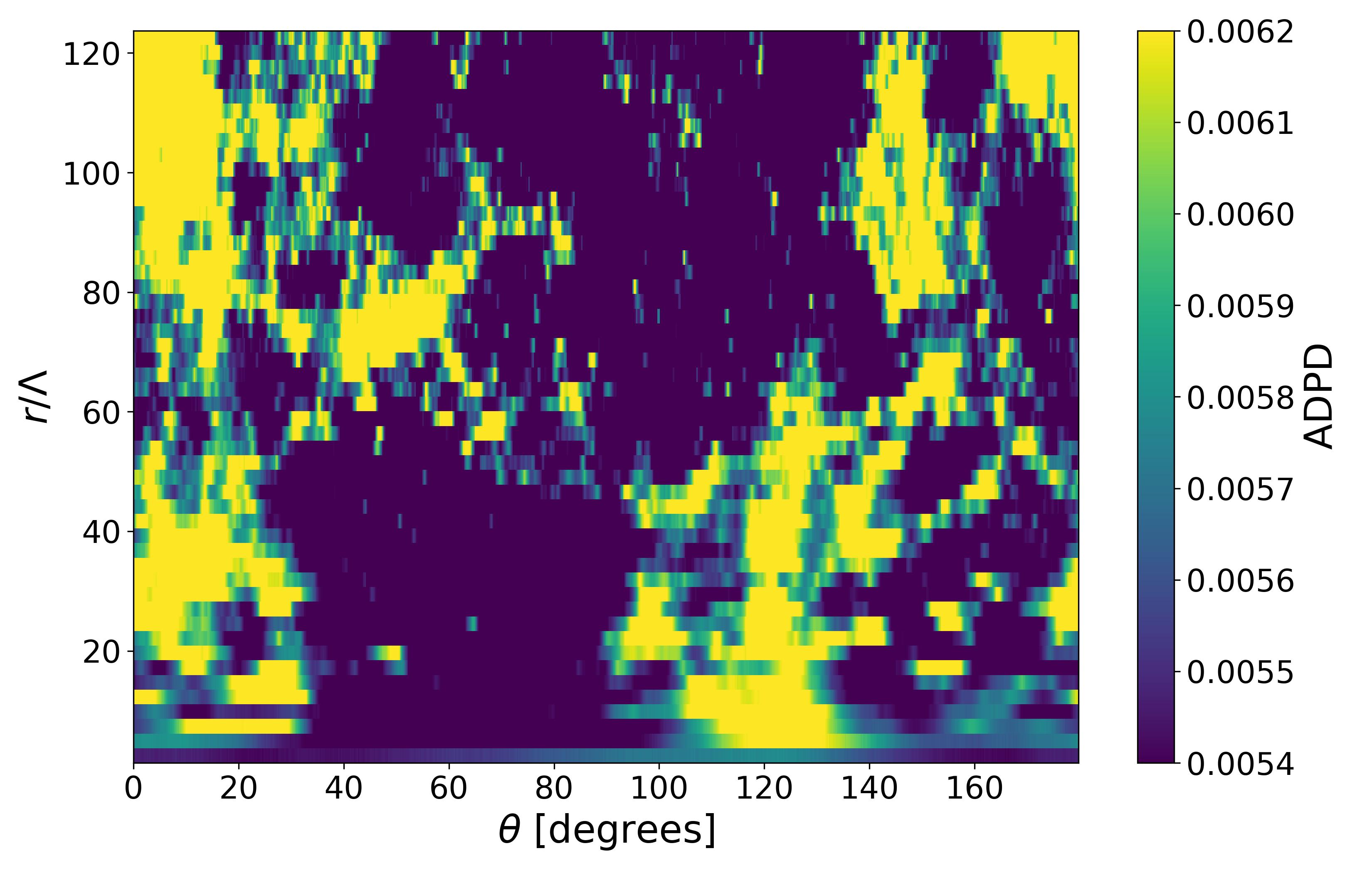} 
\includegraphics[width=0.24\textwidth]{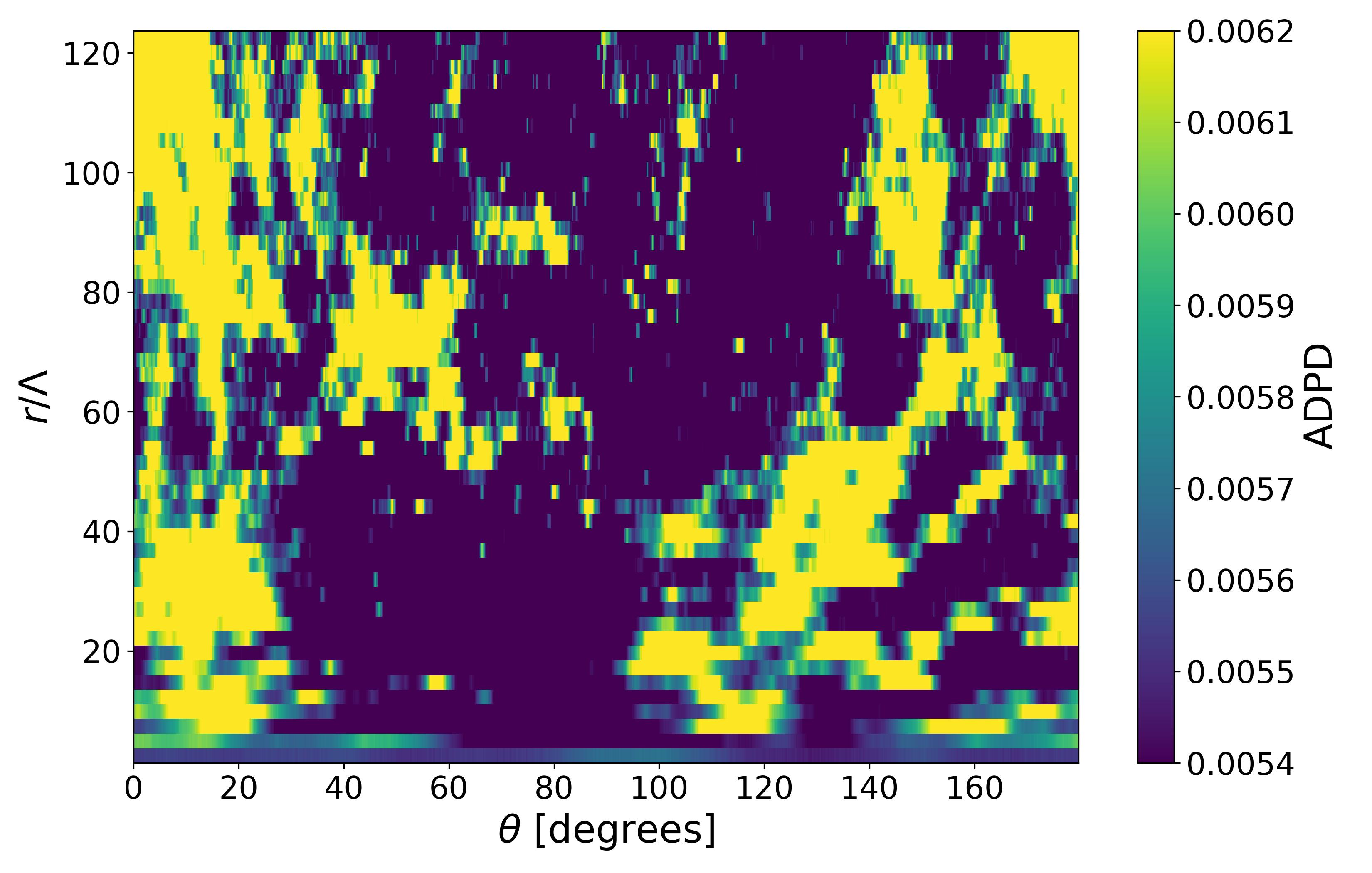} 
\caption{  Heat-map of the 
 ADPD  for 
 or a slice of thickness $\Delta z= 5$ Mpc  of the small box CDM simulation
 at different redshifts: 
 first row {  $z=9.0, 5.1, 3.7,2.7$, } 
 second row 
 $z=1.2, 0.7, 0.35, 0.05$. The ICs are shown in Fig.\ref{adpd_heatmap_small_box_IC}.
 Length scales are expressed in units of the average distance between nearest neighbors $\Lambda$.}
\label{adpd_heatmap_small_box} 
\end{figure*}

\begin{figure*} 
\includegraphics[width=0.24\textwidth]{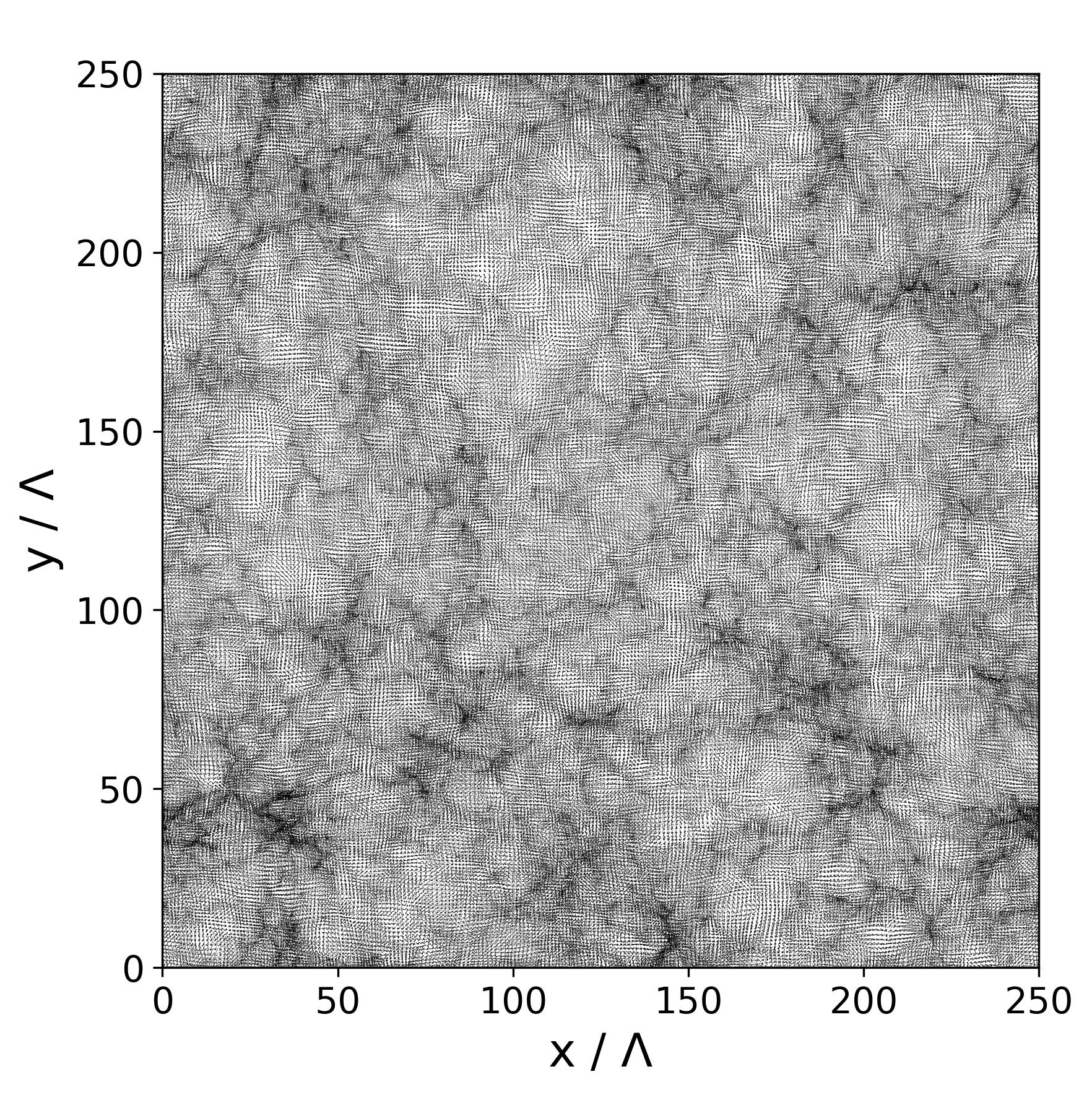} 
\includegraphics[width=0.24\textwidth]{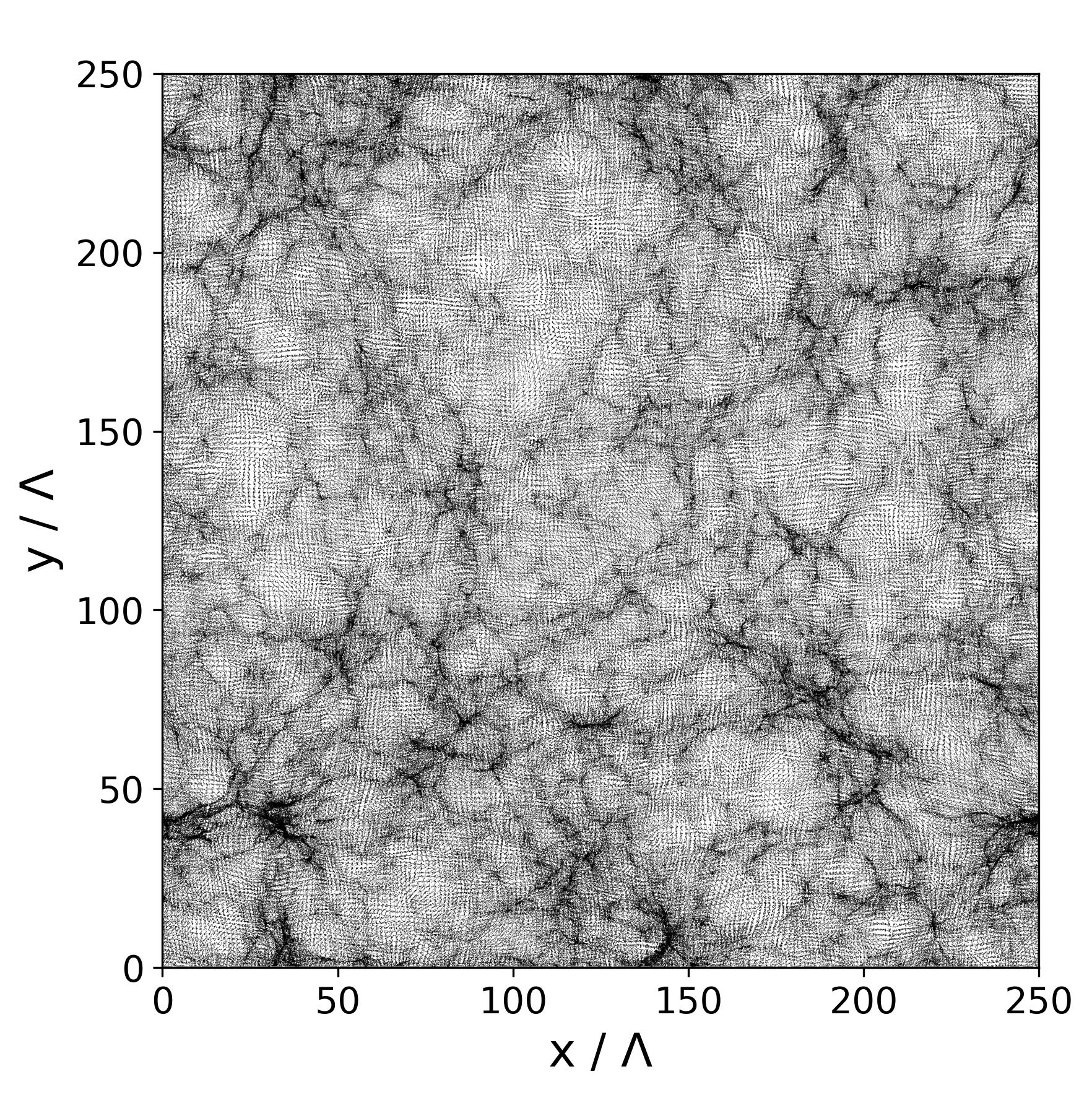} 
\includegraphics[width=0.24\textwidth]{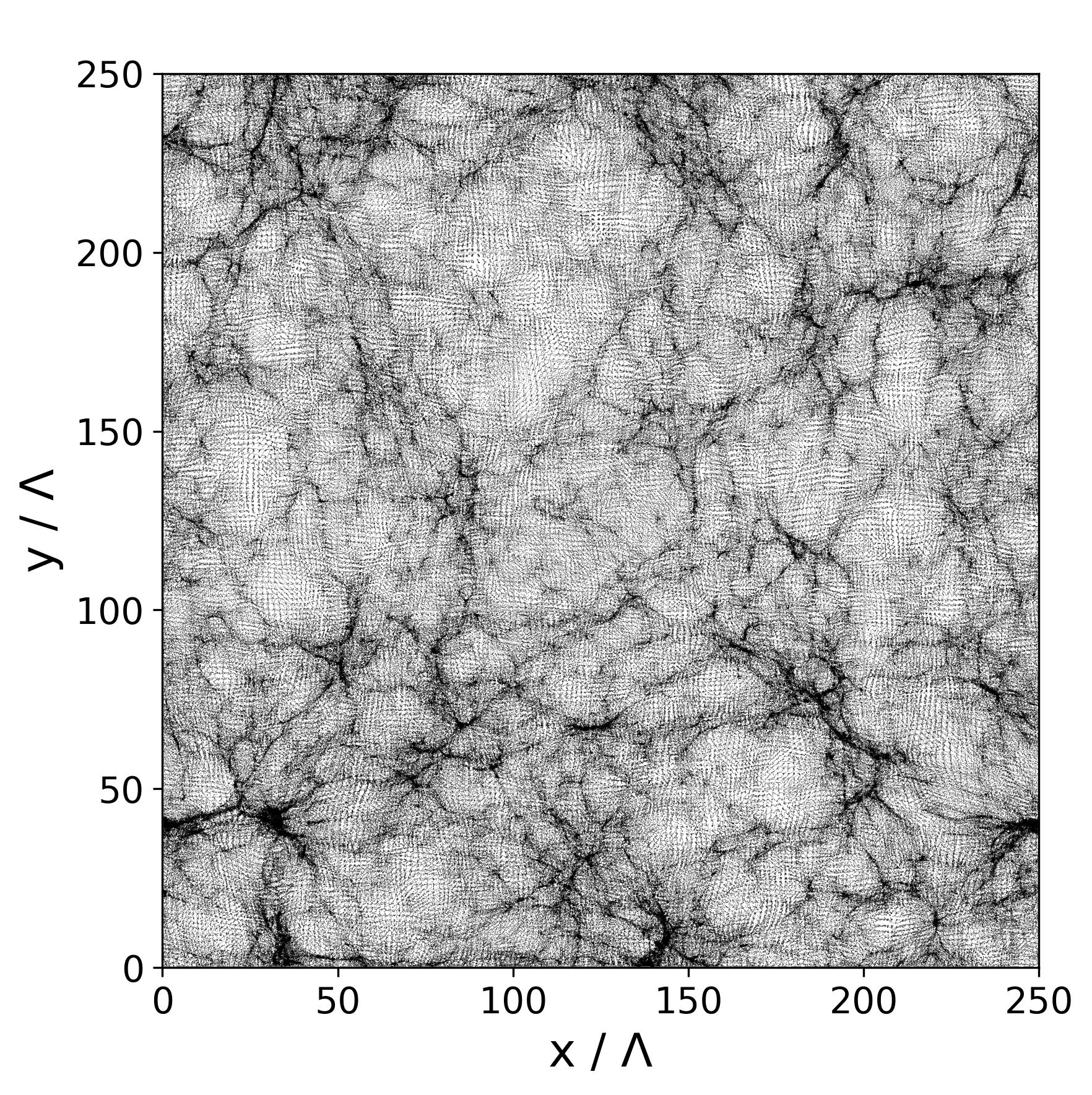} 
\includegraphics[width=0.24\textwidth]{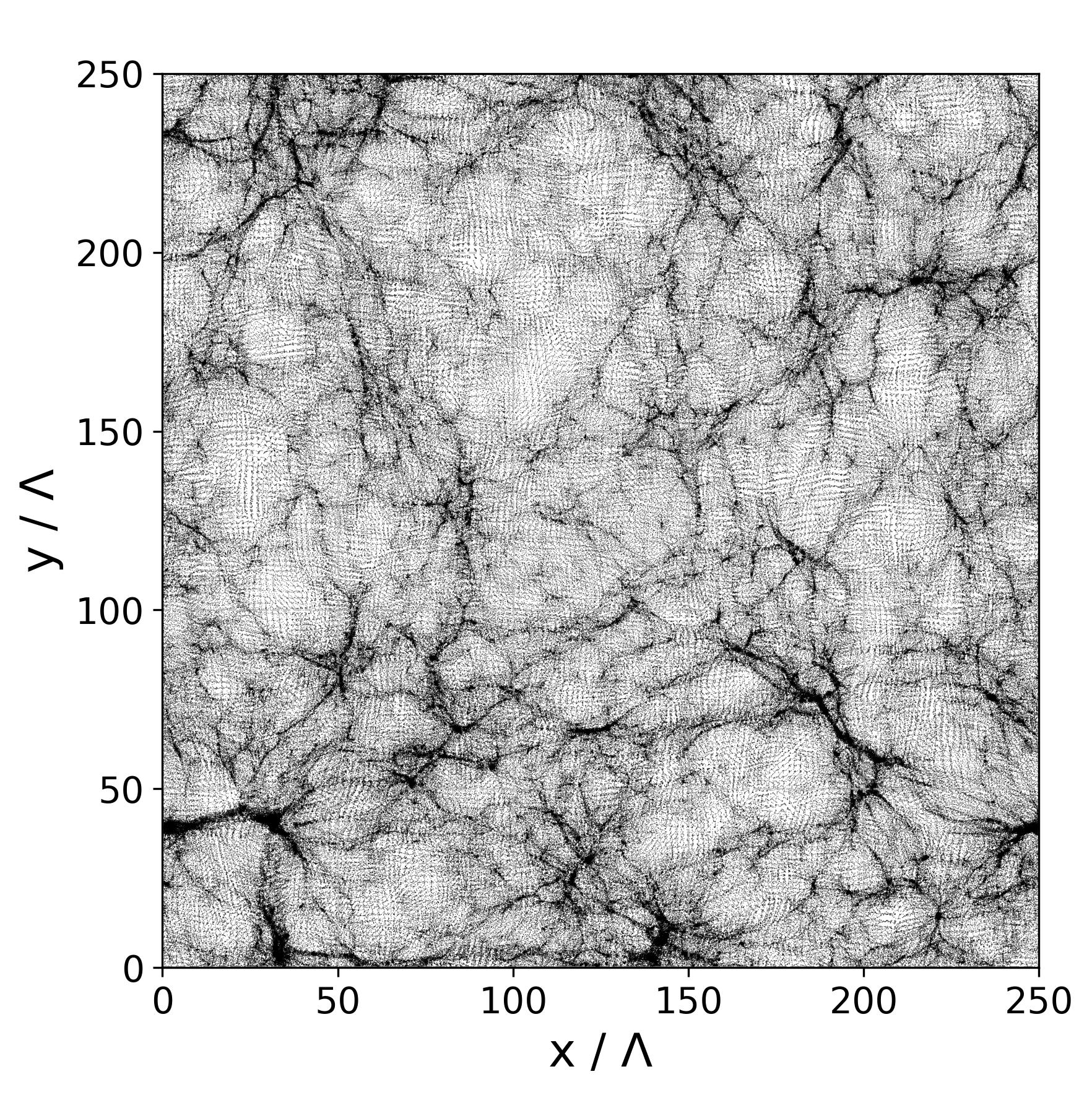}\\ 
\includegraphics[width=0.24\textwidth]{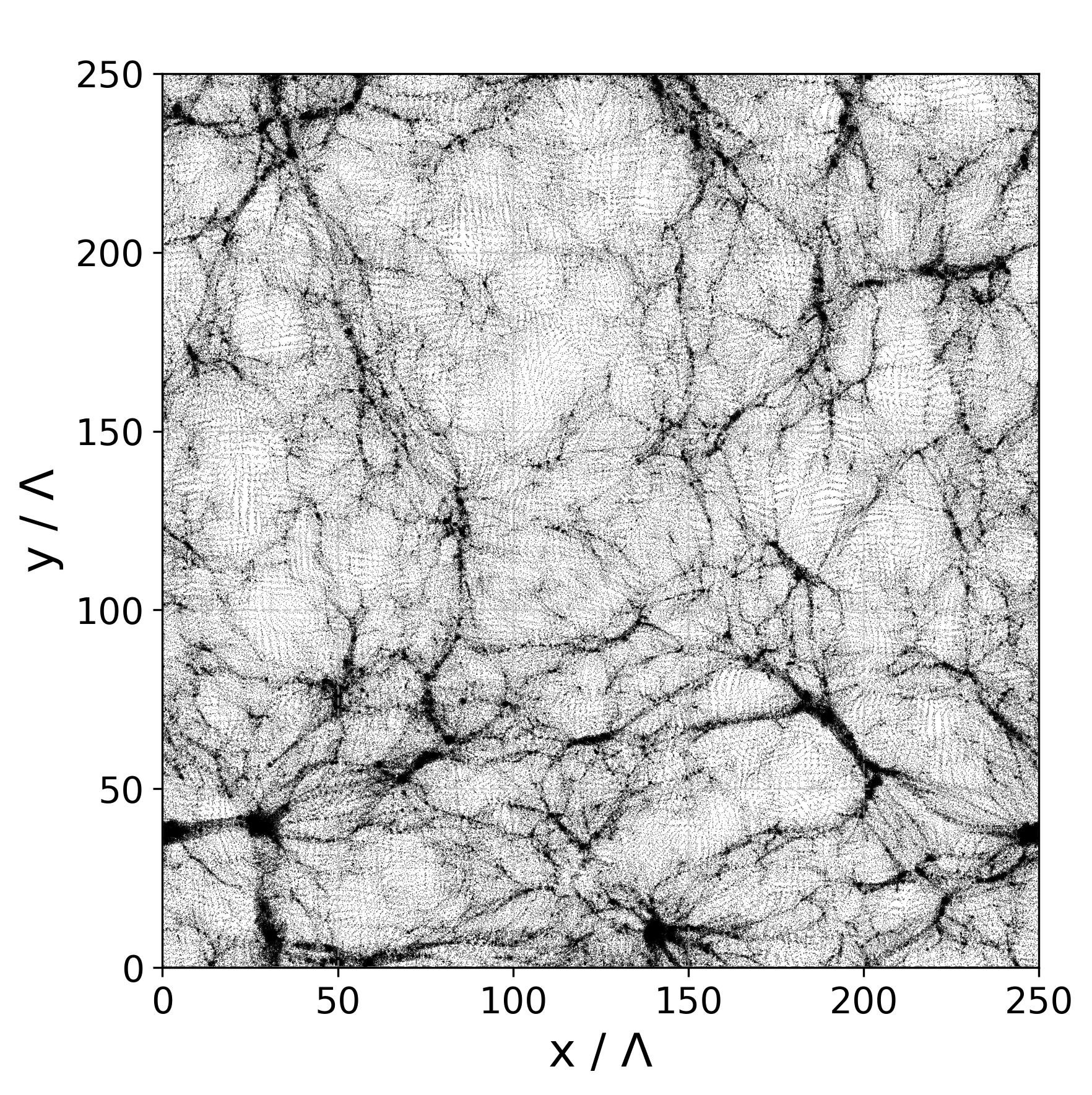} 
\includegraphics[width=0.24\textwidth]{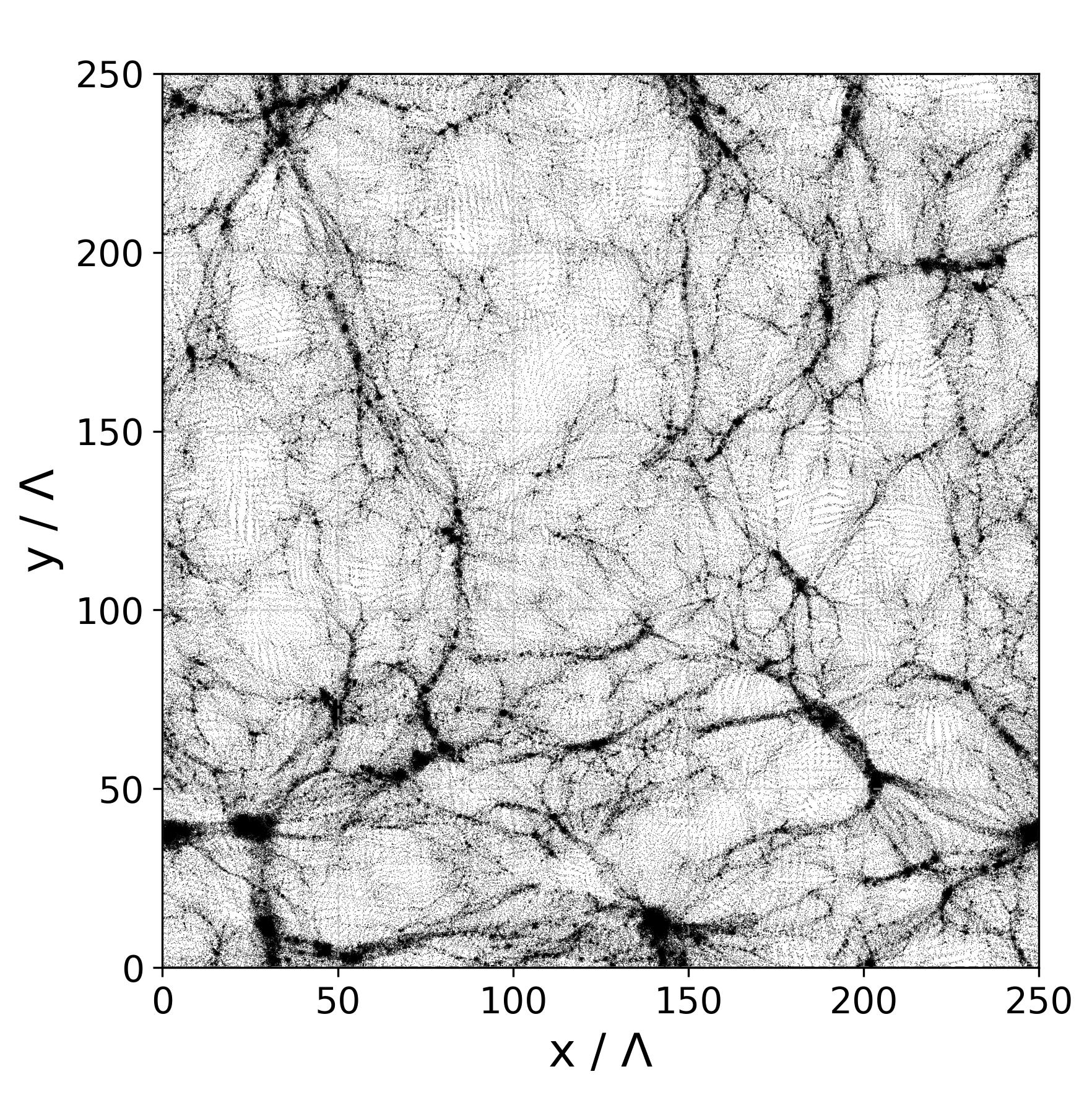} 
\includegraphics[width=0.24\textwidth]{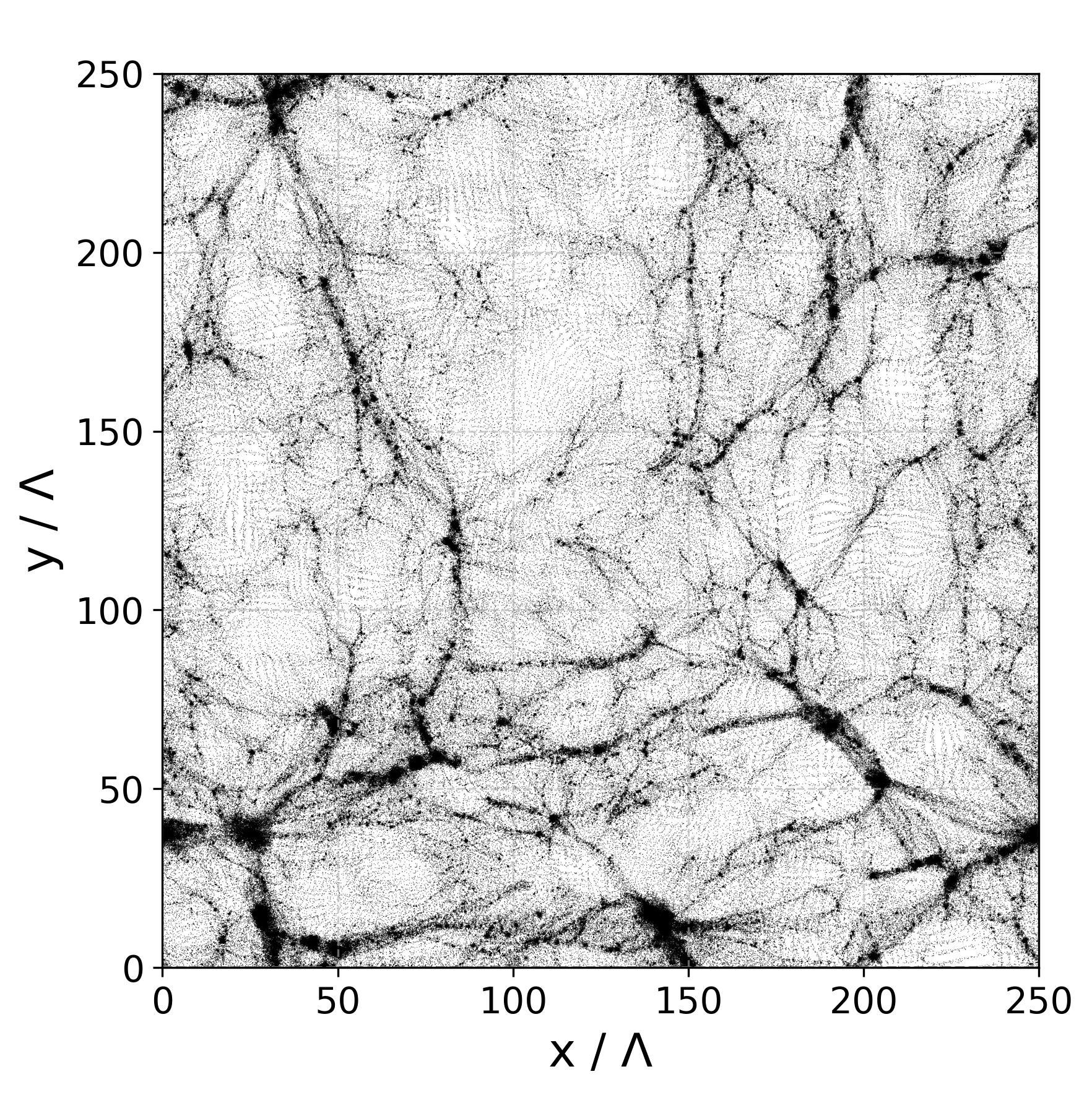} 
\includegraphics[width=0.24\textwidth]{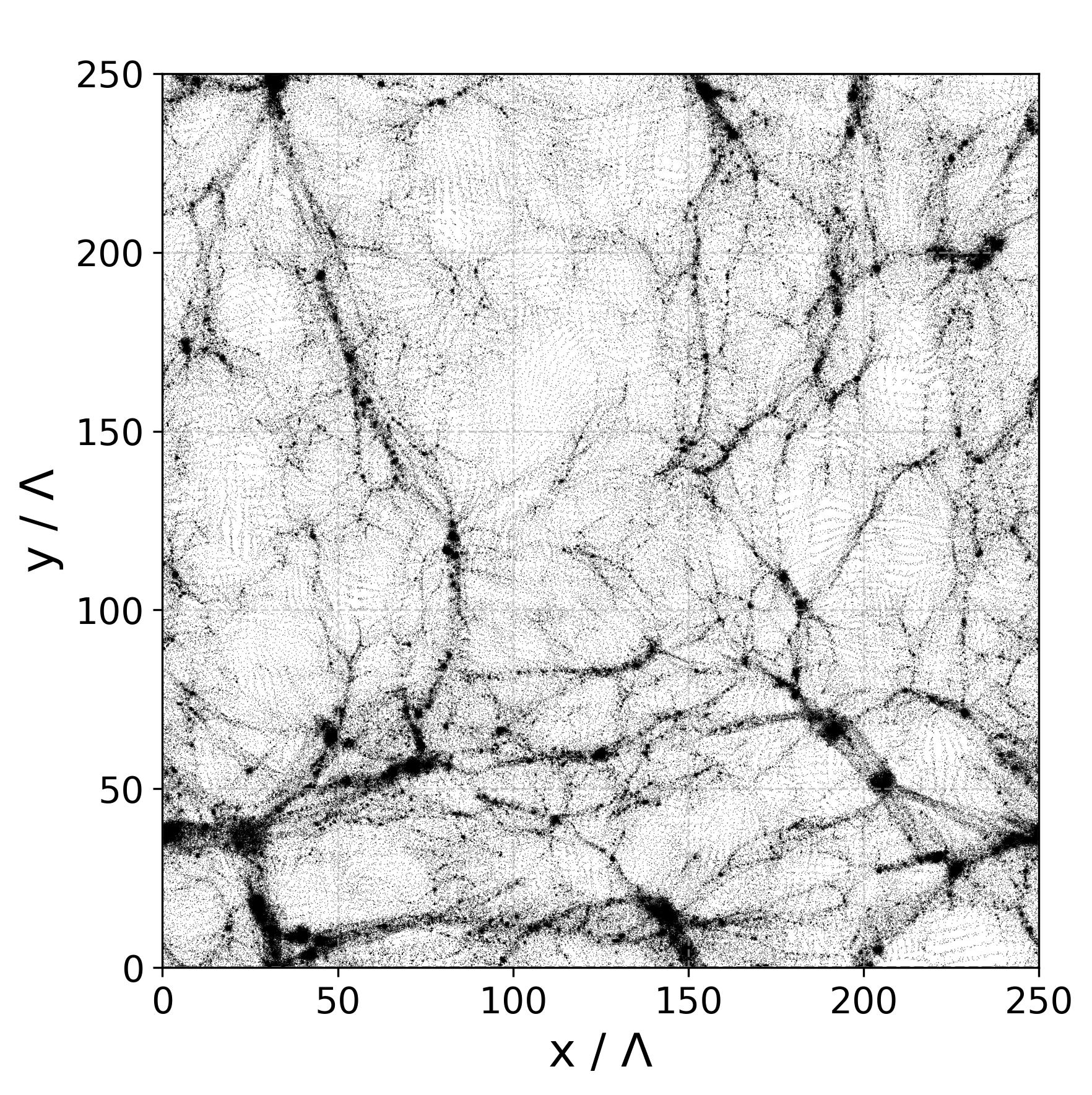} 
\caption{  Point distribution  for a  the same slices of Fig.\ref{adpd_heatmap_small_box}.  
Length scales are expressed in units of the average distance between nearest neighbors $\Lambda$.}
\label{point_small_box} 
\end{figure*}

\begin{figure*}[htbp]
  \centering
  \includegraphics[width=0.49\textwidth]{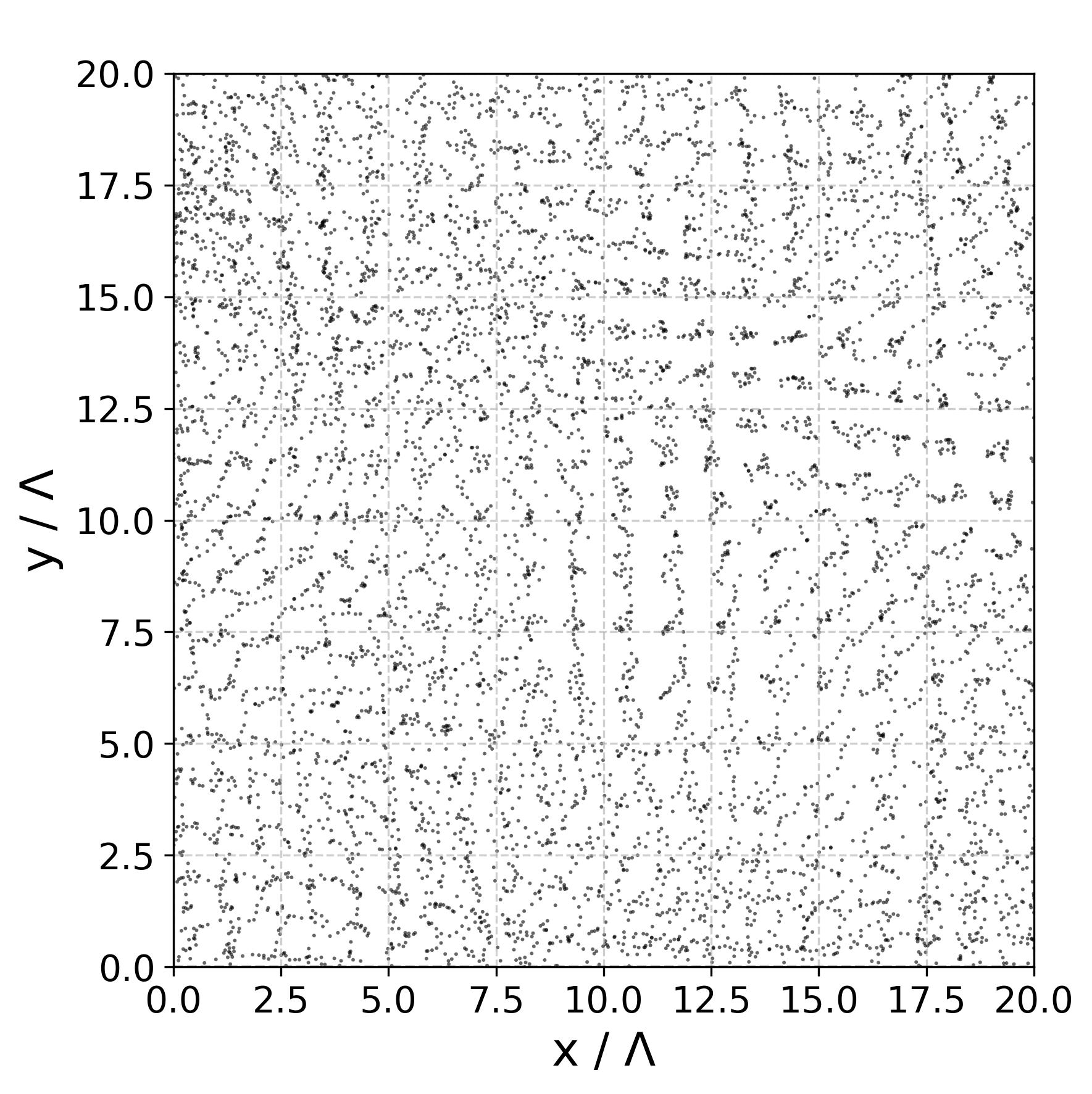}
  \includegraphics[width=0.49\textwidth]{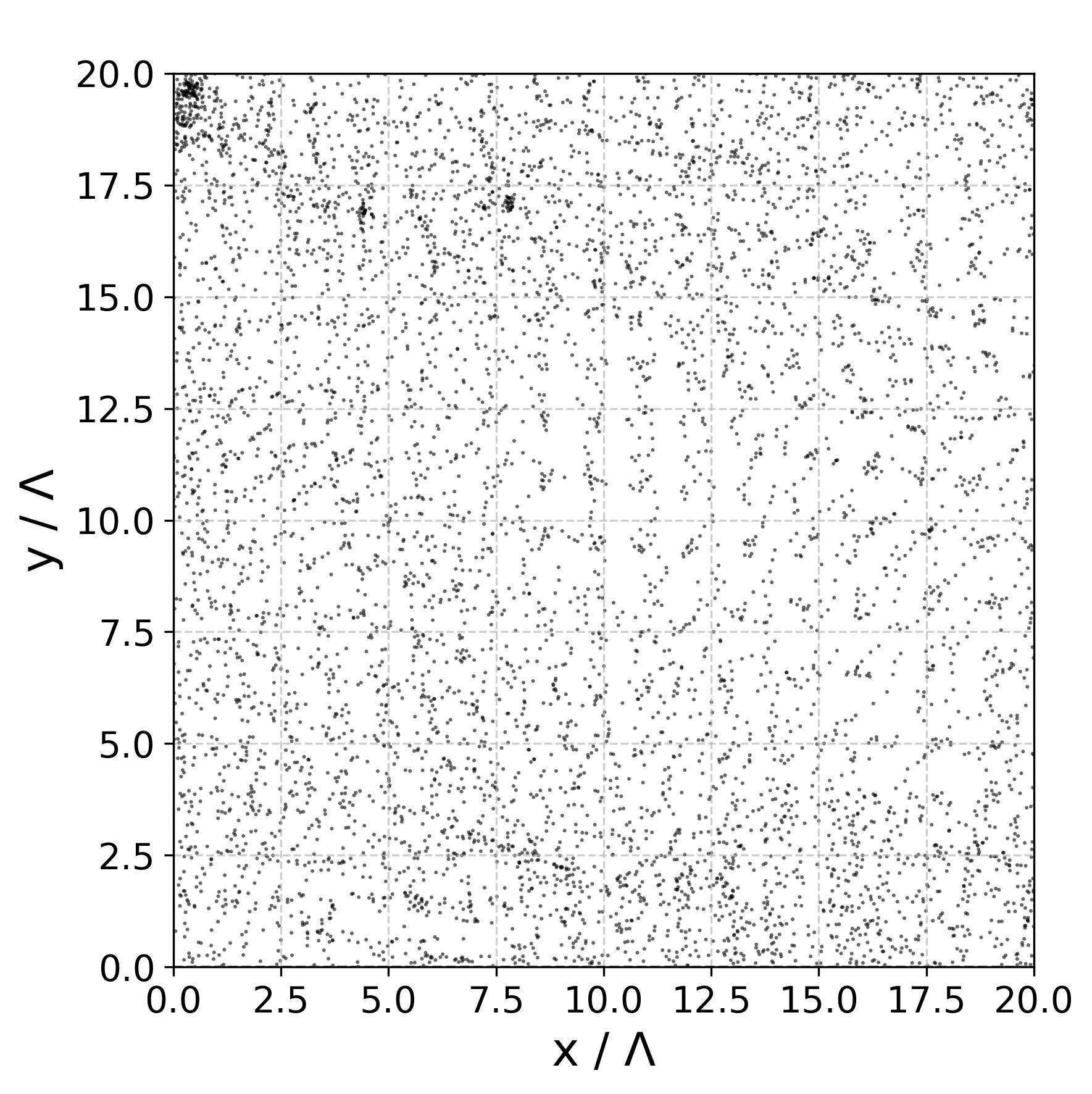}
  \caption{    
  Left panel: zoom of the point distribution for the IC of the small box at $z=9$.
  Right panel: zoom of the point distribution at $z=5.1$. 
  Length scales are expressed in units of the average distance between nearest neighbors $\Lambda$.}
  \label{point_distribution_zoom_small_box}
\end{figure*}

Fig.~\ref{adpd_heatmap_small_box} displays the evolution of the ADPD heatmap for the small-box simulation  at different redshifts. The corresponding projected particle distributions at each redshift are shown in Fig.~\ref{point_small_box}. 
{  
At redshift $z = 9$, the heatmap of the ADPD still exhibits distinct patterns, notably sharp peaks at angular positions $\theta = 0^\circ, 90^\circ, 180^\circ, 270^\circ$ for small separations ($r < 2$~Mpc$/h$), reflecting the underlying lattice structure (see Fig.\ref{adpd_heatmap_small_box_IC}). At larger separations, these peaks become broader and display secondary substructures, indicating the emergence of more complex angular correlations induced by the displacement field applied to the initial lattice. These features remain visible in the ADPD heatmap at all subsequent redshifts, with their overall amplitude progressively increasing.

Fig.~\ref{point_distribution_zoom_small_box} shows a zoomed view of the two-dimensional point distribution for the small-box simulation at redshifts $z = 9$ and $z = 5.1$. In the earlier snapshot ($z = 9$), the lattice structure---clearly visible in the ICs (see Fig.~\ref{adpd_heatmap_small_box_IC})---is still apparent, whereas in the later snapshot ($z = 5.1$) it has been partially erased by gravitational evolution. This evolution is consistent with the trends observed in the corresponding ADPD heatmaps.
}

As the simulation evolves to lower redshifts ($z < 9$), these initial anisotropies persist and gradually amplify. Even after nonlinear clustering becomes dominant at small scales ($r < 10$~Mpc$/h$), the ADPD retains the imprint of initial anisotropic features at larger scales. This behavior is consistent with the expectation that gravitational evolution amplifies coherent features present in the ICs.

To study this amplification, we examine three representative scales:
\begin{itemize}
    \item At $r \sim 10$~Mpc$/h$, the ADPD evolves rapidly. Peak amplitudes at preferred angles decrease while broadening due to particle motion. The peak structure remains visible until $z \approx 4$, and is effectively frozen by $z \approx 2$, consistent with early nonlinearity at small scales.
    \item At $r \sim 15$~Mpc$/h$, where the dynamics are weakly nonlinear, the ADPD evolves more gradually. Initial features remain discernible and become increasingly pronounced at $z < 3$.
    \item At $r \sim 30$~Mpc$/h$, the evolution is slower. Initial anisotropies persist and gradually sharpen, while new peaks emerge between $z = 4$ and $z = 2$, indicating the onset of filament formation.
\end{itemize}

Overall, the angular anisotropies present in the ICs are amplified by gravitational evolution. This is clearly demonstrated by comparing the ADPD at $z = 9$ and $z = 0$: although the anisotropic features observed at late times differ in structure, they remain correlated with those imprinted in the initial configuration. This suggests that even small-amplitude directional patterns in the ICs can act as seeds for the filamentary structures that emerge during cosmic evolution, including within the linear and quasi-linear regimes.

While these anisotropies may appear subtle, they nonetheless violate the assumption of statistical isotropy that underpins the standard $\Lambda$CDM model. Gravitational evolution, being nonlinear and sensitive to coherent perturbations, can amplify such features and render them dynamically significant. Crucially, these directional artifacts are not detectable through conventional angle-averaged diagnostics, such as the two-point correlation function $\xi(r)$ or the PS $P(k)$, yet they can influence the formation and orientation of filaments, potentially introducing systematic biases into the simulated large-scale structure.

\begin{figure} 
\includegraphics[width=0.49\textwidth]{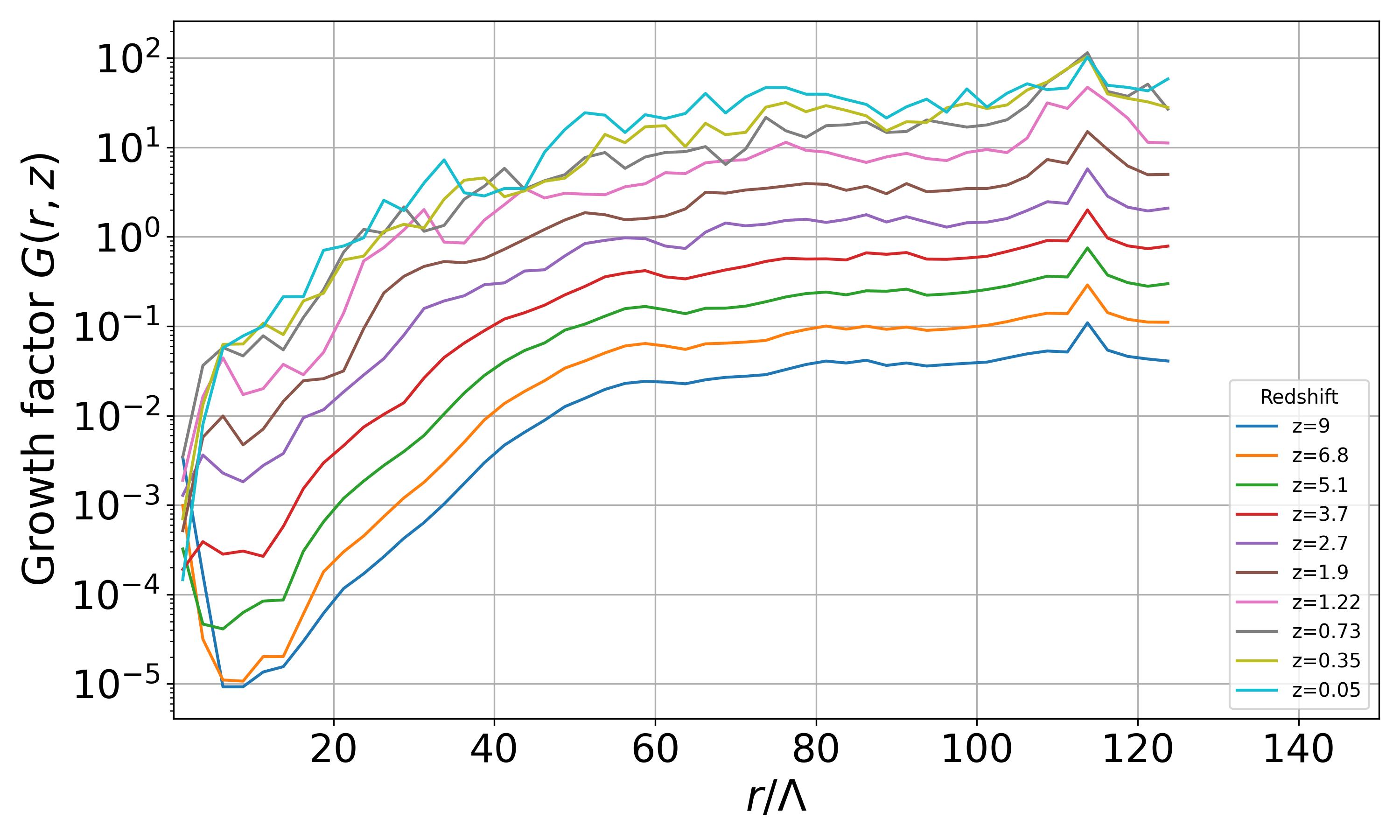} 
\caption{    
Variance growth factor for the small-box simulation, illustrating the amplification of angular anisotropies over time.
 Length scales are expressed in units of the average distance between nearest neighbors $\Lambda$.}
\label{variance_growth_factor_smallbox} 
\end{figure}

The next issue concerns how to quantify the amplification of anisotropies due to gravitational clustering during cosmic evolution. As illustrated in Fig.~\ref{adpd_heatmap_small_box}, the angular variance of the ADPD increases as redshift decreases, and the corresponding peaks become more pronounced --- indicating that anisotropic features are dynamically amplified over time.  To quantify this effect, we define a \textit{variance growth factor} as follows:
\begin{equation}
G(r, z) = \frac{\sigma^2_\theta(r, z)}{\sigma^2_\theta(r, z_{\mathrm{IC}})},
\end{equation}
where $\sigma^2_\theta(r, z)$ denotes the angular variance of the ADPD at pair separation $r$ and redshift $z$, and $z_{\mathrm{IC}}$ refers to the redshift at which the ICs are set. A value of $G(r, z) > 1$ thus reflects the amplification of anisotropic features relative to their initial amplitude. This metric provides a direct way to track the emergence and growth of directional coherence in the matter distribution throughout the simulation. By plotting $G(r, z)$ as a function of $r$, we can interpret the results as follows: If $G(r, z) \gg 1$, this indicates that initial anisotropies have been significantly amplified by gravitational clustering. If $G(r, z) \approx 1$, the angular variance likely originates from Poisson noise and has not been amplified during evolution.  As  one may note from Fig.\ref{variance_growth_factor_smallbox}  the  variance growth factor is {  of order  $10^2$ at $z=0$ } showing a very large amplification of initial anisotropies. 

We conclude that the anisotropies induced by the coupling between the isotropic displacement field and the underlying regular lattice are not mere numerical noise, but systematic artifacts that persist and grow throughout the simulation. Their presence calls into question the statistical fidelity of lattice-based ICs as fair realizations of an isotropic Gaussian field.


\subsection{Large box simulation}

\begin{figure}[htbp]
 \centering
  \includegraphics[width=0.49\textwidth]{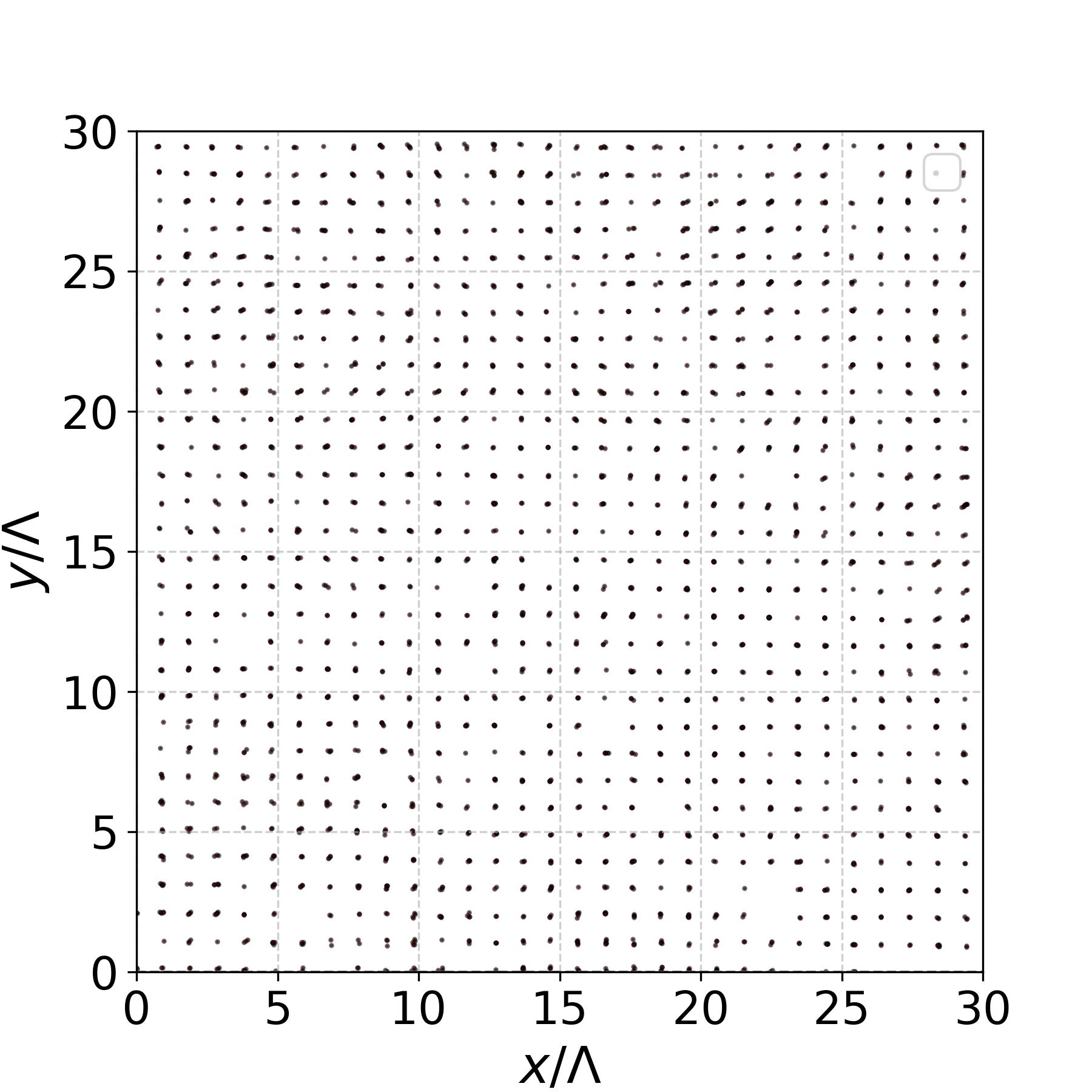}\\
  \includegraphics[width=0.49\textwidth]{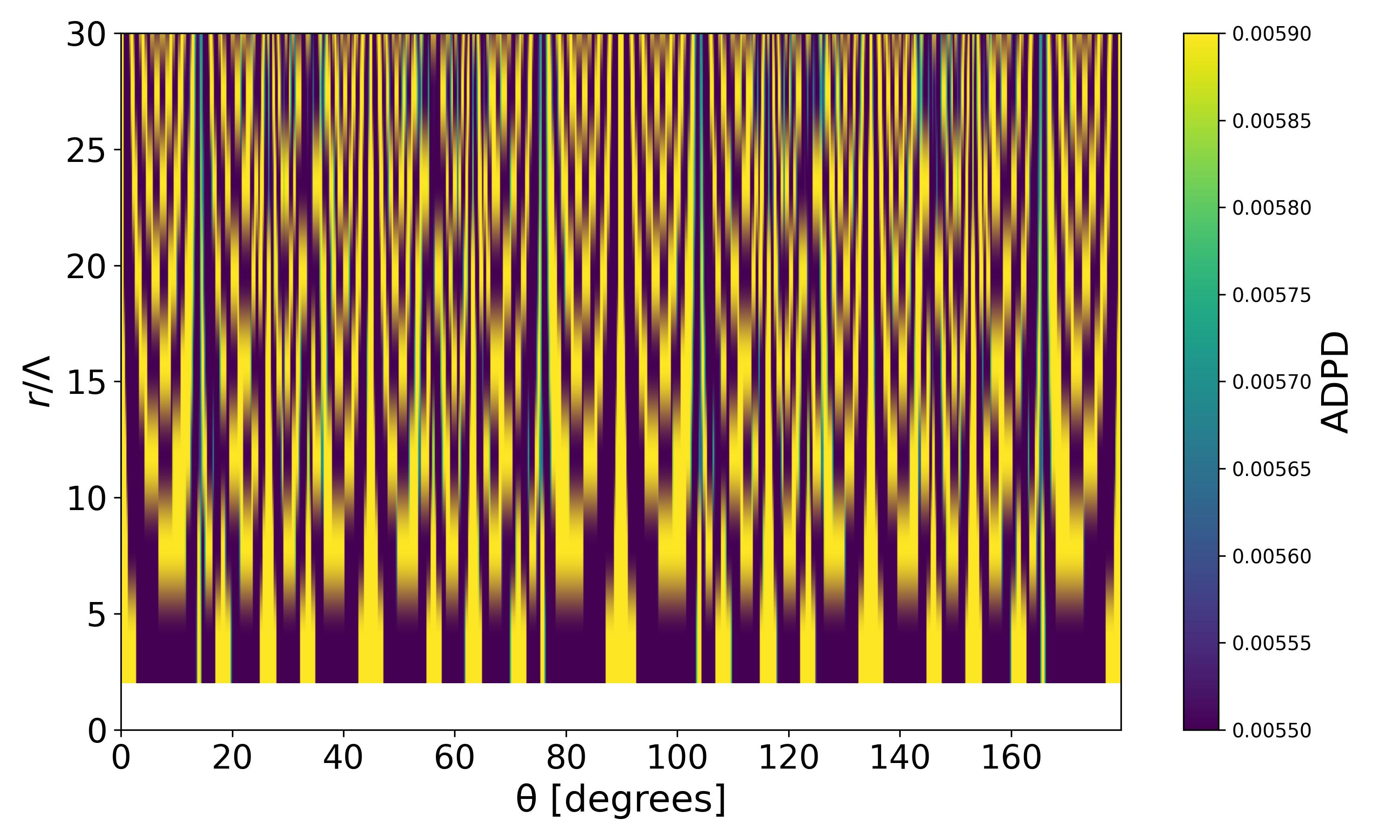}
  \caption{    
{Top panel:} Point distribution in a zoomed region of the initial conditions  for the large-box simulation at redshift $z_{\mathrm{IC}} = 49$ {(units: Mpc/$h$)}. 
{Bottom panel:} Heatmap of the Angular Distribution of Pairwise Distances   for the same region of the ICs shown above.
 Length scales are expressed in units of the average distance between nearest neighbors $\Lambda$.}
  \label{adpd_heatmap_large_box_IC} 
\end{figure}

\begin{figure*} 
\includegraphics[width=0.24\textwidth]{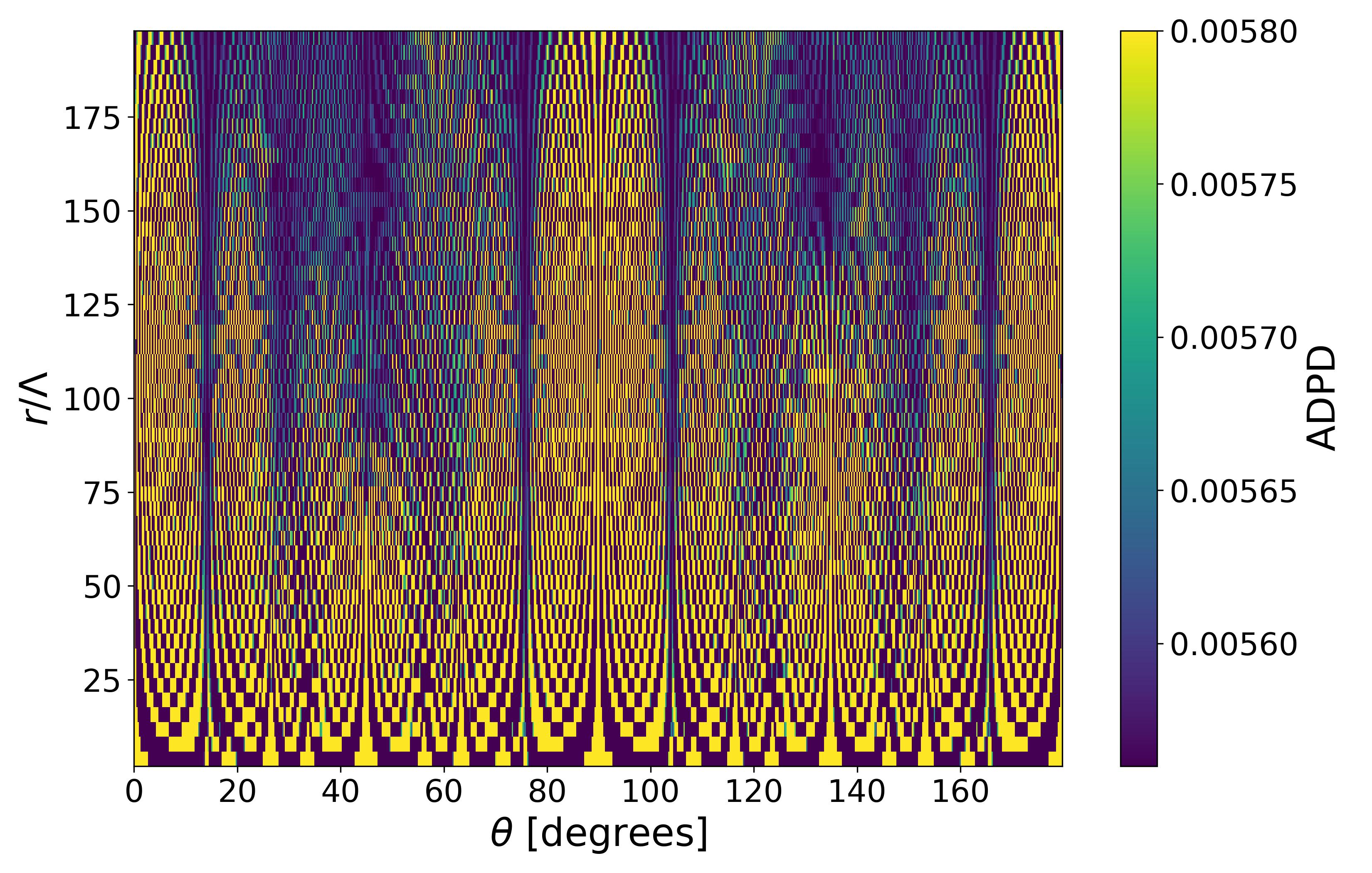}  
\includegraphics[width=0.24\textwidth]{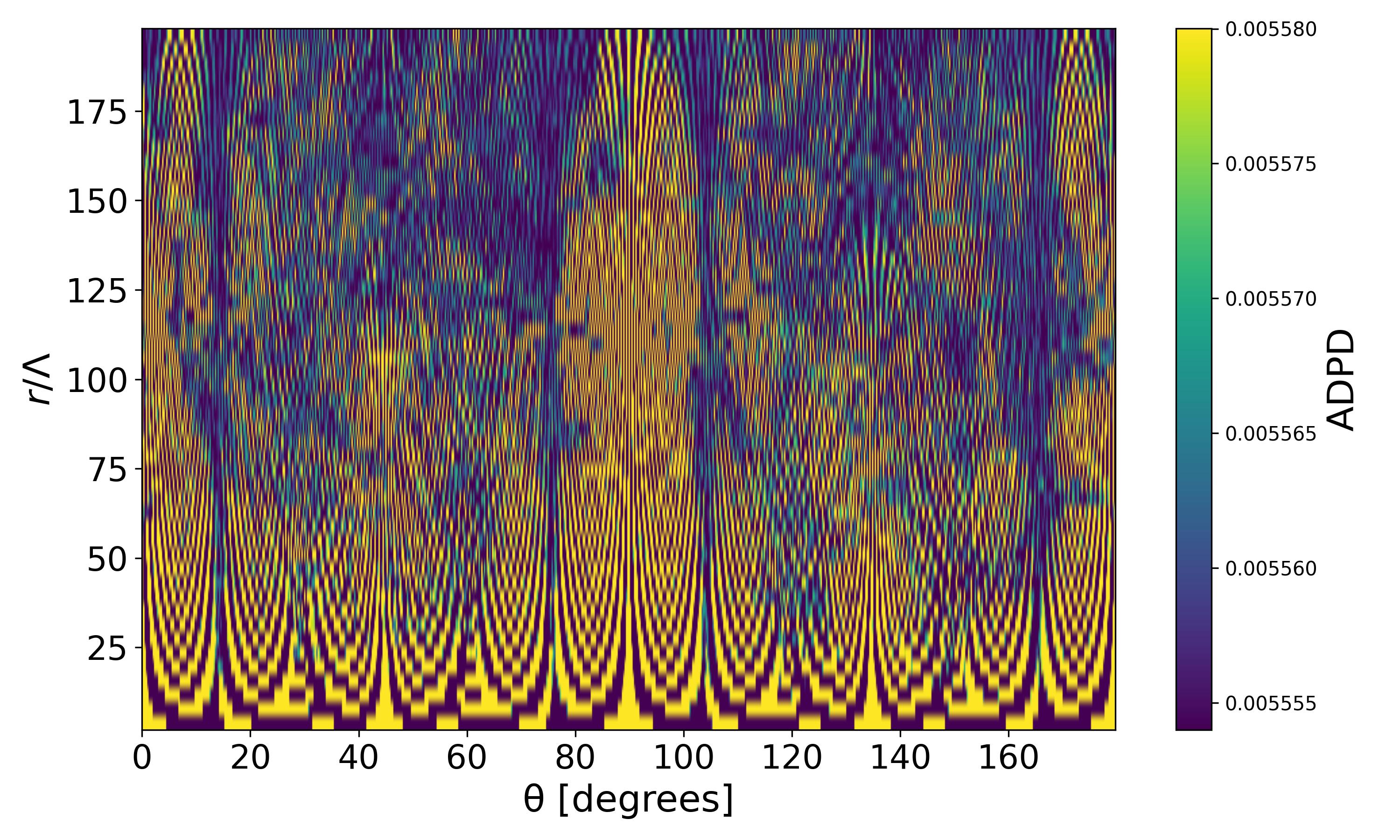}  
\includegraphics[width=0.24\textwidth]{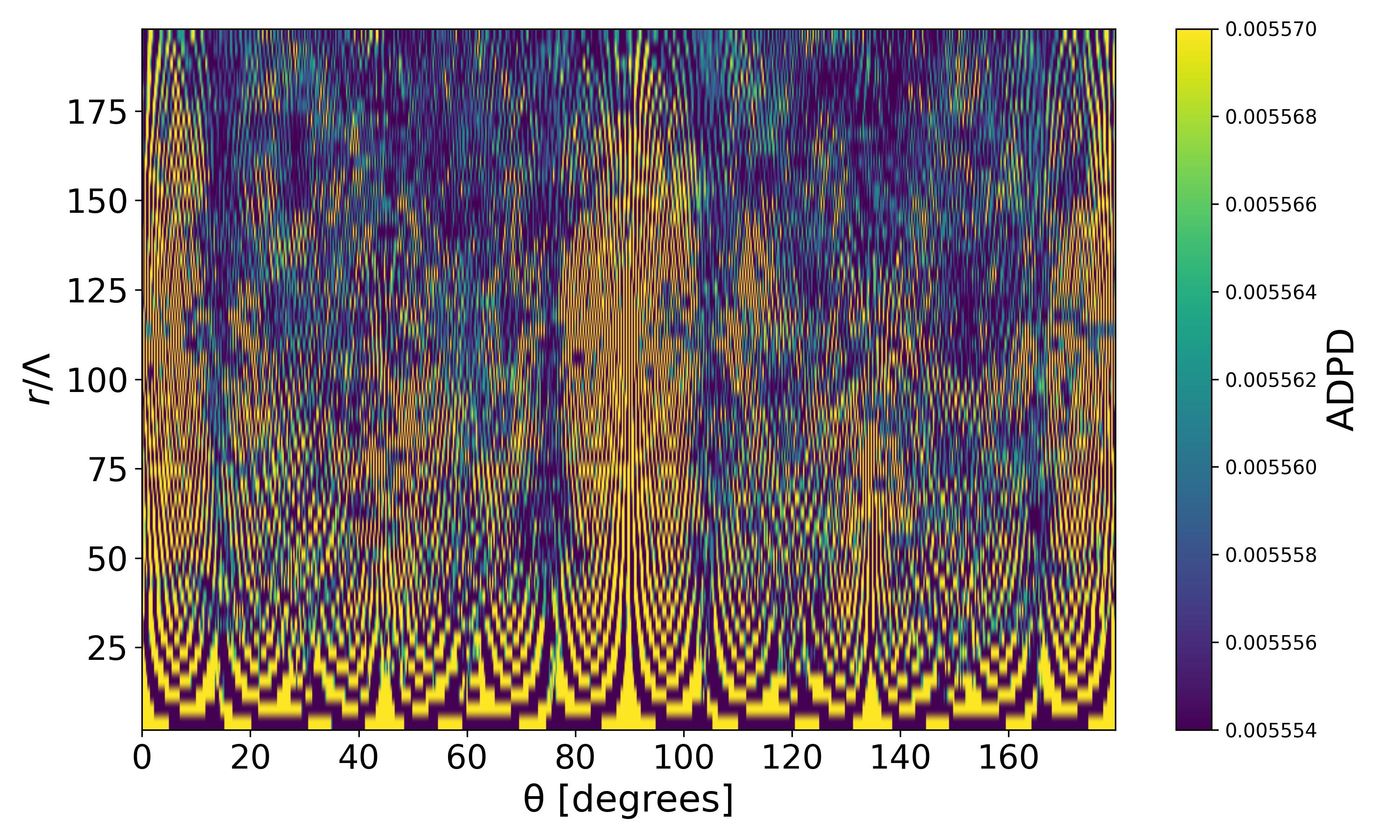}  
\includegraphics[width=0.24\textwidth]{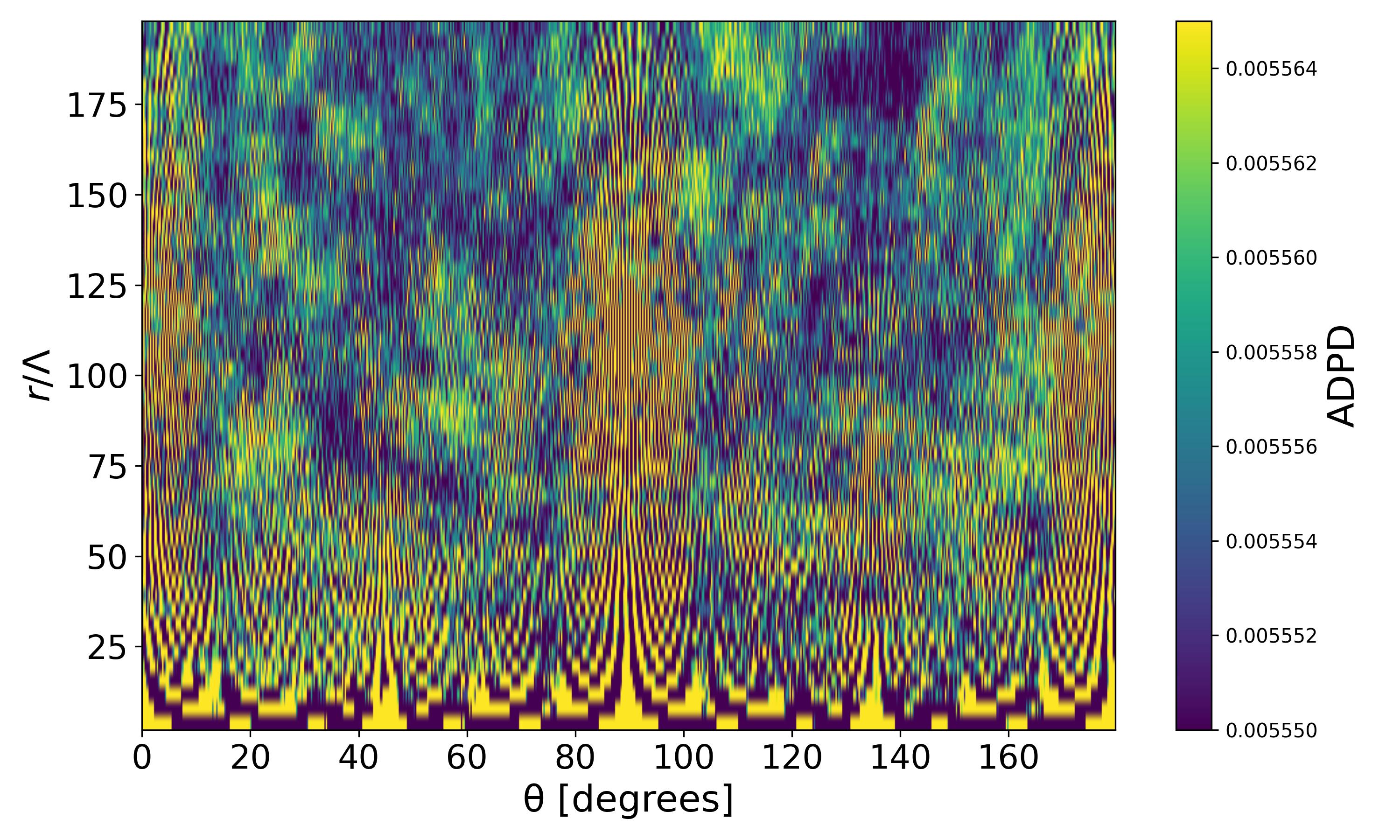}\\  
\includegraphics[width=0.24\textwidth]{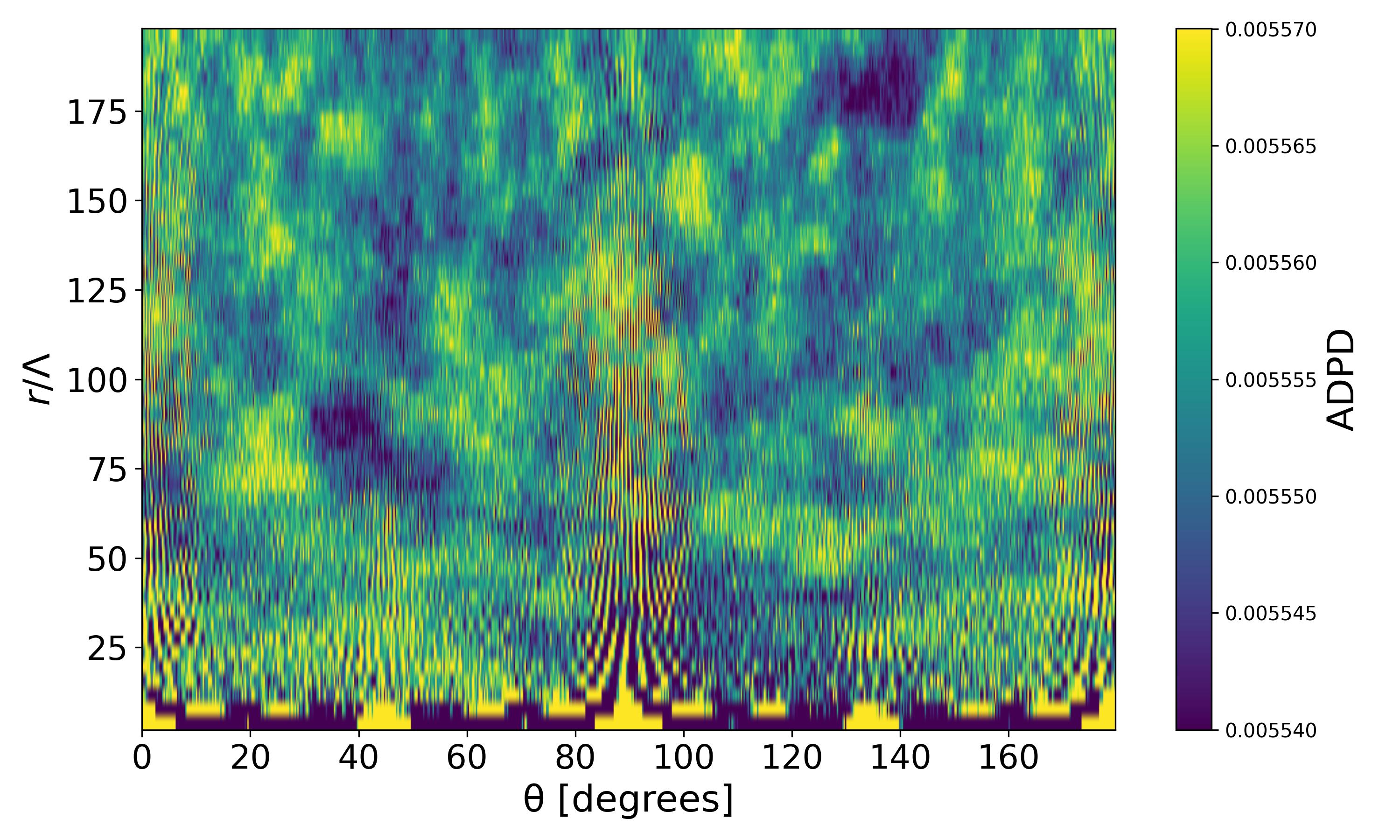}  
\includegraphics[width=0.24\textwidth]{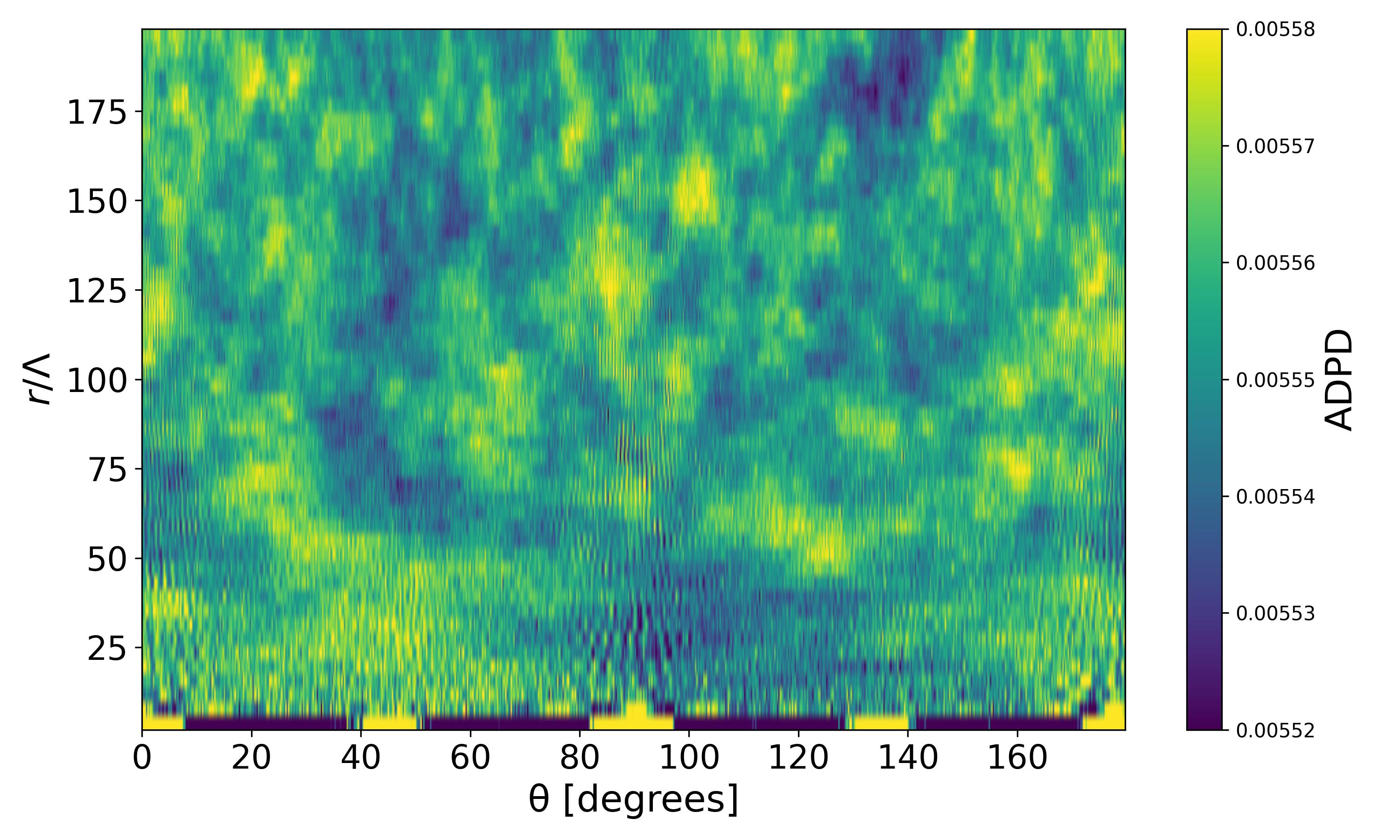}  
\includegraphics[width=0.24\textwidth]{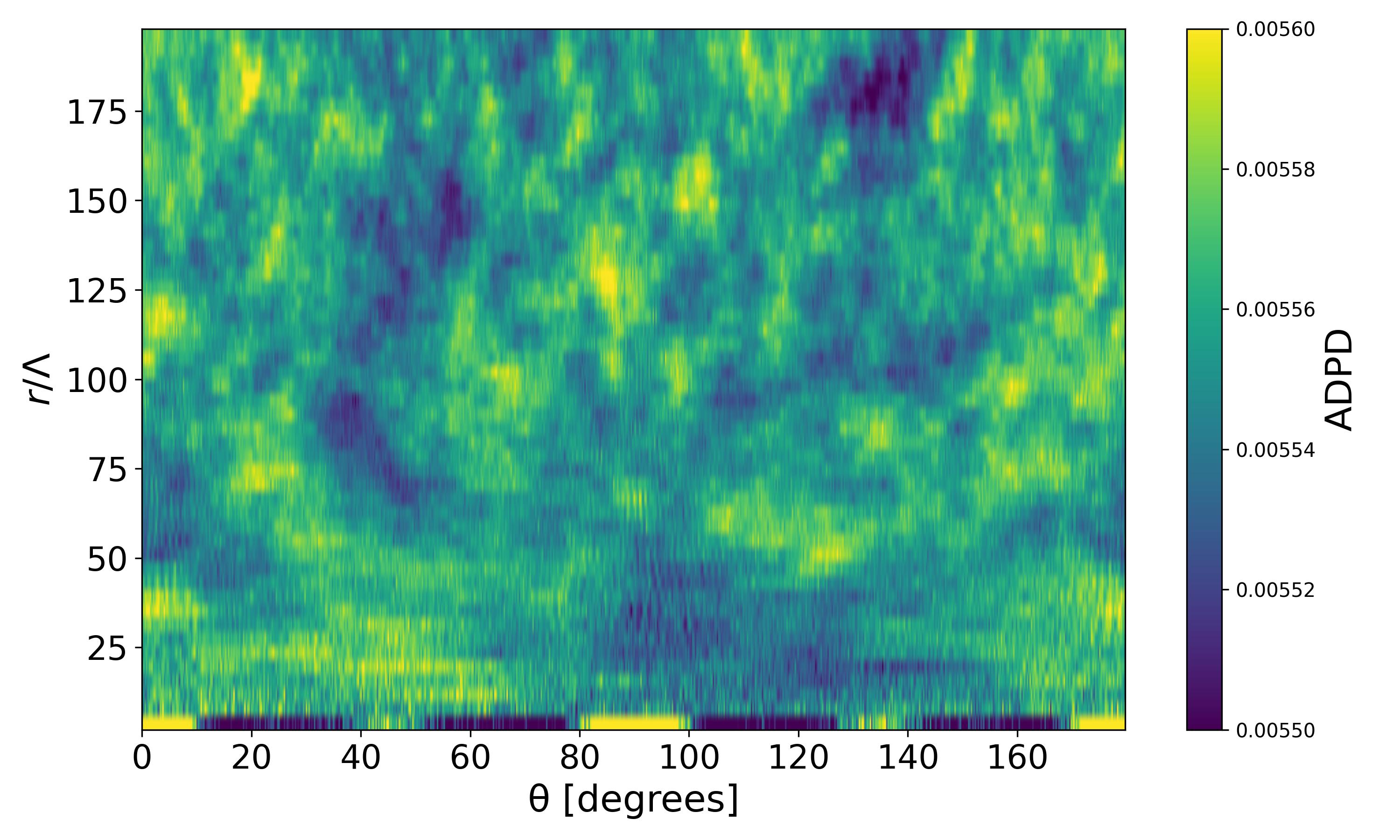}  
\includegraphics[width=0.24\textwidth]{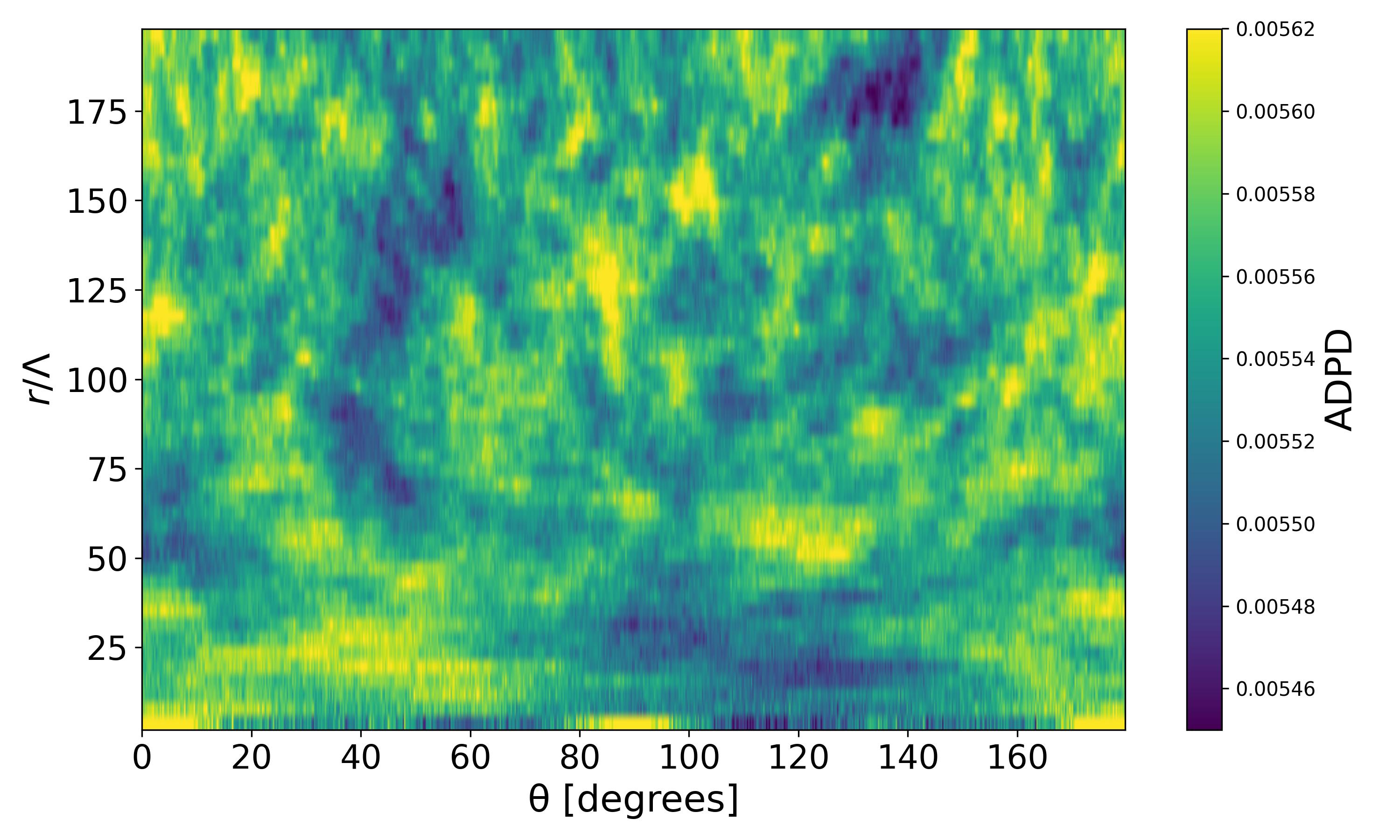}\\  
\includegraphics[width=0.24\textwidth]{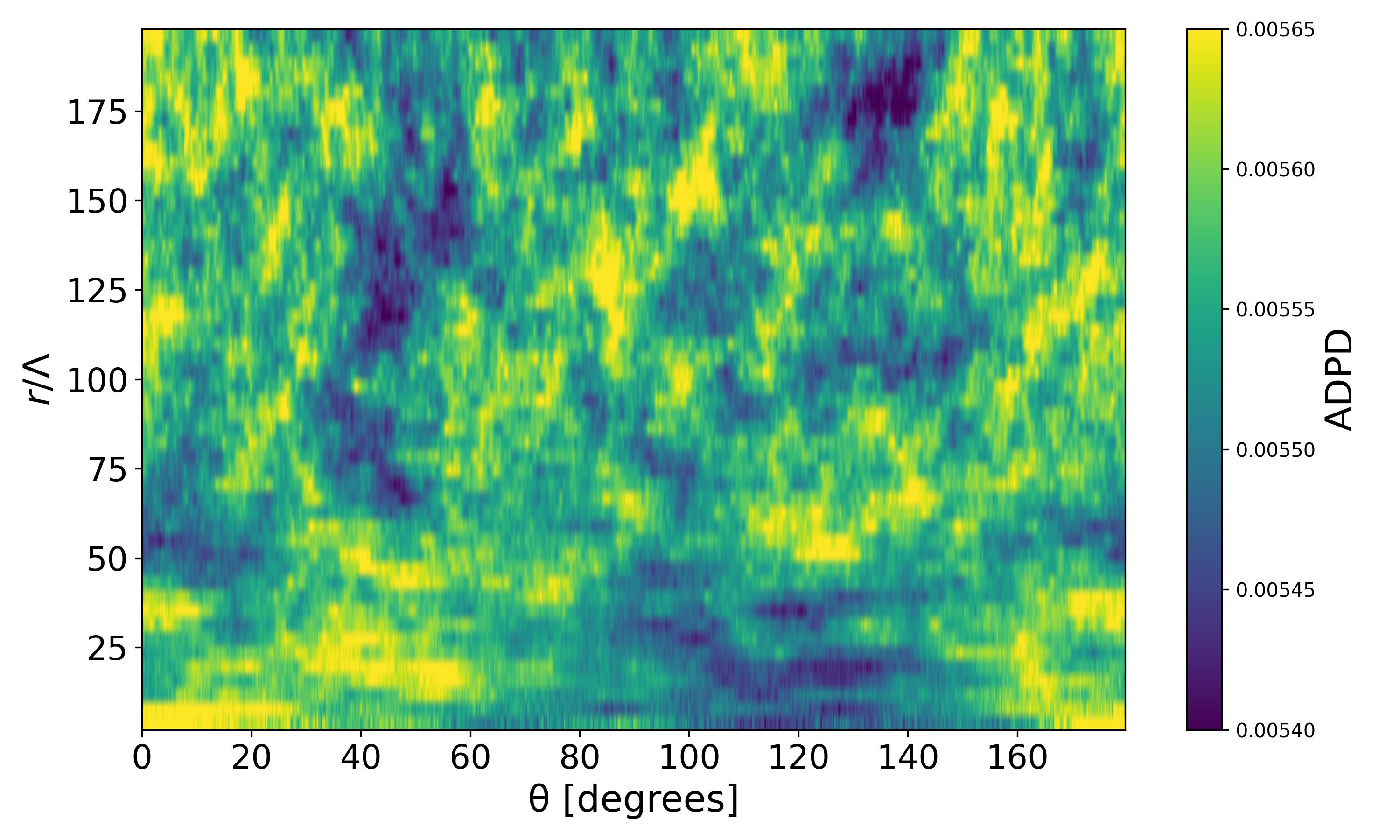}  
\includegraphics[width=0.24\textwidth]{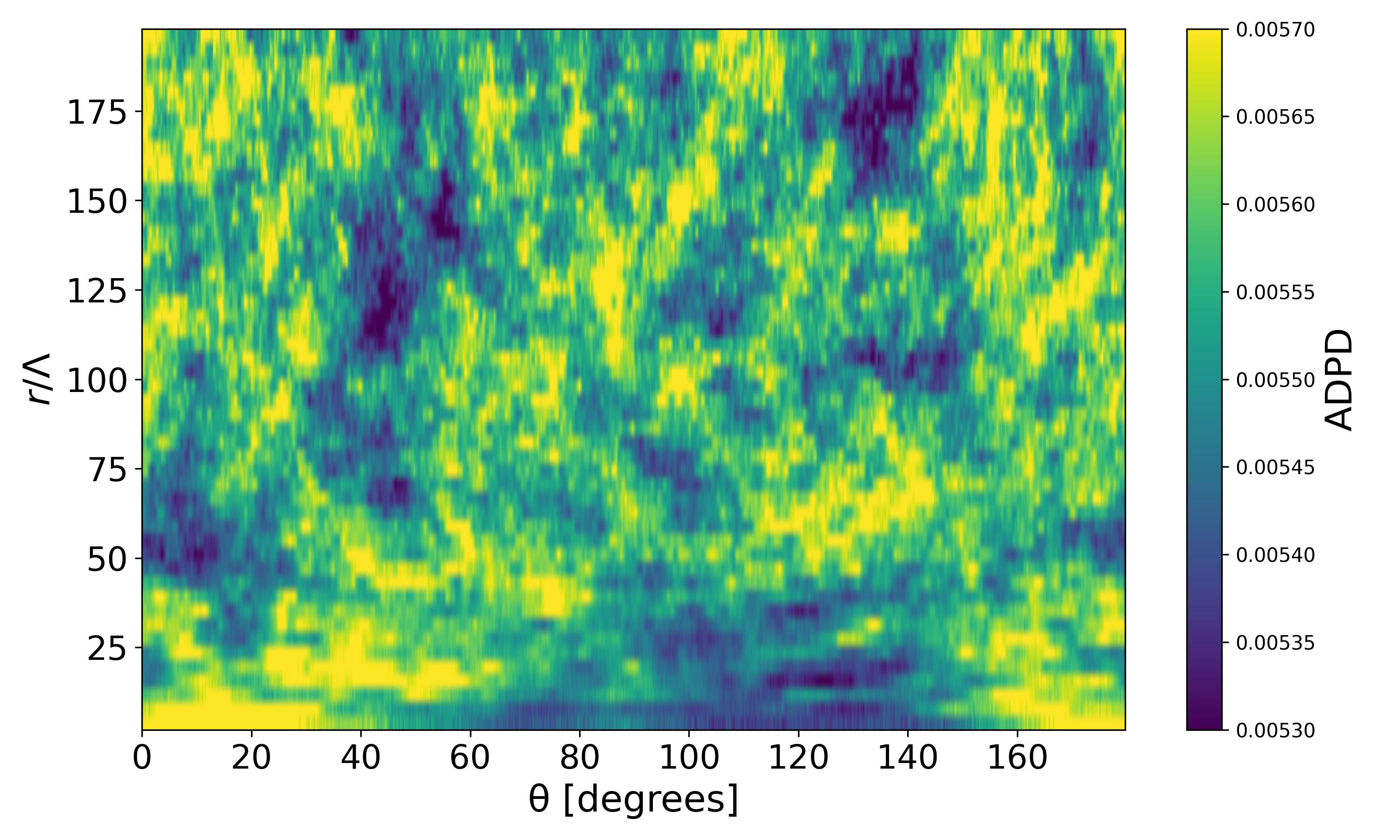}  
\includegraphics[width=0.24\textwidth]{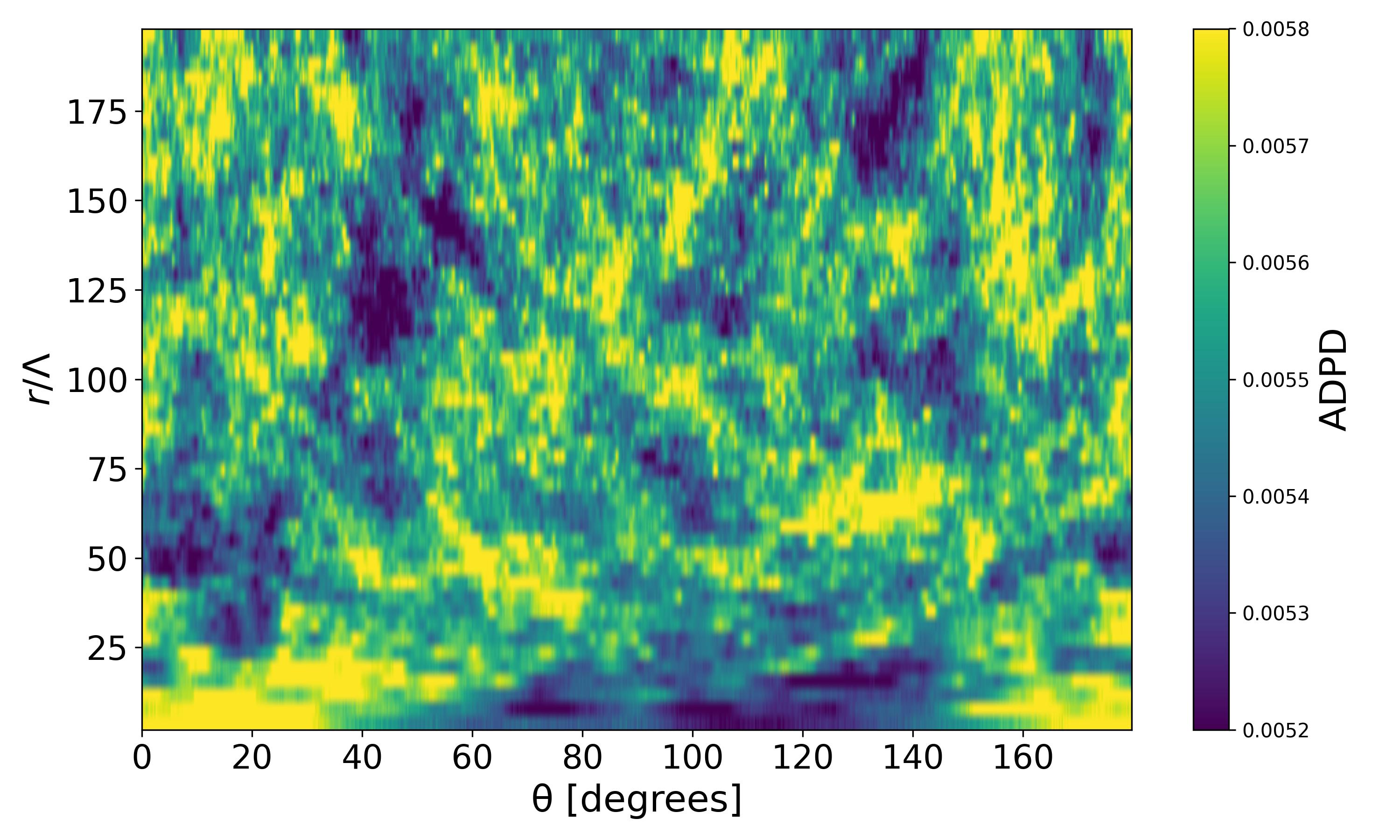}  
\includegraphics[width=0.24\textwidth]{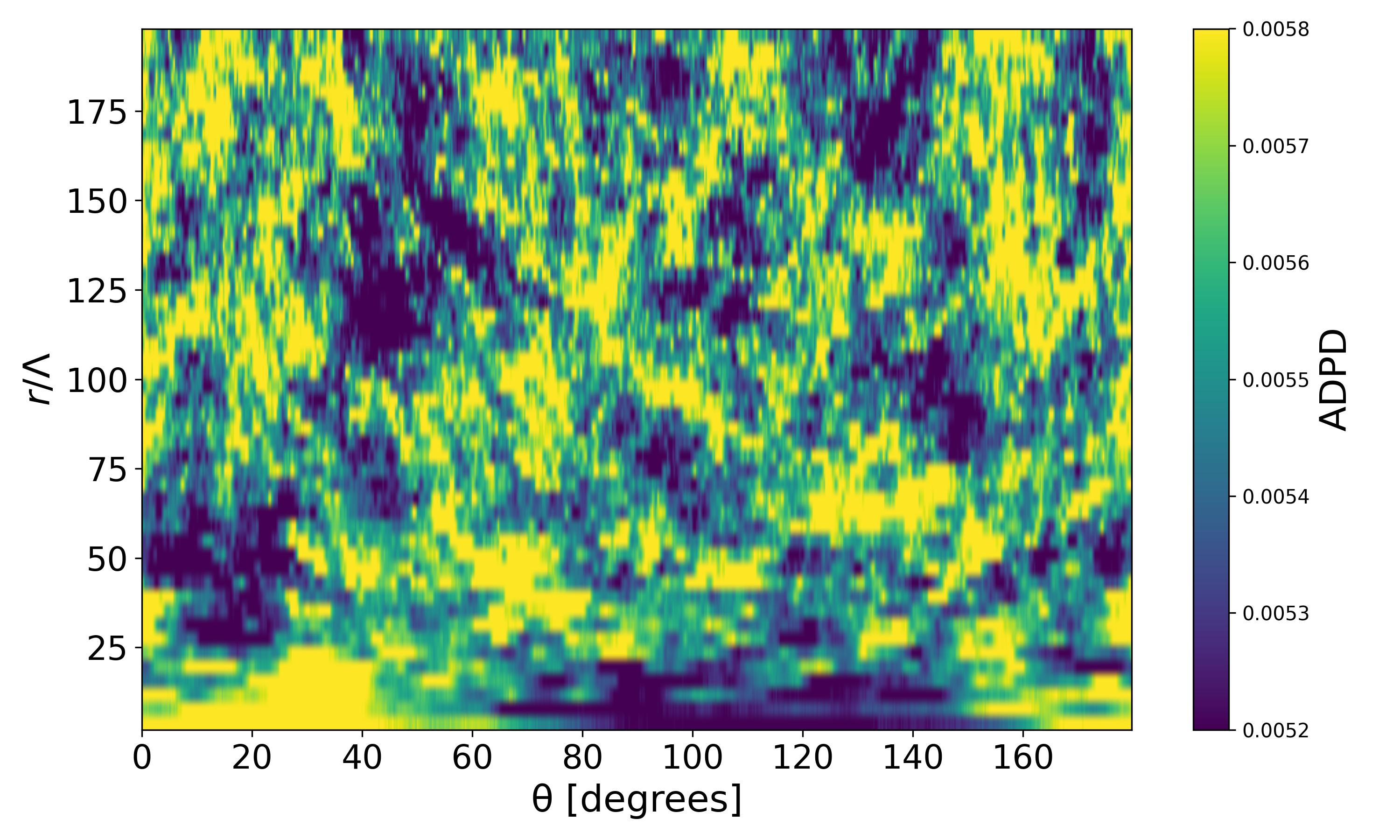}  
\caption{     Heat-map of the  ADPD 
for a slice of thickness $\Delta z= 10$ Mpc  of the large box CDM simulation
 at different redshifts {  (the ICs was shown in Fig.\ref{adpd_heatmap_large_box_IC}):} first row from right to left: $z=49, 20, 15, 10$; second row $z=7.7, 5.2, 3.6, 2.4$; third row $z=1.49, 0.84, 0.36, 0$.
  Length scales are expressed in units of the average distance between nearest neighbors $\Lambda$.}
\label{adpd_heatmap_large_box} 
\end{figure*}

\begin{figure*} 
\includegraphics[width=0.24\textwidth]{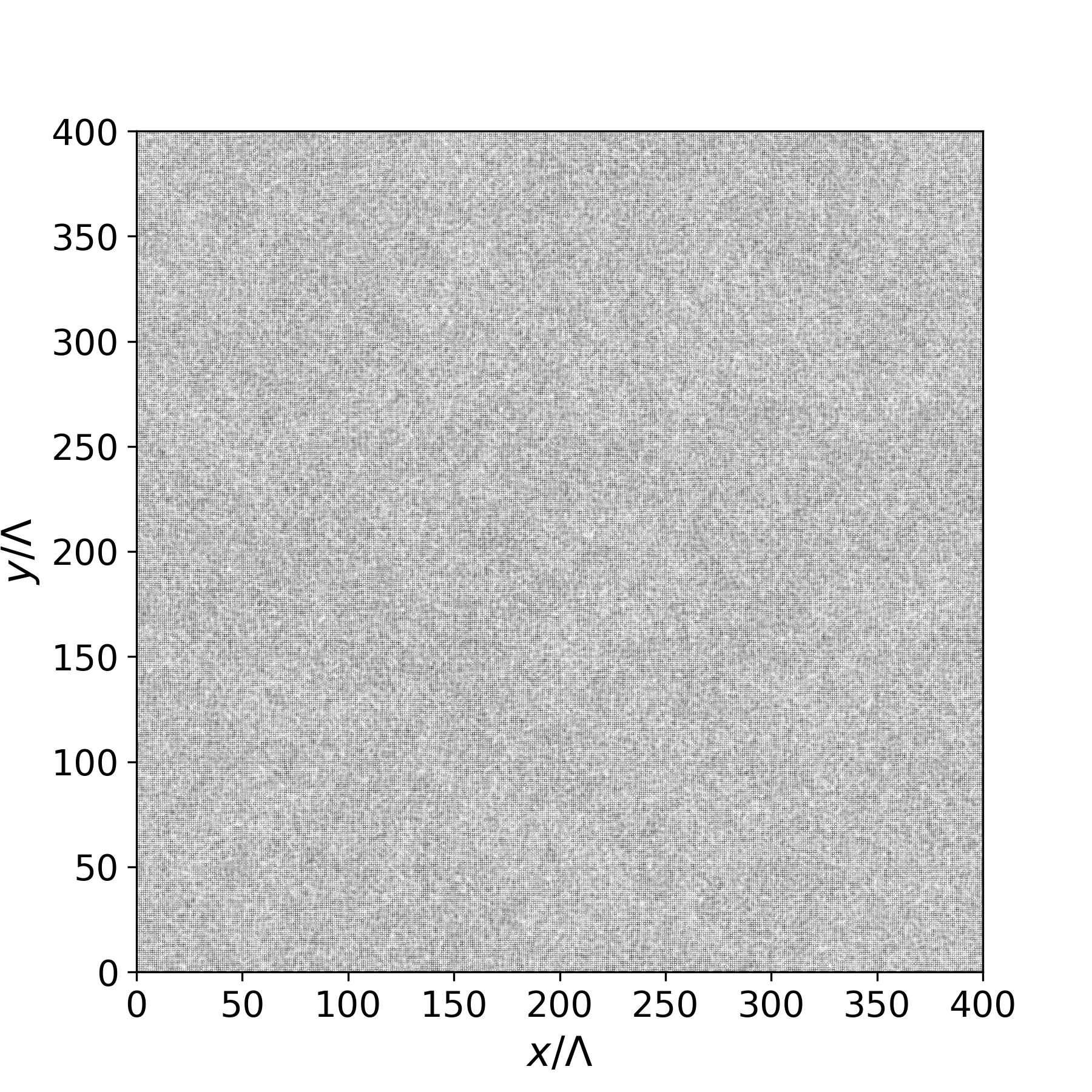}  
\includegraphics[width=0.24\textwidth]{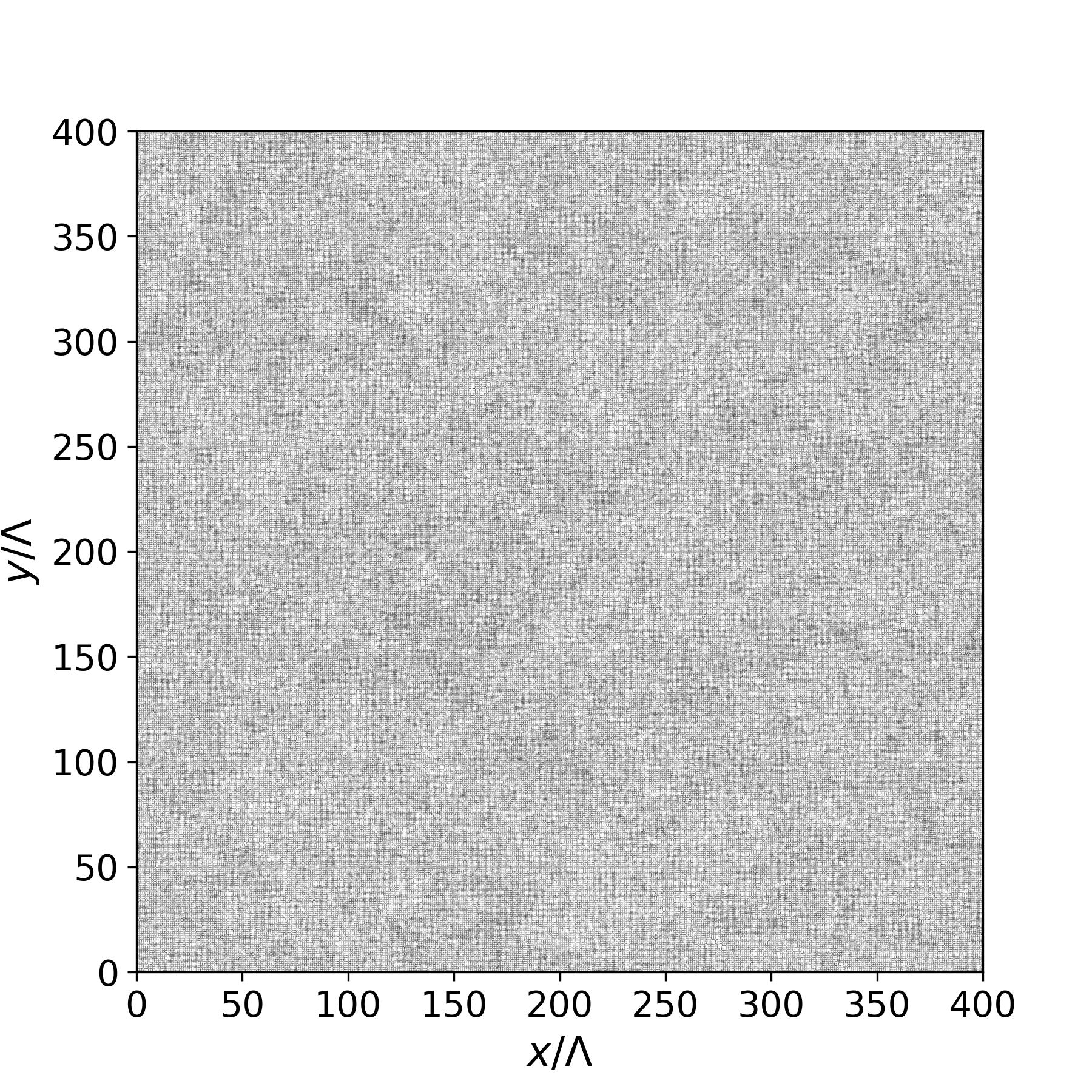}  
\includegraphics[width=0.24\textwidth]{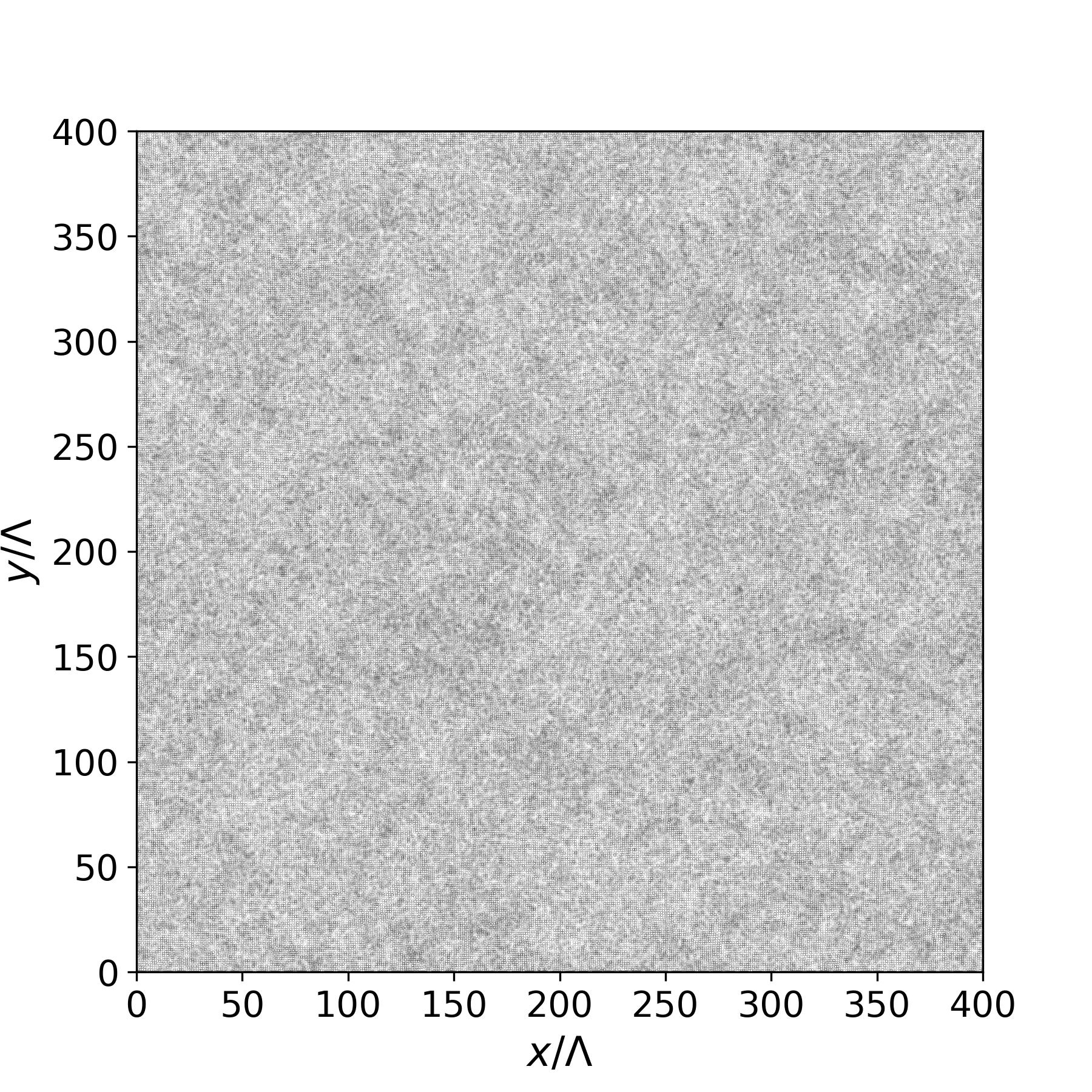}  
\includegraphics[width=0.24\textwidth]{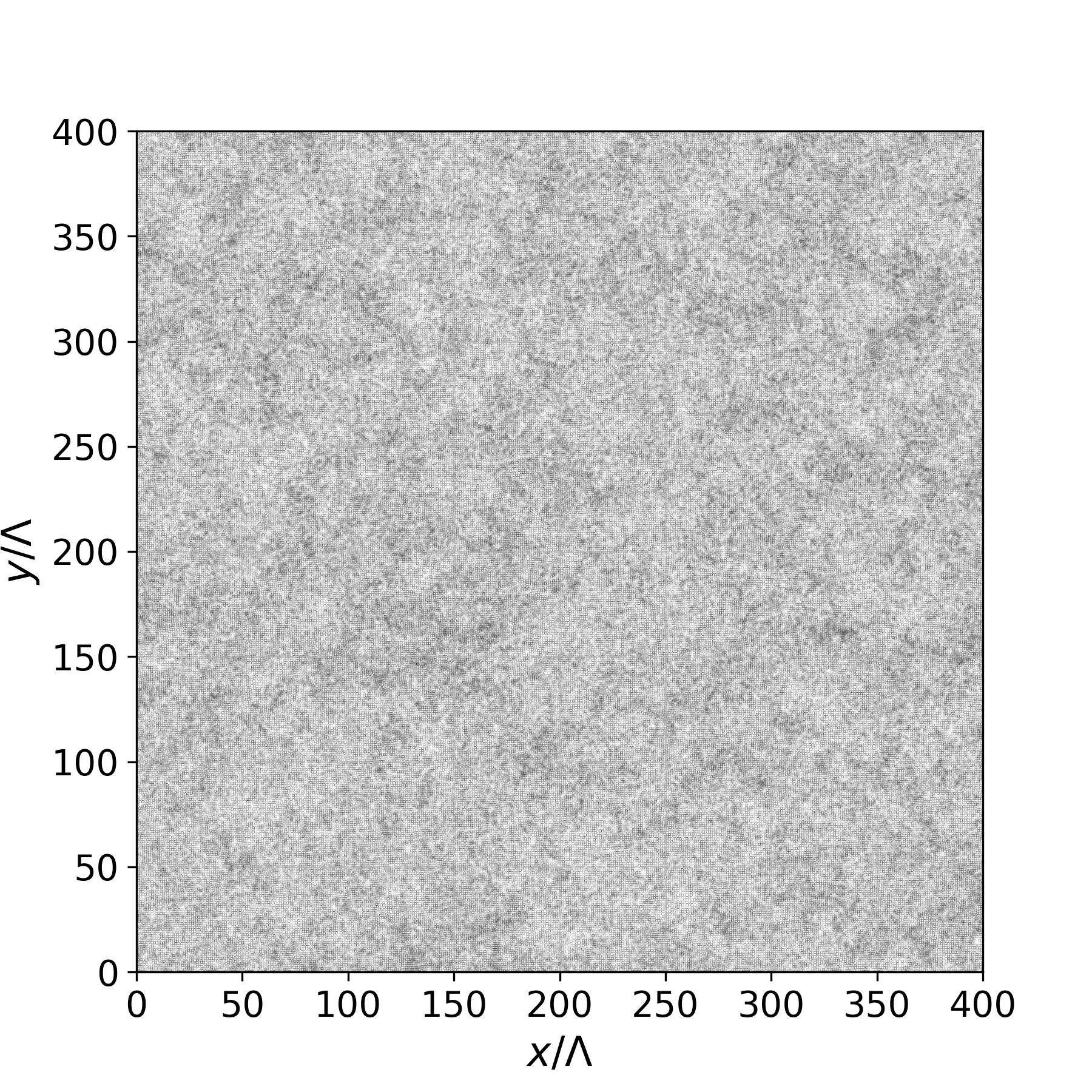}\\  
\includegraphics[width=0.24\textwidth]{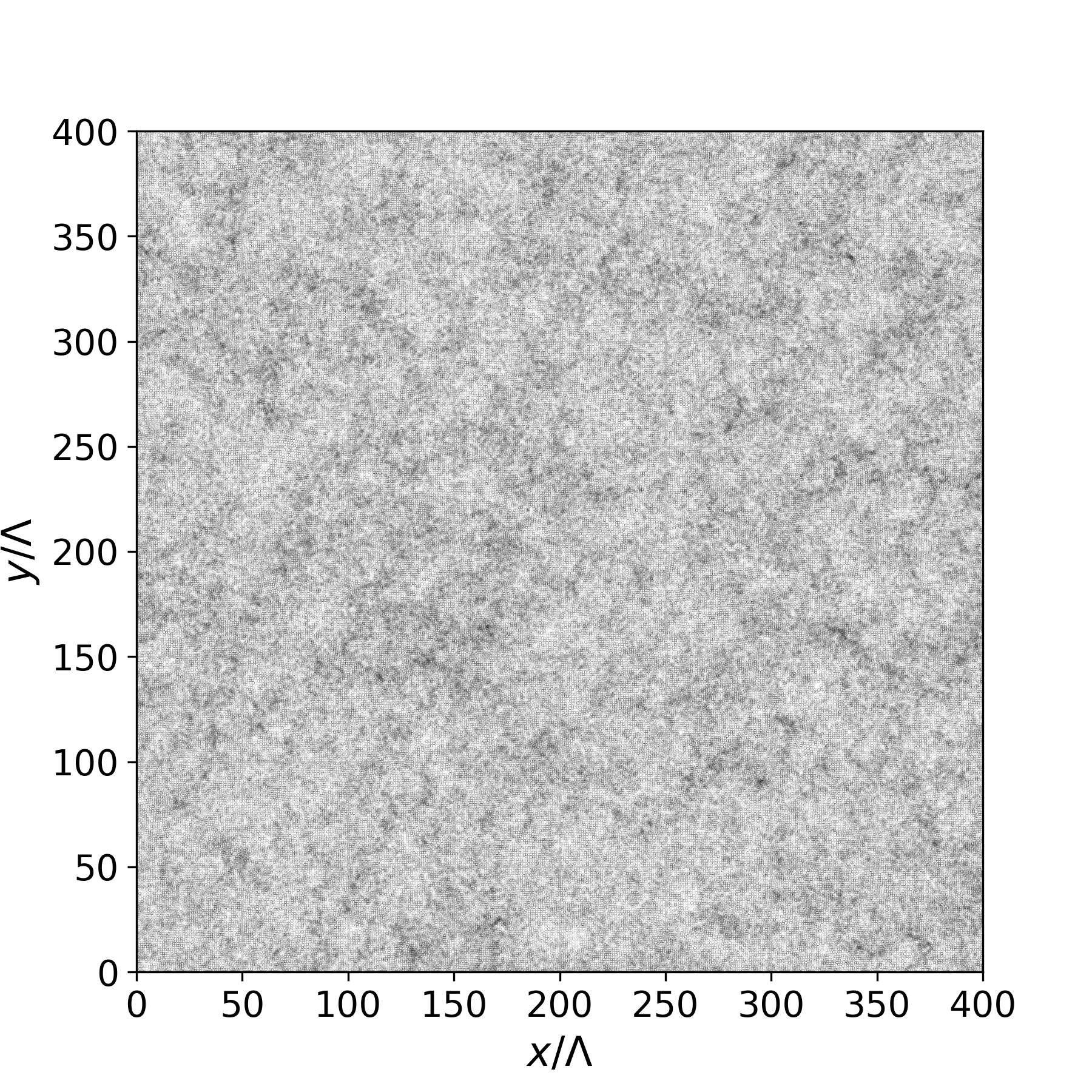}  
\includegraphics[width=0.24\textwidth]{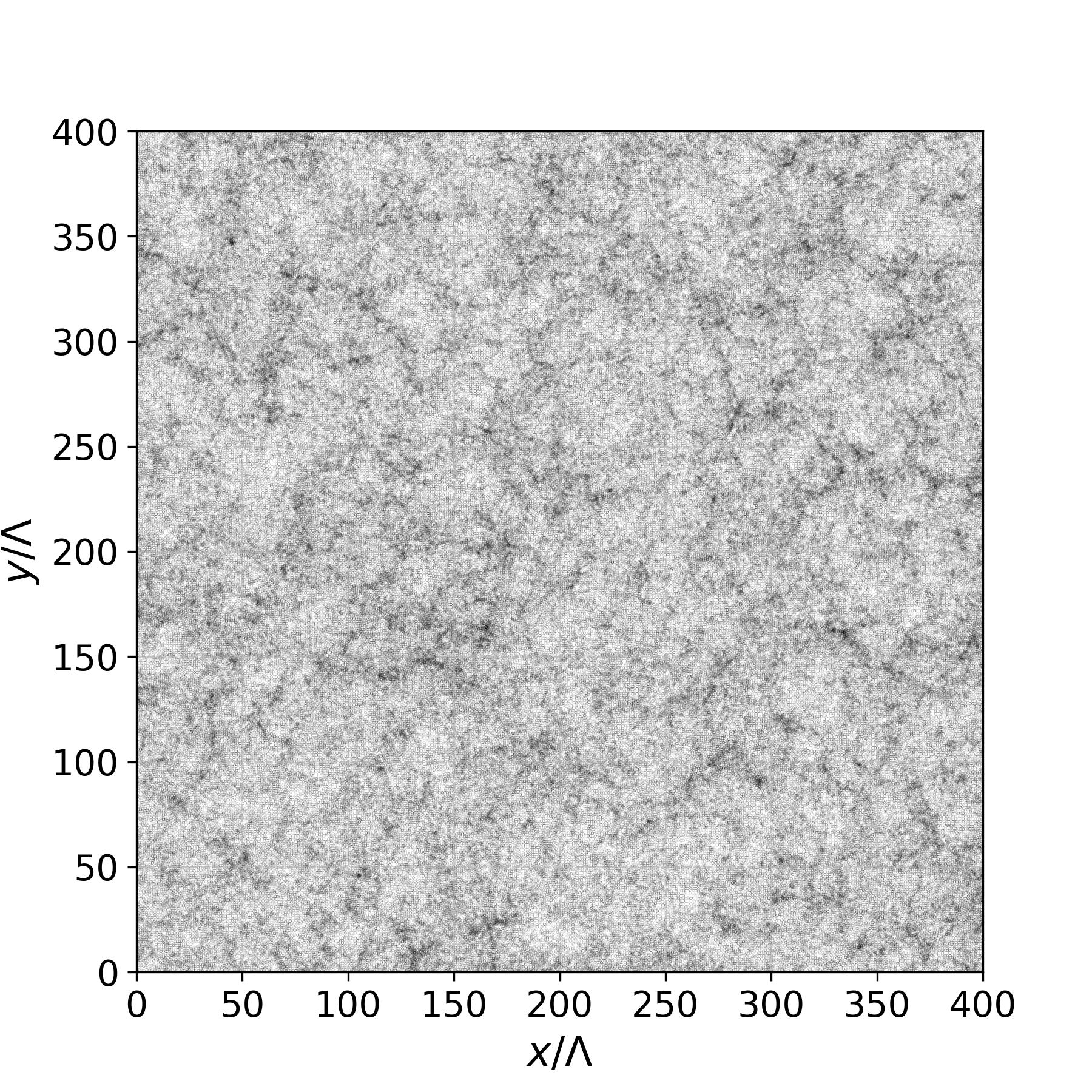}  
\includegraphics[width=0.24\textwidth]{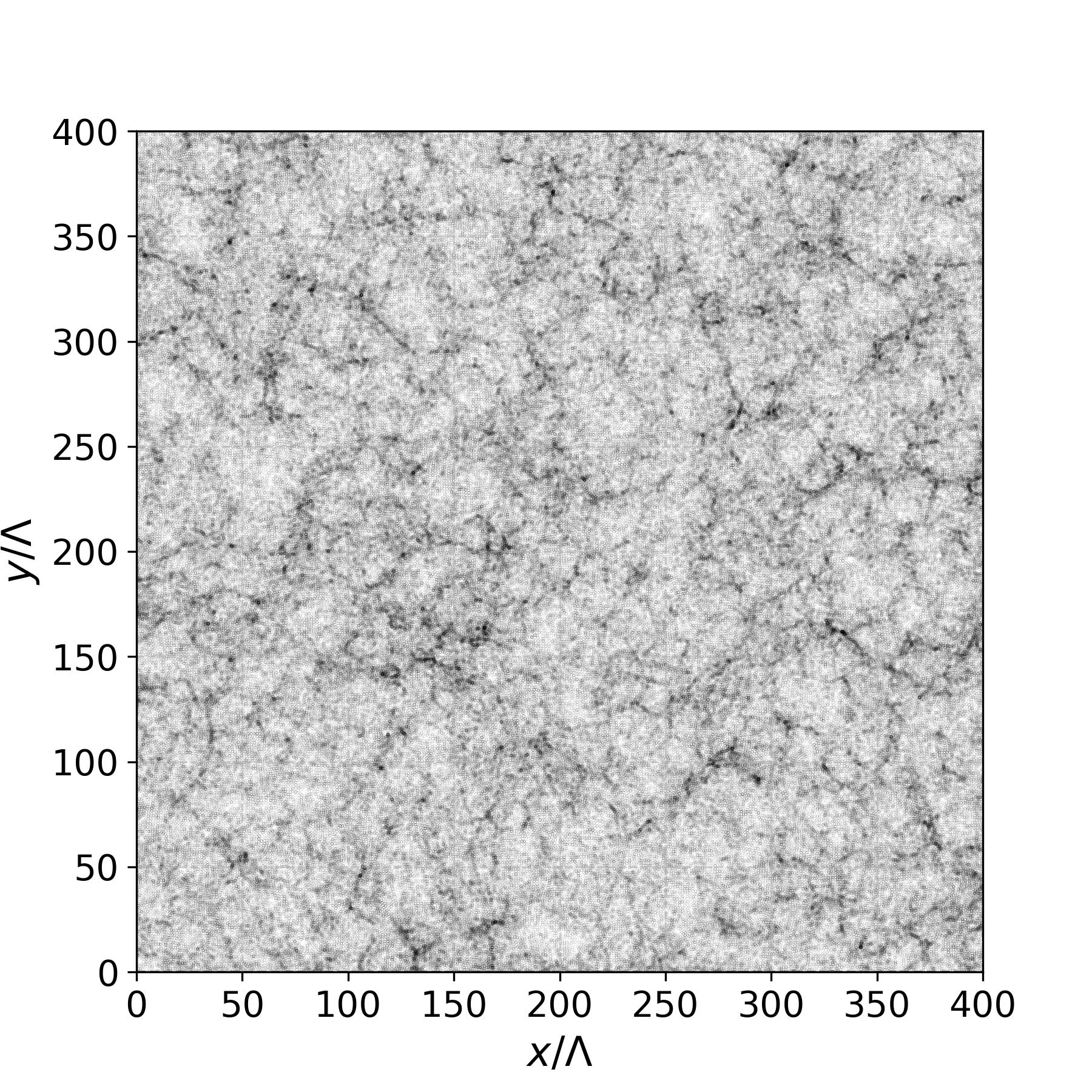}  
\includegraphics[width=0.24\textwidth]{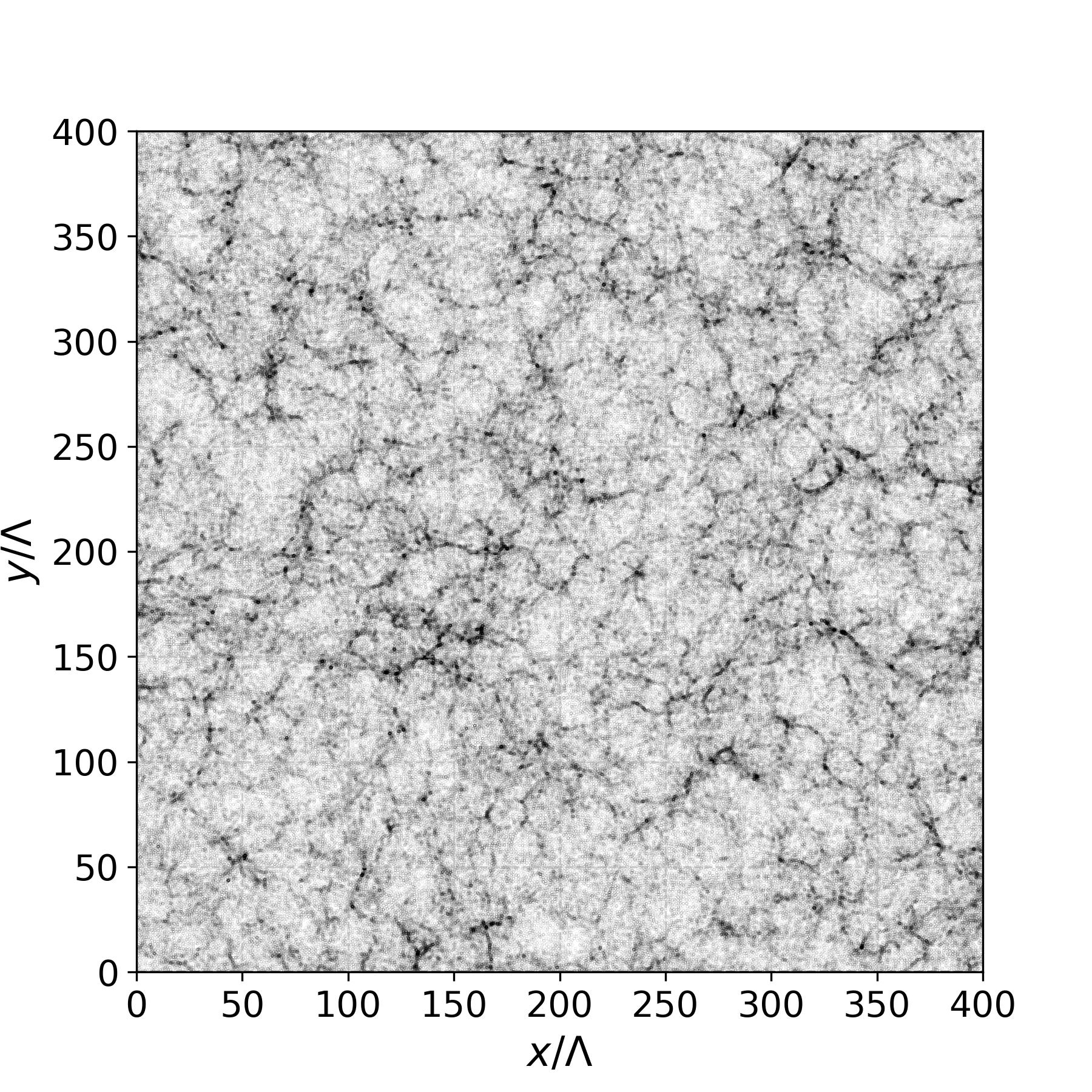}\\  
\includegraphics[width=0.24\textwidth]{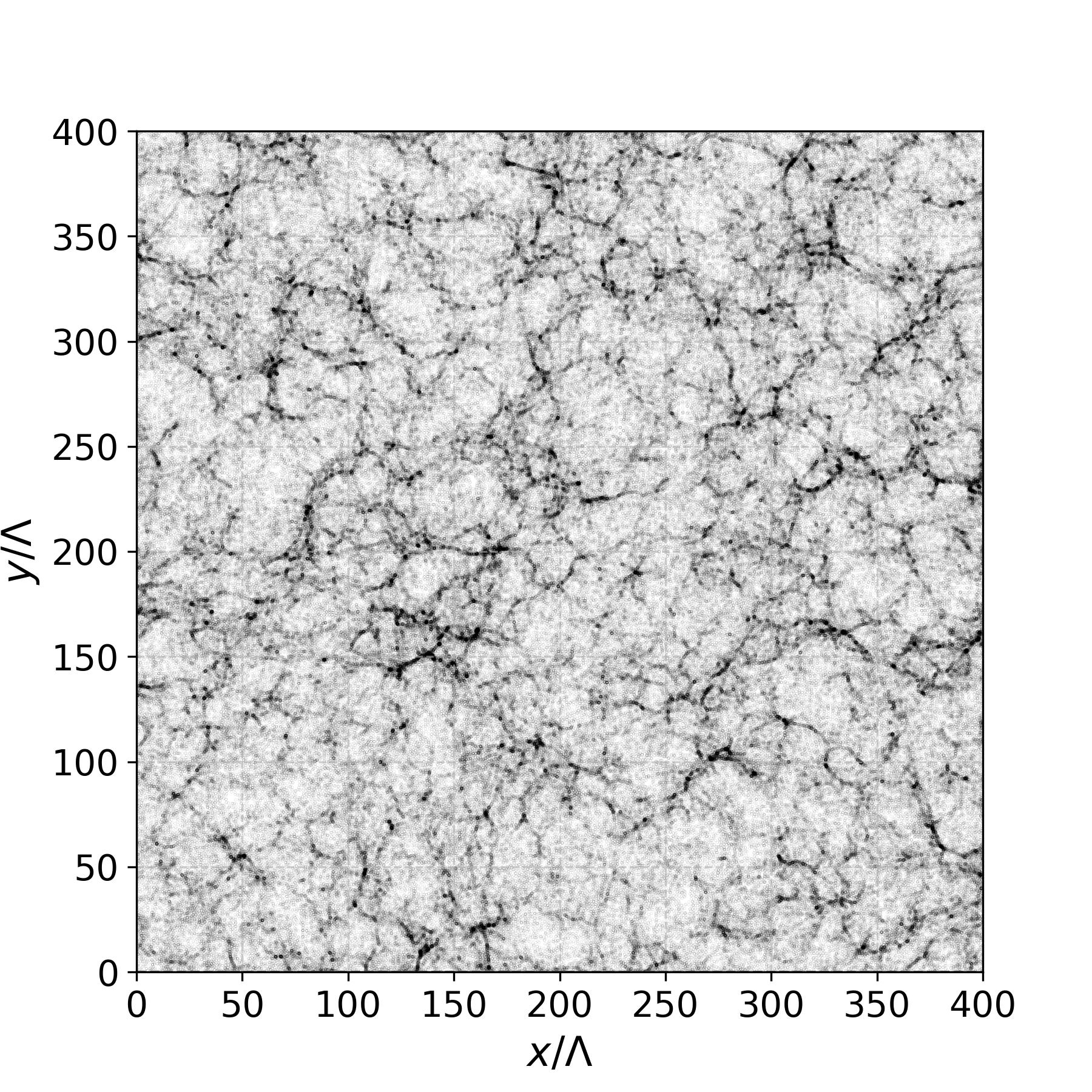}  
\includegraphics[width=0.24\textwidth]{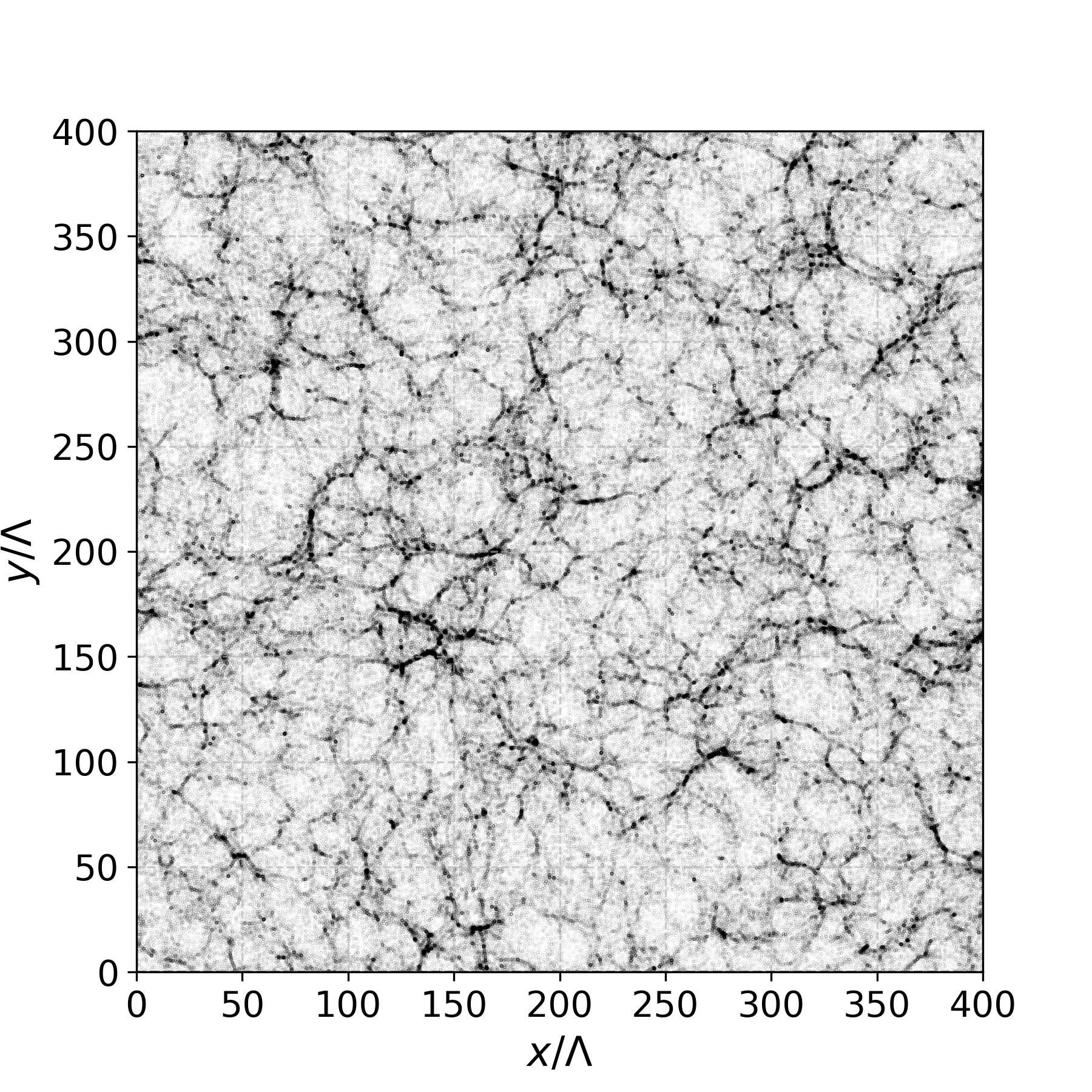}  
\includegraphics[width=0.24\textwidth]{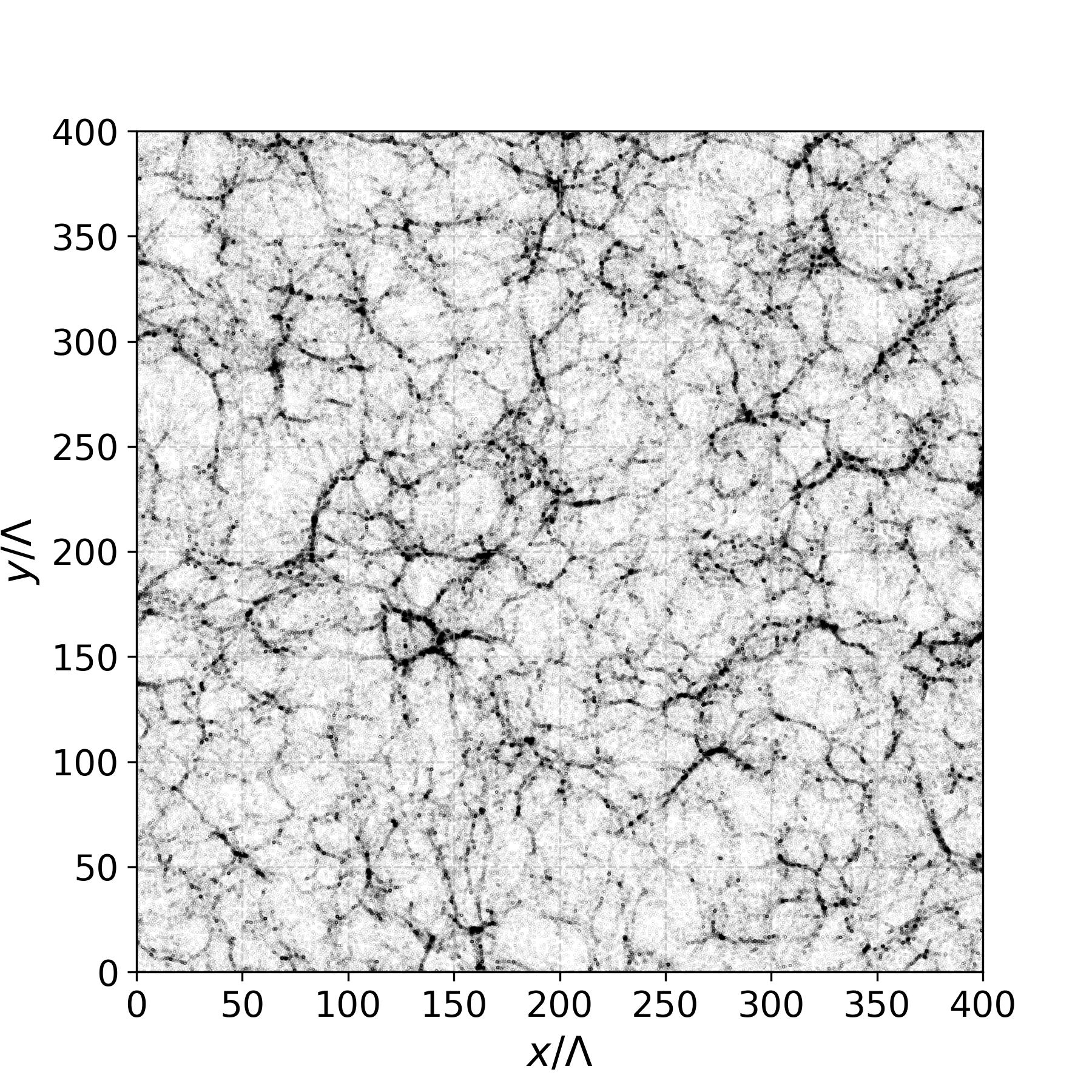}  
\includegraphics[width=0.24\textwidth]{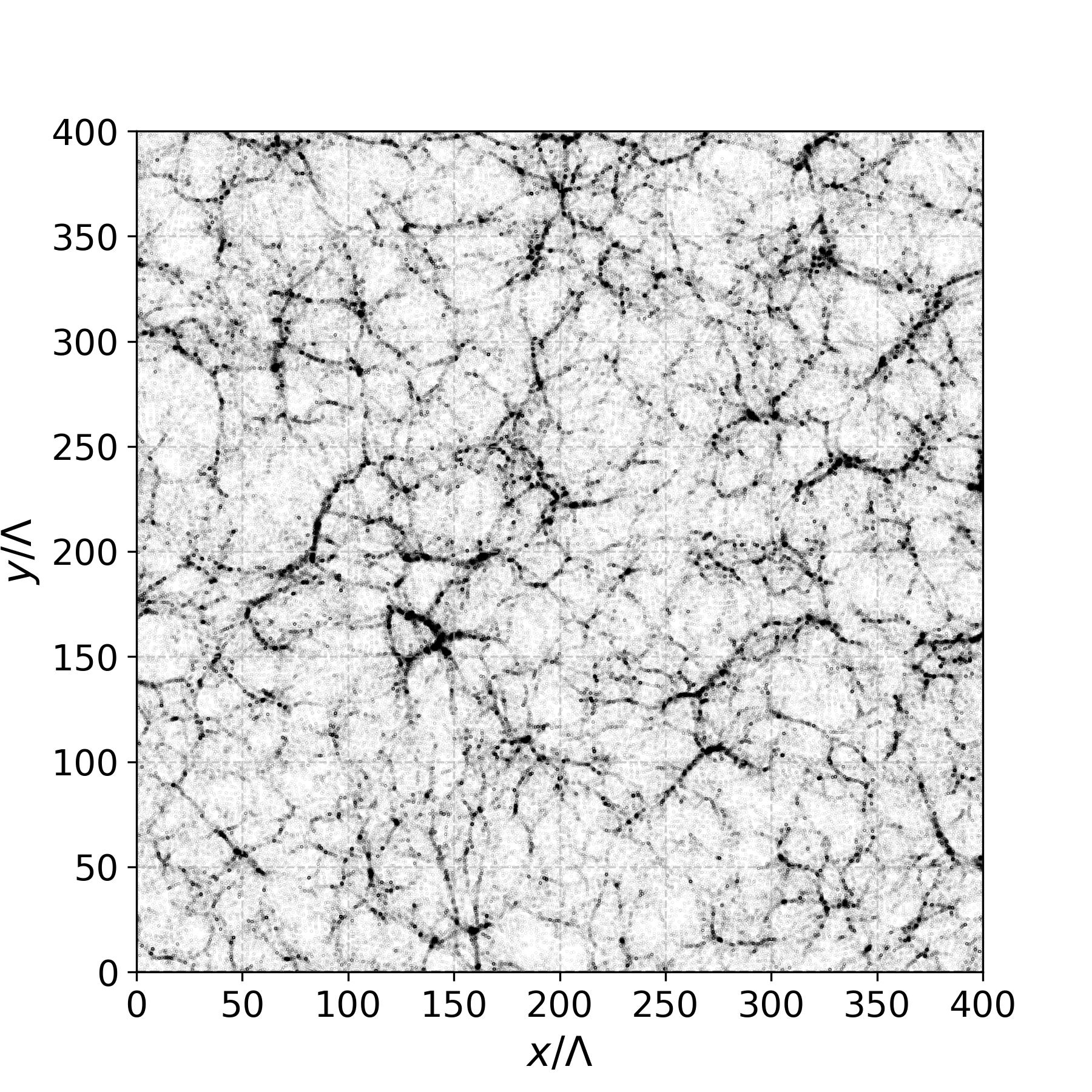}  
\caption{    Point distribution for the same slices of Fig.\ref{adpd_heatmap_large_box} (random sampling 10\%).
 Length scales are expressed in units of the average distance between nearest neighbors $\Lambda$.} 
\label{point_large_box} 
\end{figure*}

\begin{figure} 
\includegraphics[width=0.4\textwidth]{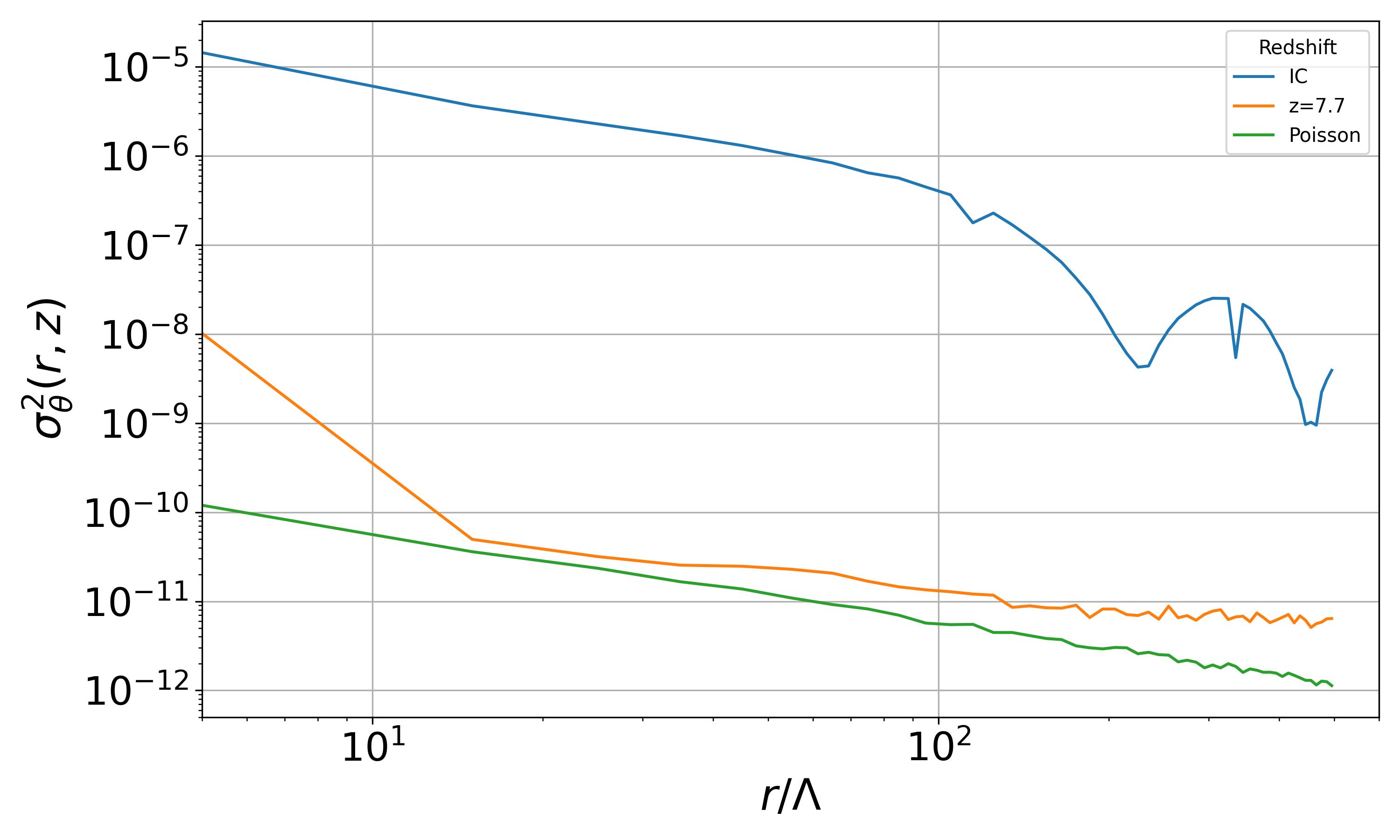}  
\caption{    Angular variance of the ADPD for the large-box simulation at redshifts $z_{\mathrm{IC}} = 49$ 
and $z = 7$, shown alongside the corresponding result for a Poisson distribution with the same number of particles.
 Length scales are expressed in units of the average distance between nearest neighbors $\Lambda$.} 
\label{variance_poisson_plot_largebox} 
\end{figure}

\begin{figure} 
\includegraphics[width=0.49\textwidth]{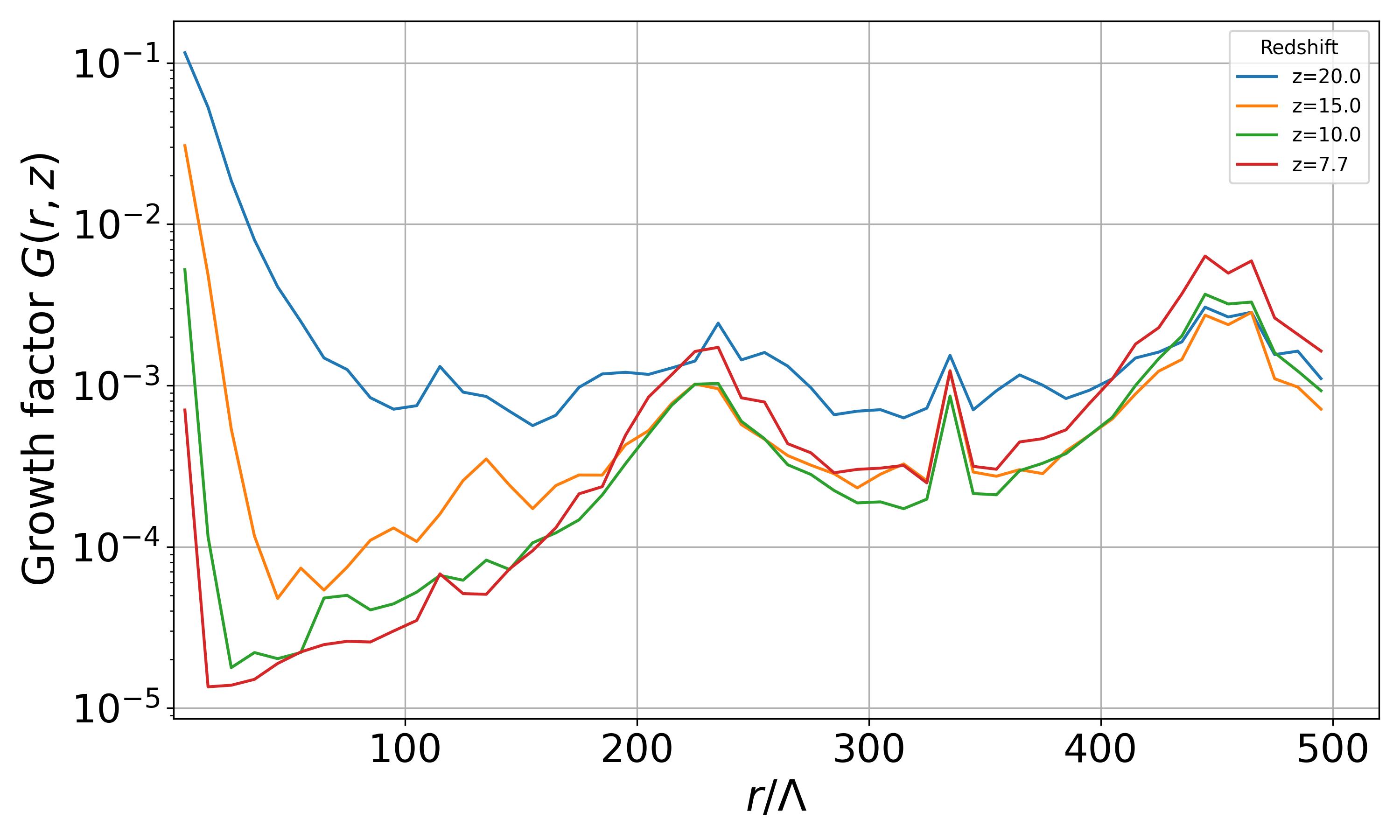}  
\includegraphics[width=0.49\textwidth]{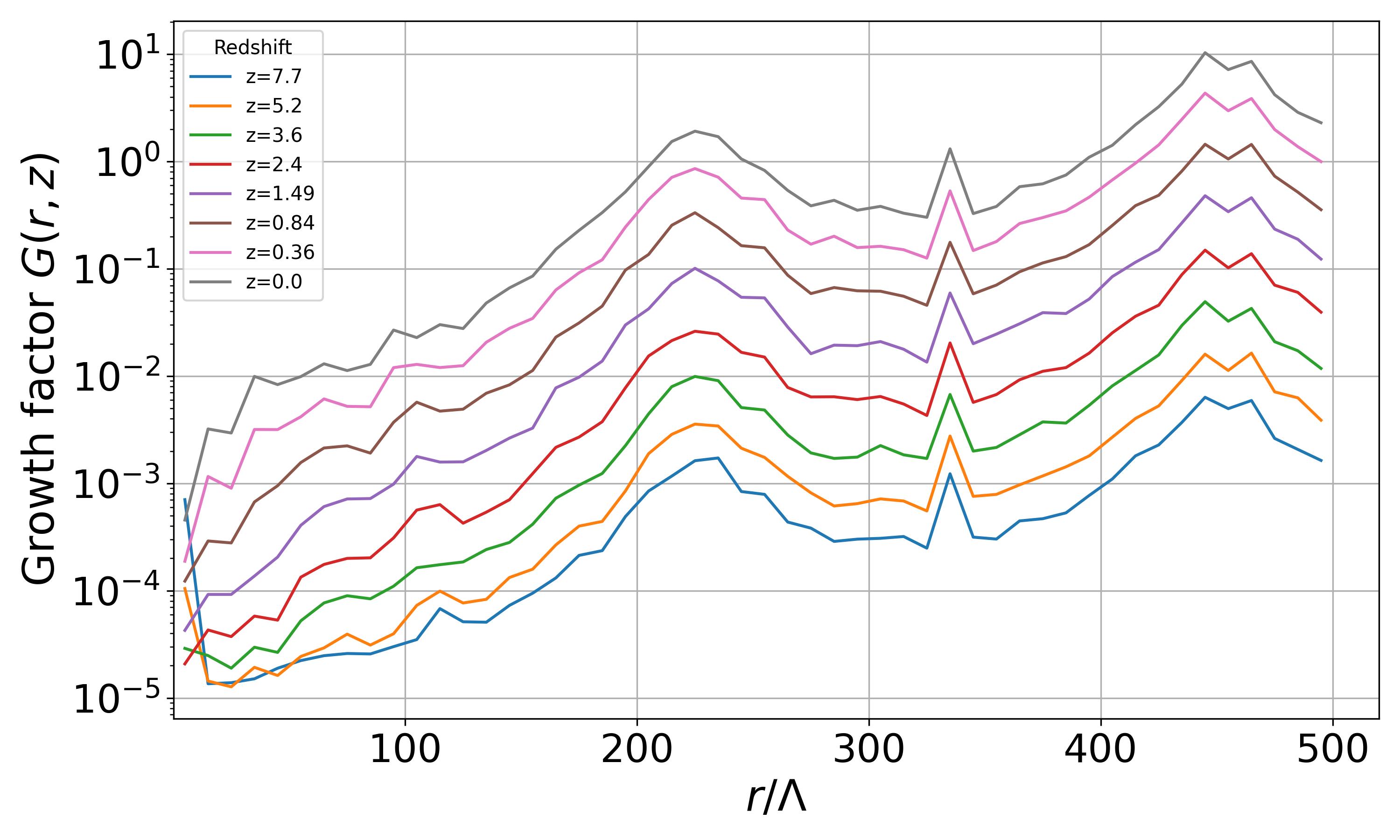} 
\caption{    Variance growth factor for the large-box simulation. {Top panel:} From redshift $z = 20$ to $z = 7$, the growth factor decreases as the imprint of the initial lattice is progressively erased, particles have moved, on average, by a distance comparable to the pre-initial lattice spacing. {Bottom panel:} After reaching a minimum at $z \approx 7$, the growth factor increases significantly, indicating the  amplification of anisotropic structures seeded by residual features in the initial conditions. 
In this case the non-linear regime, occurs at very small scales, i.e. $r<10$ Mpc/$h$, compared to the simulation box.
 Length scales are expressed in units of the average distance between nearest neighbors $\Lambda$.} 
\label{variance_growth_factor_largebox} 
\end{figure}

{ 
Fig.~\ref{adpd_heatmap_large_box_IC} displays the ADPD heatmap for the initial particle configuration at redshift $z_{\mathrm{IC}} = 49$ measured in a slice of thickness $\Delta z = 10$~Mpc$/h$. Similar to the case of the small-box simulation (see Fig.~\ref{adpd_heatmap_small_box_IC}), the ADPD heatmap of the ICs exhibits patterns that directly reflect the intrinsic anisotropy of the underlying cubic lattice, characterized by preferred directions along the Cartesian axes and diagonals. Since the initial displacements are small, their contribution to the ADPD is negligible at this stage, and the anisotropic lattice signature remains dominant.
}

{  
The redshift evolution of the ADPD measured in a slice of thickness $\Delta z = 10$~Mpc$/h$ 
is shown  in Fig.~\ref{adpd_heatmap_large_box}  whereas the corresponding point distributions in these slices are shown in Fig.~\ref{point_large_box}. 
The redshift slices in the range $49 < z < 10$ (first row of Fig.~\ref{adpd_heatmap_large_box}) allow us to track the progressive washing out of the initial lattice structure. This temporal evolution is clearly visible in the ADPD heatmaps of the large-box simulation, whereas for the small-box case this was not shown (see Fig.~\ref{adpd_heatmap_small_box}).
}

At $z \approx 7$  the average particle displacement becomes comparable to the lattice spacing, effectively scrambling the initial grid configuration. The sharp lattice-induced striping observed in the ICs is largely washed out, indicating that the pairwise angular distribution is no longer governed by the pre-ICs. Instead, the ADPD begins to reflect the geometry of the displacement field, including nascent  features such as filaments and walls.

\begin{figure*} 
\includegraphics[width=0.49\textwidth]{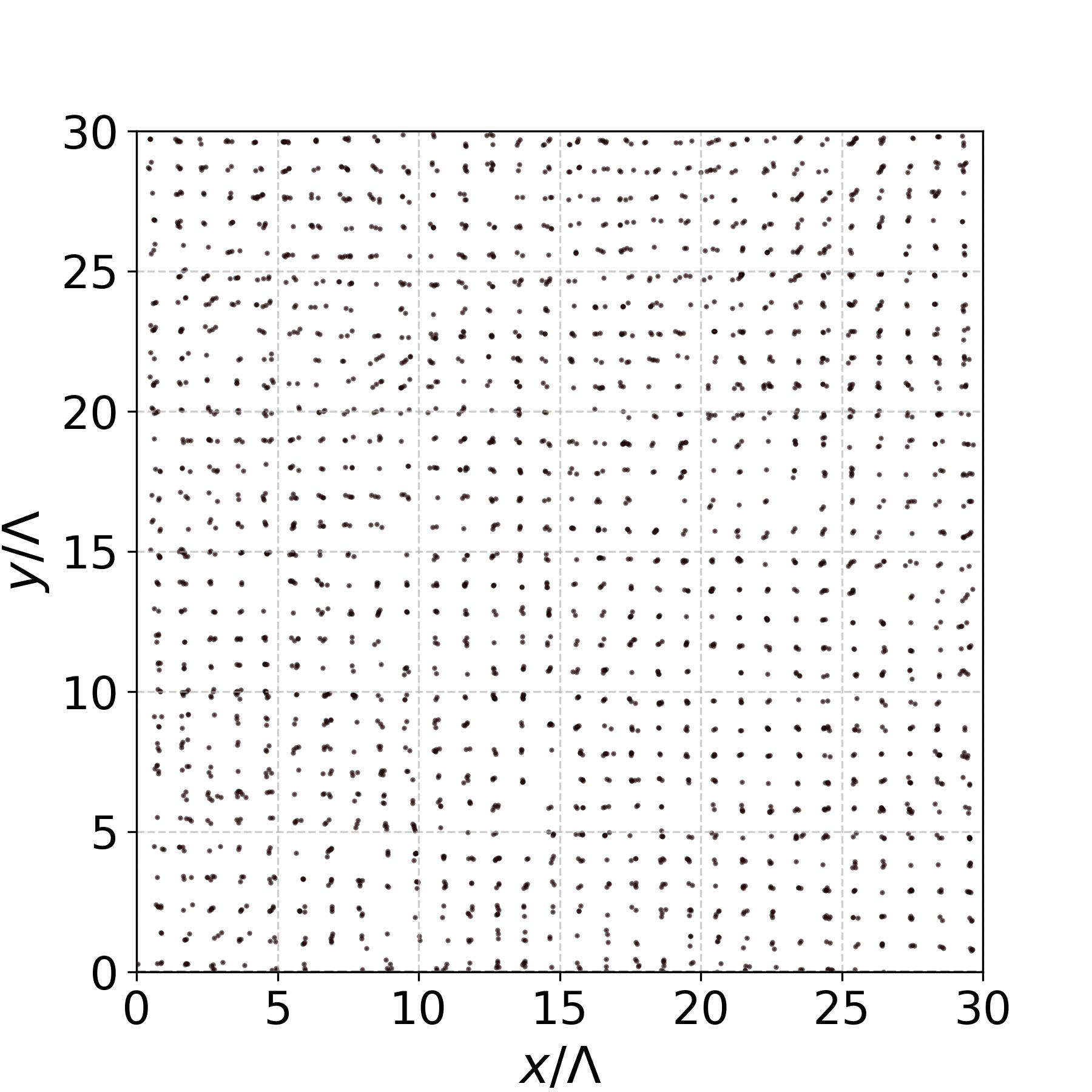}  
\includegraphics[width=0.49\textwidth]{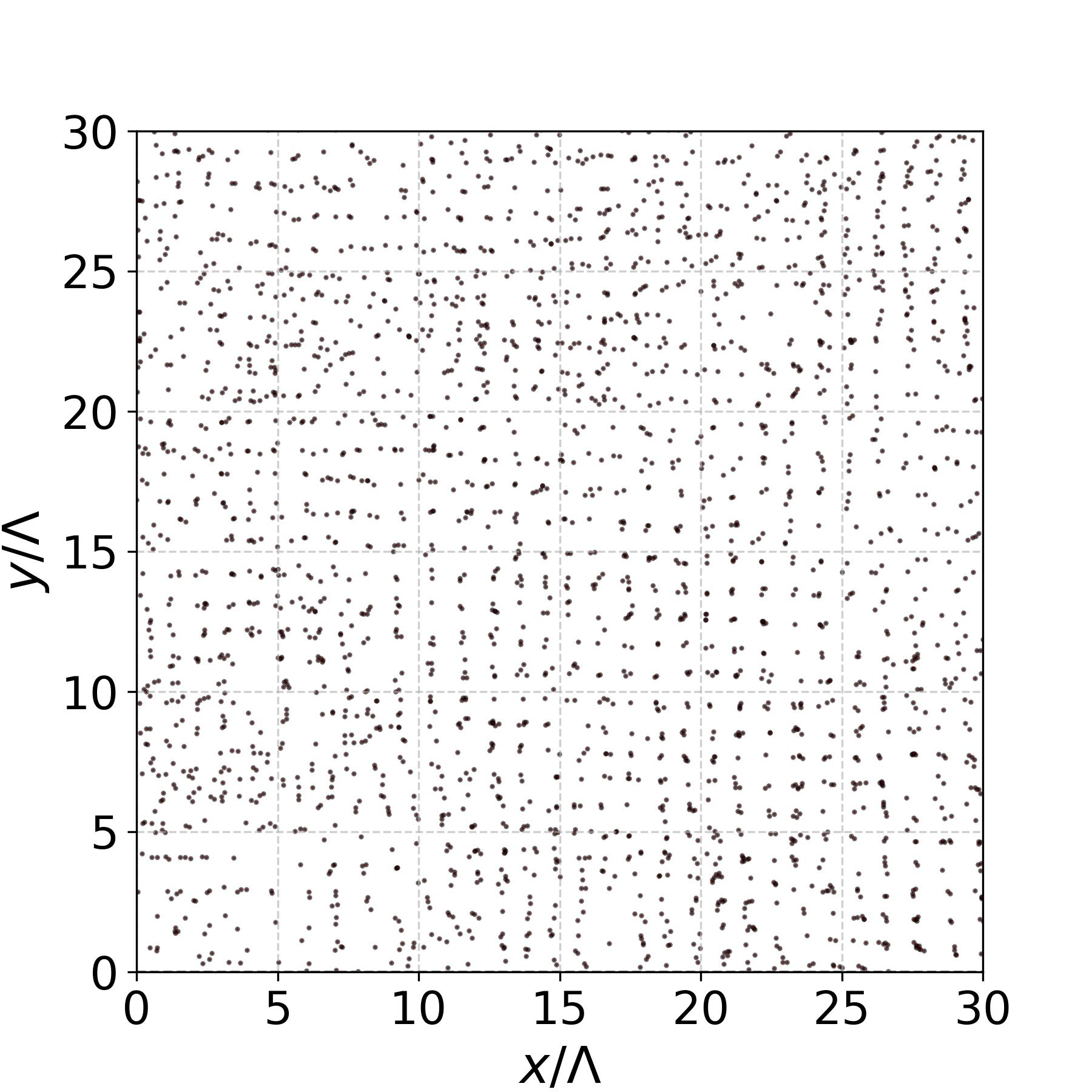} 
\caption{  
Left panel: zoom of the point distribution for the ICs of the large box at $z=20$.
Right panel:  zoom of the point distribution  at $z=7.7$.
 Length scales are expressed in units of the average distance between nearest neighbors $\Lambda$.} 
\label{point_distribution_zoom_large_box} 
\end{figure*}

{  
Fig.~\ref{point_distribution_zoom_large_box} shows a zoomed view of the two-dimensional point distribution in the large-box simulation at redshifts $z = 20$ and $z = 7.7$. In the early snapshot, the lattice structure---clearly visible in the ICs (see Fig.~\ref{adpd_heatmap_large_box_IC})---remains largely preserved, reflecting the small amplitude of the displacement field at such high redshift. By contrast, in the later snapshot, the imprint of the lattice has been completely erased by gravitational evolution. This transition is consistent with the behavior observed in the corresponding ADPD heatmaps.
 }

Despite the fading of the lattice imprint, the ADPD continues to display anisotropic features that are diffuse yet coherent. These structures are not stochastic artifacts but reflect patterns seeded by the initial correlated displacement field. If this field contains long-wavelength anisotropic modes, they can be amplified during gravitational evolution, appearing as directional excesses in the ADPD. To quantify this transition, we compute the angular variance $\sigma^2_\theta(r)$ as a function of pair separation $r$ and redshift $z$ (see Fig.~\ref{variance_poisson_plot_largebox}), and define a growth factor that captures the amplification of anisotropies over time (Fig.~\ref{variance_growth_factor_largebox}).

At $z \approx 7$, anisotropies seeded by the displacement field become dominant. Notably, the angular variance at this epoch exceeds that of a Poisson distribution with the same number of particles, confirming that the observed anisotropies are not statistical fluctuations, but systematic deviations induced by the coupling between an isotropic displacement field and a non-isotropic pre-initial lattice. Even in this case (see Fig.\ref{variance_poisson_plot_largebox}), 
{  as for the small-box case (see Fig.\ref{variance_growth_factor_smallbox}), }
we note that $\sigma^2_{\theta,\text{IC}}(r) \approx \text{const.}$, whereas the theoretical expectation scales as $\propto r^{-4}$ (see Eq.~\ref{sigma_theta_2D}).

As in the small-box case, we find that even in the large-box simulation the imposition of an isotropic displacement field onto a regular lattice introduces anisotropies across all scales. These are absent in the theoretical cosmological model, which assumes statistical isotropy beyond $\sim 10$--$20$ Mpc$/h$. In such a regime, the angular variance of the ADPD should be flat and consistent with that of a homogeneous, isotropic random field. However, our results show that:
\begin{itemize}
  \item the ADPD exhibits persistent anisotropies not predicted by theory,
  \item these anisotropies exceed those from Poisson noise,
  \item the angular variance increases with time, approximately tracking the linear growth factor.
\end{itemize}

These findings imply that even low-amplitude initial anisotropies can significantly influence the formation of large-scale structures, particularly in the linear and quasi-linear regimes where dynamics are still sensitive to ICs. While non-linear evolution at small scales eventually erases this memory, the early imprint of anisotropies remains relevant at scales $r>10 $ Mpc$/h$.

In summary, applying a correlated displacement field to a regular lattice does not yield a statistically isotropic particle distribution. The resulting anisotropies are systematic artifacts stemming from discreteness and finite-size effects. Although not part of the physical model, they can seed filamentary structures and bias the outcome of cosmological simulations.


\section{Conclusions}
\label{conclusions} 

Standard cosmological $N$-body simulations assume statistically isotropic initial conditions (ICs), consistent with the symmetries of the $\Lambda$CDM model. However, our analysis reveals that spurious anisotropies can arise when the displacement field --- although isotropic by construction --- is applied to a non-isotropic reference configuration such as a regular cubic lattice. These anisotropies manifest as coherent directional features in the evolved matter distribution, particularly in the quasi-linear regime, where gravitational dynamics are still sensitive to initial perturbations.

The Angular Distribution of Pairwise Distances (ADPD) proves to be a powerful diagnostic for identifying such anisotropies. Unlike angle-averaged statistics such as the power spectrum $P(k)$ and the two-point correlation function $\xi(r)$, which are inherently insensitive to directional structure, the ADPD directly probes angular coherence at fixed spatial scales. In our simulations, we find that the angular variance $\sigma^2_\theta(r)$ derived from the ADPD systematically exceeds that of a Poisson distribution and reveals anisotropic features that persist across a wide range of scales and redshifts.

These anisotropies are not merely transient or restricted to early times. Even when particles are displaced by more than the lattice spacing, the imprint of the initial configuration persists in the ADPD. This confirms that the anisotropic signature originates from the coupling of a structured (non-isotropic) pre-ICs with an isotropic displacement field, producing a statistically anisotropic realization. As such, these features should be interpreted as numerical artifacts, discreteness effects not present in the theoretical model, but capable of influencing the morphology and clustering properties of the evolved matter distribution.

{  
To demonstrate the impact of these artifacts, we compared two simulations: one with a small box ($L = 62.5$~Mpc$/h$, $N = 256^3$, $z_{\mathrm{IC}} = 49$) and another with a large box ($L = 1000$~Mpc$/h$, $N = 1024^3$, $z_{\mathrm{IC}} = 49$). In both cases, the amplitude of the initial displacement is smaller than the lattice spacing. As a result, the ADPD clearly retains the imprint of the underlying grid, indicating that spurious anisotropies persist when the particle displacements from their pre-initial positions are too small to effectively erase the lattice-induced anisotropy.
}

We have shown that the initial spurious anisotropies generated by the coupling of an isotropic displacement field with the non-isotropic lattice structure of the pre-initial particle distribution result in a complex angular pattern that is inherently anisotropic. The observed structure comprises multiple subcomponents with varying orientations and scales, which do not align with the symmetries of a perfect lattice but instead reflect a rich and irregular configuration. These anisotropies, introduced at the initial stage, persist throughout the simulation and are amplified by gravitational dynamics into filamentary structures that remain in the linear regime even at the present epoch. Since such features are not predicted by the theoretical model --- which assumes statistical isotropy --- they must be regarded as numerical artifacts arising from the discretization procedure.

{  
These results highlight the limitations of lattice-based  ICs in accurately modeling a statistically isotropic universe. They underscore the importance of employing diagnostics that are sensitive to directional features --- such as the ADPD   --- and motivate the adoption of alternative pre-initial configurations, particularly glass-like realizations. Glasses, constructed through reversed-sign gravitational relaxation, are specifically designed to minimize artificial symmetries and suppress long-range correlations, making them preferable choice for  pre-ICs  of cosmological simulations.

Quantifying the anisotropy induced by different types of pre-ICs, and understanding its evolution during cosmic structure formation, is a highly relevant and timely question. This issue warrants a systematic investigation, especially in view of its potential impact on late-time observables beyond the two-point statistics --- such as the bispectrum, measures of filamentarity, and topological descriptors.

A detailed analysis will be presented in a forthcoming work, which will involve the generation and analysis of simulations based on a variety of pre-initial configurations designed to reduce artificial symmetries. In particular,  alternatives to the standard simple cubic (SC) lattice, include: the following possibilities. 
(i) {Face-Centered Cubic (FCC):} A highly symmetric and dense lattice, often regarded as the densest sphere packing in 3D, potentially providing a better approximation to isotropy than the SC lattice.
(ii) {Body-Centered Cubic (BCC):} A configuration with an additional point at the center of each cube, offering higher packing efficiency than SC.
 (iii) {Hexagonal Close-Packed (HCP):} Composed of stacked hexagonal layers, this configuration is efficient in packing but exhibits reduced 3D symmetry.
 (iv) {Tetrahedral and Octahedral Lattices:} Based on repeating Platonic solids, these structures can introduce distinctive anisotropic signatures.
 (v) {Quasi-crystalline Lattices:} Aperiodic yet ordered structures that break translational invariance while maintaining long-range coherence.
 (vi) {Voronoi and Delaunay Tessellations:} Derived from regular or Poisson seed distributions, these methods are of particular interest in mesh-based modeling approaches.
  (vii) {Glass-like Configurations:} For instance, gravitational glasses, which are explicitly designed to be statistically isotropic and devoid of preferred directions.
}

{  Our main conclusion is that} the  role of anisotropies introduced by pre-ICs is non-negligible. They represent a subtle but significant source of systematic error that can impact the fidelity of cosmological simulations. The ADPD provides a novel and robust method for identifying, characterizing, and ultimately mitigating these artifacts, paving the way for more accurate modeling of large-scale structure formation.

{ 
As a final remark, we note that the ADPD is a versatile tool that can be applied to a variety of problems in cosmology and astrophysics. Its applications range from probing anisotropic features in the large-scale distribution of galaxies to identifying bar-like structures in individual galaxies, among other contexts.
}
\bigskip 

\begin{acknowledgments}
I wish to thank Benedikt Diemer for kindly providing the simulation data used in this study and for clarifying several aspects of the generation of ICs. I am also grateful to Marco Bruni, Roberto Capuzzo-Dolcetta,  Andrea Gabrielli, {  Marco Galoppo} and Michael Joyce  for insightful discussions and valuable comments that helped improve the interpretation and presentation of the results.
\end{acknowledgments}
\vspace{0.2cm}

\section*{Data Availability Statement} 
The data that support the findings of this article are openly available at: 

 {\tt https://www.astro.umd.edu/$\sim$diemer/erebos/}


%

\end{document}